\documentclass[pra, 
longbibliography, 
twocolumn, 
showpacs, 
nofootinbib, 
superscriptaddress, 
notitlepage]{revtex4-1}
\usepackage{amsmath}
\usepackage{amssymb,bm}
\usepackage{amsthm}

\newtheorem{innercustomgeneric}{\customgenericname}
\providecommand{\customgenericname}{}
\newcommand{\newcustomtheorem}[2]{%
  \newenvironment{#1}[1]
  {%
   \renewcommand\customgenericname{#2}%
   \renewcommand\theinnercustomgeneric{##1}%
   \innercustomgeneric
  }
  {\endinnercustomgeneric}
}

\newcustomtheorem{customthm}{Theorem}
\newcustomtheorem{definitionBob}{Definition}

\usepackage{color,dsfont} 
\usepackage{graphicx} 

\usepackage[bf, font=scriptsize]{caption}
\usepackage{ragged2e}
\usepackage[colorlinks=true,
hyperindex,
breaklinks,
linkcolor=blue,
urlcolor=blue,
citecolor=blue]{hyperref} 
\usepackage[normalem]{ulem}
\usepackage[capitalise]{cleveref}
\usepackage{mathrsfs}
\usepackage{subcaption}
\usepackage{mathtools}
\usepackage{float}

 \newtheorem{definition}{Definition}

\usepackage{titlesec}
\titleformat{\subsubsection}[runin]
{\normalfont\bfseries}{}{0pt}{}[.]
\titlespacing*{\subsubsection}{0pt}{5pt}{5pt}
\DeclareMathOperator{\relu}{relu}
\DeclareMathOperator{\diag}{diag}
\newtheorem*{rmk}{Remark}
\renewcommand{\epsilon}{\varepsilon} 

\newcommand{\ket}[1]{|#1\rangle}  
 
\newcommand{\codepar}[1]{\ensuremath{[\![#1]\!]}}

\newcommand{\License}[1]{\begingroup
  \renewcommand\thefootnote{}
  \footnote{#1}%
  \addtocounter{footnote}{-1}%
  \endgroup
}

\begin{document}

\title{Deep neural decoders for near term fault-tolerant experiments}

\author{Christopher Chamberland}
\email{c6chambe@uwaterloo.ca}
\affiliation{
    Institute for Quantum Computing and Department of Physics and Astronomy,
    University of Waterloo,
    Waterloo, Ontario, N2L 3G1, Canada
    }
\author{Pooya Ronagh}
\email{pooya.ronagh@uwaterloo.ca}
\homepage[\\]{https://github.com/pooya-git/DeepNeuralDecoder}
\affiliation{
    Institute for Quantum Computing and Department of Physics and Astronomy,
    University of Waterloo,
    Waterloo, Ontario, N2L 3G1, Canada
    }
\affiliation{
    Perimeter Institute for Theoretical Physics,
    Waterloo, Ontario, N2L 2Y5, Canada}
\affiliation{
    1QBit, 
    Vancouver, British Columbia, V6C 2B5, Canada}
\License{Both authors contributed equally to this work.}

\begin{abstract}
Finding efficient decoders for quantum error correcting codes adapted to
realistic experimental noise in fault-tolerant devices represents a significant
challenge. In this paper we introduce several decoding algorithms complemented
by deep neural decoders and apply them to analyze several fault-tolerant error
correction protocols such as the surface code as well as Steane and Knill error
correction. Our methods require no knowledge of the underlying noise model
afflicting the quantum device making them appealing for real-world experiments.
Our analysis is based on a full circuit-level noise model. It considers both
distance-three and five codes, and is performed near the codes pseudo-threshold
regime. Training deep neural decoders in low noise rate regimes appears to be a
challenging machine learning endeavour. We provide a detailed description of our
neural network architectures and training methodology. We then discuss both the
advantages and limitations of deep neural decoders. Lastly, we provide a
rigorous analysis of the decoding runtime of trained deep neural decoders and
compare our methods with anticipated gate times in future quantum devices.
Given the broad applications of our decoding schemes, we believe that the
methods presented in this paper could have practical applications for near term
fault-tolerant experiments.
\end{abstract}

\pacs{03.67.Pp}

\maketitle
\License{All files in this repository are released under the license agreement
provided in \tt{LICENSE.md}.}

\section{Introduction}
\label{sec:Intro}

Recently, significant progress has been made in building small quantum devices
with enough qubits allowing them to be potential candidates for several quantum
information experiments \cite{Intel, IBM, Rigetti, 2017arXiv170906678N}. 
Fault-tolerant quantum computing is one such avenue that has so far had a very 
limited experimental analysis \cite{Vuillot5qubit}.
Given the sensitivity of quantum devices to noise, quantum error correction will
be crucial in building quantum devices capable of reliably performing long
quantum computations. However, quantum error correction alone is insufficient
for achieving the latter goal. Since gates themselves can introduce additional
errors into a quantum system, circuits need to be constructed carefully in order
to prevent errors that can be corrected by the code from spreading into
uncorrectable errors. Fault-tolerant quantum computing provides methods for
constructing circuits and codes that achieves this goal. However this is
at the expense of a significant increase in the number of qubits and the space-
time overhead of the underlying circuits (although some methods use very few
qubits but have a large space-time overhead and vice-versa).

In recent years, several fault-tolerant protocols for both error correction and
universal quantum computation have been proposed, each with their own tradeoffs 
\cite{CS96, SteaneCSS, Knill04, Knill05, FMMC12, CR17v1, CR17v2, CB17, KLZ96, 
JL14, PR13, ADP14, Bombin14, YTC16, Yoder2017surfacecodetwist}. 
One important aspect of quantum error correcting codes is in finding efficient
decoders (the ability to identify the most likely errors which are afflicting
the system) that can optimally adapt to noise models afflicting quantum systems
in realistic experimental settings. Better decoders result in higher thresholds,
and can thus tolerate larger noise rates making near term devices more
accessible to fault-tolerant experiments. In \cite{CWBL16}, a hard decoding
algorithm was proposed for optimizing thresholds of concatenated codes afflicted
by general Markovian noise channels. In \cite{DP16,DP18}, tensor network
algorithms were used for simulating the surface code and obtaining efficient
decoders for general noise features. However, the above schemes are not adapted
to fault-tolerant protocols where gate and measurement errors plays a significant
role. Furthermore, some knowledge of the noise is required in order for the
decoding protocols to achieve good performance. This can be a significant
drawback since it is often very difficult to fully characterize the noise in
realistic quantum devices.

The above challenges motivate alternative methods for finding efficient decoders
which can offer improvements over more standard methods such as minimum weight
perfect matching for topological codes \cite{Edmonds65,FowlerAutotune} and
message passing for concatenated codes \cite{Poulin06}. One interesting idea
is using deep neural networks for constructing decoders which are both
efficient and can tolerate large noise rates. The hope is that even if the
underlying noise model is completely unknown, with enough experimental data,
deep neural networks could learn the probability density functions of the
different possible errors corresponding to the sequences of measured syndromes.
Note that due to measurement and gate errors, it
is often necessary to repeat the syndrome measurements in fault-tolerant protocols as will be explained in
\cref{sec:FaultTolerantProtocols}.

The first work in which machine learning was used for decoding was in a paper by
Torlai and Melko \cite{TorlaiNeural}. In this paper, a Boltzmann machine was
trained to correct phase-flip errors of a 2-dimensional toric code.
Krastanov and Jiang obtained a neural network decoder applicable to general
stabilizer codes and applied it to the 2-D toric code obtaining a higher 
code-capacity threshold than previous results. Varsamopoulos, Criger and Bertels
used a feed-forward neural network to decode the surface code \cite{CrigerNN17}.
They also applied their decoding scheme to the distance three surface code under
a full circuit level noise model. Baireuther, O'Brien, Tarasinski and Beenakker
used a recurrent neural network that could be trained with experimental data
\cite{Baireuther2018machinelearning}. They applied their decoding scheme to
compare the lifetime of qubits encoded in a distance-three surface code. The
analysis was based on a full circuit level noise model, albeit with a modified
CNOT gate error model. 
Breuckmann and Ni \cite{BreuckmannNiNeural} gave a scalable neural
decoder applicable to higher dimensional codes by taking advantage of the fact
that these codes have local decoders. To our knowledge, these methods could not
be applied to codes of dimensions less than four.
Lastly, while preparing the updated version of our manuscript, Maskara, Kubica and Jochym-O'Connor used neural-network decoders to study the code capacity thresholds of color codes \cite{TomasColorCodeNeural}.

Despite the numerous works in using neural networks for decoding, there are
still several open questions that remain:

\begin{enumerate}
    \item What are the fastest possible decoders that can be achieved using
    neural networks and how does the decoding time compare to gate times in
    realistic quantum devices?

    \item Can neural networks still offer good performance beyond distance three
    codes in a full circuit level noise model regime? If so, what are the
    limitations?

    \item How well do neural networks perform near and below typical thresholds
	of fault-tolerant schemes under full circuit level noise models? 
\end{enumerate}

In this paper we aim to address the above questions. We apply a plethora of
neural network methods to analyze several fault-tolerant error correction
schemes such as the surface code as well as the CNOT-exRec gate using Steane
error correction (EC) and Knill-EC, and consider both distance-three
\textit{and} distance-five codes. We chose the CNOT-exRec circuit since (in most
cases) it limits the threshold of the underlying code when used with Steane and
Knill-EC units \cite{AGP06}. Our analysis is done using a full circuit level
noise model. Furthermore our methods are designed to work with experimental
data; i.e. no knowledge of the underlying noise model is required.
 
Lastly, we provide a rigorous analysis of the decoding times of the neural
network decoders and compare our results with expected gate delays in future
superconducting quantum devices. We suspect that even though inference from a
trained neural network is a simple procedure comprising only of matrix
multiplications and arithmetic operations, state-of-the-art parallel processing
and high performance computing techniques would need to be employed in order for
the inference to provide a reliable decoder given the anticipated gate times in
future quantum devices.

The deep neural decoders (DND) we design in this paper assist a \emph{baseline}
decoder. For the baseline decoders, we will use both lookup table and
\emph{naive} decoding schemes which will be described in
\cref{sec:FaultTolerantProtocols}. The goal
of the deep neural decoder is to determine whether to add logical corrections to
the corrections provided by the baseline decoders. Although the lookup table
decoder is limited to small codes, the naive decoder can efficiently be
implemented for arbitrary distance codes.

We stress that to offer a proper analysis of the performance of neural network
decoders, the neural network should be trained for \textit{all} considered
physical error rates. We believe that from an experimental point of view, it is
not realistic to apply a network trained for large physical error rates to lower
rate noise regimes. The reason is simply that the network will be trained based
on the hardware that is provided by the experimentalist. If the experimentalist
tunes the device to make it noisier so that fewer non-trivial training samples
are provided to the neural network, the decoder could be fine tuned to a
different noise model than what was present in the original device. As will be
shown, training neural networks at low error rates is a difficult task for
machine learning and definitely an interesting challenge.

Our goal has been to compose the paper in such a way that makes it accessible
to both the quantum information scientists and machine learning experts.
The paper is structured as follows. 

In \cref{sec:FaultTolerantProtocols} we
begin by providing a brief review of stabilizer codes followed by the 
fault-tolerant error correction criteria used throughout this paper as well as the
description of our full circuit level noise model. In
\cref{subsec:RotatedSurfaceCode}, we review the rotated surface code and provide
a new decoding algorithm that is particularly well adapted for deep neural
decoders. In \cref{subsec:SteaneEC,subsec:KnillEC}, we review the Steane
and Knill fault-tolerant error correction methods and give a description of the
distance-three and five color codes that will be used in our analysis of Steane
and Knill-EC. In \cref{subsec:NaiveDecoder} we give a description of the naive
decoder and in \cref{sec:LookpNaiveComplexity} we discuss the decoding
complexity of both the lookup table and naive decoders.

\cref{sec:DND} focuses on the deep neural decoders constructed, trained and
analyzed in this paper. In \cref{subsec:DLOverview} we give an overview of deep
learning by using the application of error decoding as a working example. We
introduce three widely used architectures for deep neural networks: (1) simple
feedforward networks with fully connected hidden layers, (2) recurrent neural
networks,  and (3) convolutional neural networks. We introduce hyperparameter
tuning as a commonly used technique in machine learning and an important
research tool for machine learning experts. In
\cref{subsec:DNDSteaneKnill,subsec:DNDSurfaceCode} we introduce the deep neural
network architectures we designed for decoding the CNOT-exRec circuits in the
case of Steane- and Knill-EC, and for multiple rounds of EC in the case of
the rotated surface code.

In \cref{sec:NumericalExperiments} we provide our numerical results by
simulating the above circuits under a full circuit level depolarizing noise
channel, and feeding the
results as training and test datasets for various deep neural decoders.

Finally, in \cref{sec:performance} we address the question of practical
applicability of deep neural decoders in their inference mode for fault-tolerant
quantum error correction. We will address several hardware and software
considerations and recommend a new development in machine learning known as
network quantization as a suitable technology for decoding quantum error 
correcting codes.

\section{Fault-tolerant protocols}
\label{sec:FaultTolerantProtocols}

In this section we will describe the fault-tolerant protocols considered in this
paper. The surface code will be described in \cref{subsec:RotatedSurfaceCode}
while Steane and Knill error correction will be described in
\cref{subsec:SteaneEC,subsec:KnillEC}. For each protocol, we will also describe
the baseline decoder used prior to implementing a deep neural decoder (DND).
Since we are focusing on near term fault-tolerant experiments, we will first
describe decoding schemes using lookup tables which can be implemented extremely
quickly for small distance codes. In \cref{sec:NumericalExperiments} we will
show that the lookup table decoding schemes provide very competitive
pseudo-thresholds. With existing computing resources and the code families
considered in this paper, the proposed decoders can be used for distances $d \le
7$. For example, the distance-nine color code would require 8.8 exabytes of
memory to store the lookup table. Lastly,
in \cref{subsec:NaiveDecoder} we will describe a naive decoder which is scalable
and can be implemented efficiently while achieving competitive logical failure
rates when paired with a deep neural decoder.

Before proceeding, and in order to make this paper as self contained as
possible, a few definitions are necessary. First, we define the $n$-qubit Pauli
group $\mathcal{P}_{n}^{(1)}$ to be group containing $n$-fold tensor products of
the identity $I$ and Pauli matrices $X,Y$ and $Z$. The weight of an error $E \in
\mathcal{P}_{n}^{(1)}$ ($\text{wt}(E)$) is the number of non-identity Pauli
operators in its decomposition. For example, if $E = IXYIZ$, then $\text{wt}(E)
= 3$.

A $\codepar{n,k,d}$ quantum error correcting code, which encodes $k$ logical
qubits into $n$ physical qubits and can correct $t = \lfloor (d-1)/2 \rfloor$
errors, is the image space $C_q$ of the injection $\xi: \mathcal{H}_2^k \to C_q
\subset \mathcal{H}_{2}^{n}$ where $\mathcal{H}_2$ is the two-dimensional
Hilbert space. Stabilizer codes are codes $C_q$ which form the unique subspace
of $\mathcal{H}_{2}^{n}$ fixed by an Abelian stabilizer group $\mathcal{S}
\subset \mathcal{P}_{n}^{(1)}$ such that for any $s \in \mathcal{S}$ and any
codeword $\ket{c} \in C_q$, $s\ket{c} = \ket{c}$. Any $s \in \mathcal{S}$ can
be written as $s = g_{1}^{p_1} \cdots g_{n-k}^{p_{n-k}}$ where the stabilizer
generators $g_i$ satisfy $g_i^{2}=I$ and mutually commute. Thus $S = \ \langle
g_1, \cdots , g_{n-k} \rangle$. We also define $N(\mathcal{S})$ to be the
normalizer of the stabilizer group. Thus any non-trivial logical operator on
codewords belongs to $N(\mathcal{S}) \setminus \mathcal{S}$. The distance $d$ of
a code is the lowest weight operator $P \in N(\mathcal{S}) \setminus
\mathcal{S}$. For more details on stabilizer codes see
\cite{Gottesman97, Gottesman2010}.

For a given stabilizer group $\mathcal{S} = \ \langle g_1, \cdots , g_{n-k}
\rangle$, we define the error syndrome $s(E)$ of an error $E$ to be a bit string
of length $n-k$ where the $i$-th bit is zero if $[E, g_i] = 0$ and one
otherwise. We say operators $E$ and $E'$ are logically equivalent, written as $E
\sim E'$, if and only if $E' \propto gE$ for some $g \in \mathcal{S}$.

The goal of an error correction protocol is to find the most likely error
$E$ afflicting a system for a given syndrome measurement $s(E)$. However, the gates used to
perform a syndrome measurement can introduce more errors into the system. If not
treated carefully, errors can spread leading to higher weight errors which are
non longer correctable by the code. In order to ensure that correctable errors
remain correctable and that logical qubits have longer lifetimes than their 
un-encoded counterpart (assuming the noise is below some threshold), 
an error correction protocol needs to be implemented fault-tolerantly. More
precisely, an error correction protocol will be called fault-tolerant if the
following two conditions are satisfied \cite{AGP06, Gottesman2010, CB17}:

\begin{definition}[Fault-tolerant error correction]
	
  For $t = \lfloor (d-1)/2\rfloor$, an error correction protocol using a
  distance-$d$ stabilizer code $C$ is $t$-fault-tolerant if the following two
  conditions are satisfied:
	\begin{enumerate}
    \item For an input codeword with error of weight $s_{1}$, if $s_{2}$ faults
    occur during the protocol with $s_{1} + s_{2} \le t$, ideally decoding the
    output state gives the same codeword as ideally decoding the input state.
    \item For $s$ faults during the protocol with $s \le t$, no matter how many
    errors are present in the input state, the output state differs from a
    codeword by an error of at most weight $s$.
	\end{enumerate}
	\label{Def:FaultTolerantDef}
\end{definition}

A few clarifications are necessary. By ideally decoding, we mean performing
fault-free error correction. In the second condition of
\cref{Def:FaultTolerantDef}, the output state can differ from \textit{any}
codeword by an error of at most weight $s$, not necessarily by the same codeword
as the input state. It is shown in \cite{AGP06, CB17} that both conditions are
required to guarantee that errors do not accumulate during multiple error
correction rounds and to ensure that error correction extends the lifetime of
qubits as long as the noise is below some threshold.

\begin{figure}
\center
\includegraphics[scale=.3]{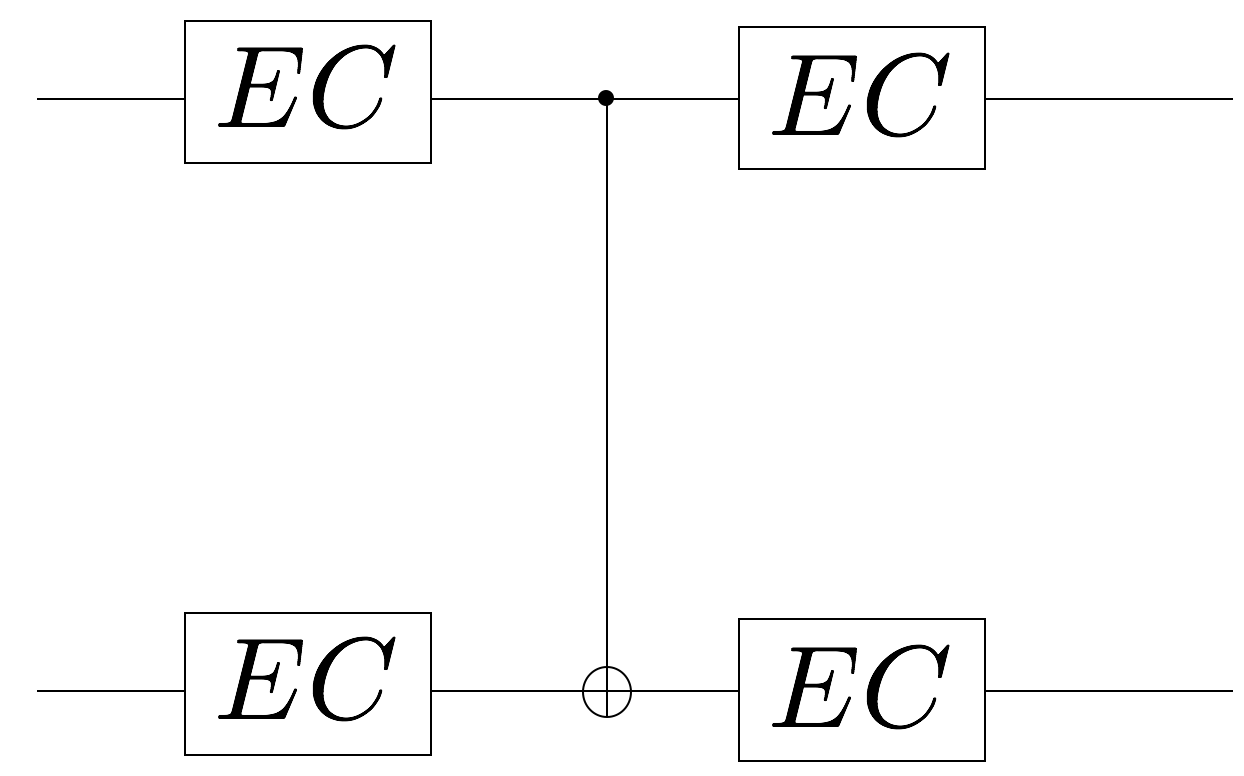}
\caption{Illustration of an extended rectangle (exRec) for a logical CNOT gate.
The EC box consists of performing a round of fault-tolerant error correction.
The error correction rounds prior to applying the logical CNOT gate are
referred to as leading-EC's (LEC) and the error correction rounds after the CNOT
are referred to as trailing-EC's (TEC).}
\label{fig:ExRecFig}
\end{figure}

In this paper we focus on small distance codes which could potentially be
implemented in near term fault-tolerant experiments. When comparing the
performance of fault-tolerant error correction protocols, we need to consider a
full \emph{extended rectangle} (exRec) which consists of leading and trailing
error correction rounds in between logical gates. Note that this also applies
to topological codes. An example of an exRec is given in \cref{fig:ExRecFig}.
We refer the reader to \cite{AGP06, AGP08} for further details on exRec's.

In constructing a deep neural decoder for a fault-tolerant error correction
protocol, our methods will be devised to work for unknown noise models which
would especially be relevant to experimental settings. However, throughout
several parts of the paper, we will be benchmarking our trained decoder against
a full circuit level depolarizing noise channel since these noise processes can
be simulated efficiently by the Gottesman-Knill theorem \cite{Gottesman98}. A
full circuit level depolarizing noise model is described as follows:

\begin{enumerate}
  \item With probability $p$, each two-qubit gate is followed by a two-qubit
  Pauli error drawn uniformly and independently from 
  $\{I,X,Y,Z\}^{\otimes2}\setminus \{I\otimes I\}$.
  \item With probability $\frac{2p}{3}$, the preparation of the $\ket{0}$ state
  is replaced by $\ket{1}=X\ket{0}$. Similarly, with probability $\frac{2p}{3}$,
  the preparation of the $\ket{+}$ state is replaced by $\ket{-}=Z\ket{+}$.
  \item With probability $\frac{2p}{3}$, any single qubit measurement has its
  outcome flipped.
  \item Lastly, with probability $p$, each resting qubit location is
  followed by a Pauli error drawn uniformly and independently from $\{ X,Y,Z
  \}$.
\end{enumerate}

\subsection{Rotated surface code}
\label{subsec:RotatedSurfaceCode}

\begin{figure}
\center
\includegraphics[scale=.3]{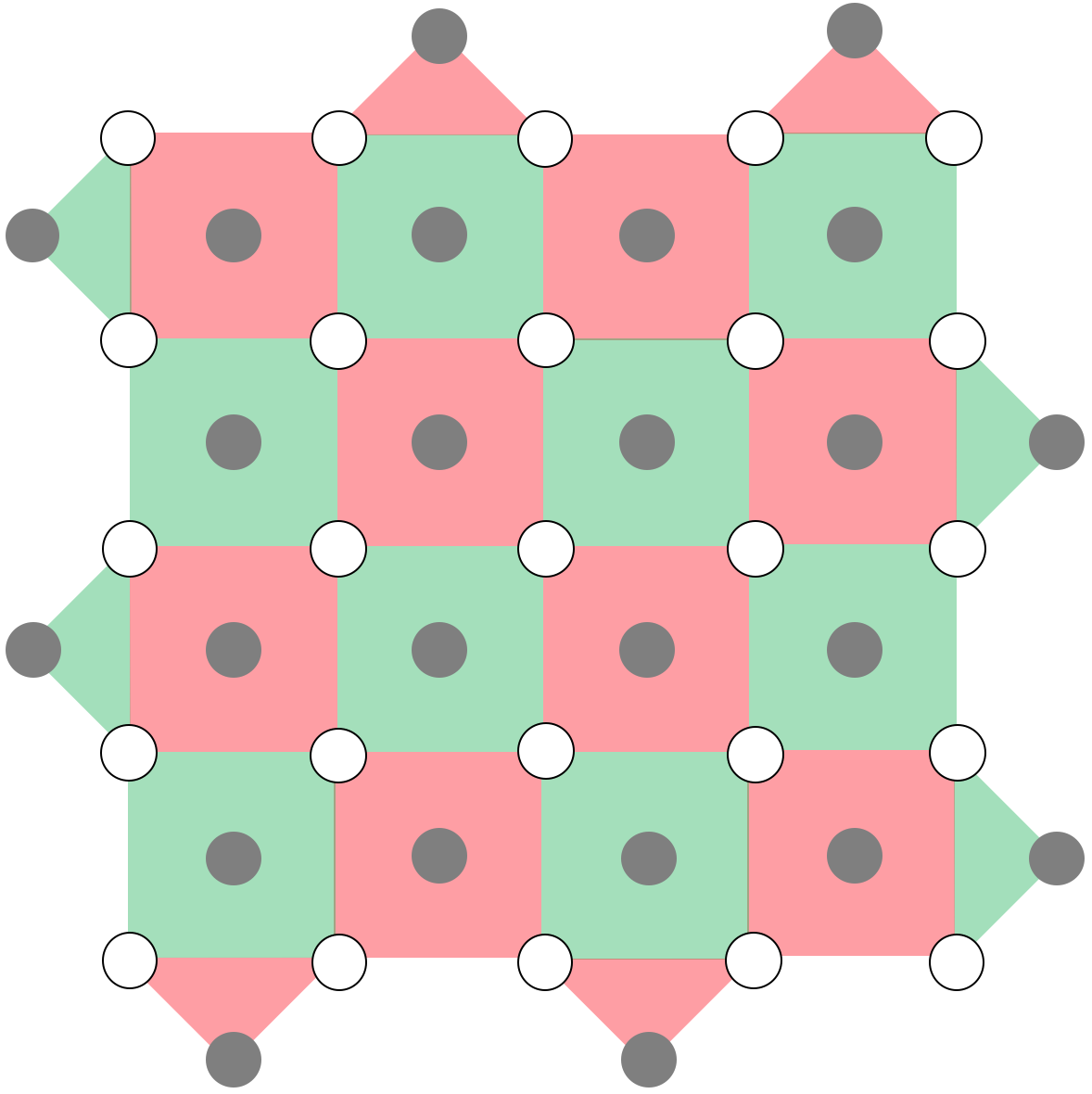}
\caption{Illustration of the $d=5$ rotated surface code. Data qubits are located
at the white circles and the ancilla qubits used to measure the stabilizers are
located on the black circles of the lattice. Green squares measure the $Z$
stabilizers and red squares measure $X$ stabilizers. }
\label{fig:Distance5SurfaceCodeLattice}
\end{figure}

\begin{figure}
\centering
\begin{subfigure}{0.35\textwidth}
\includegraphics[width=\textwidth]{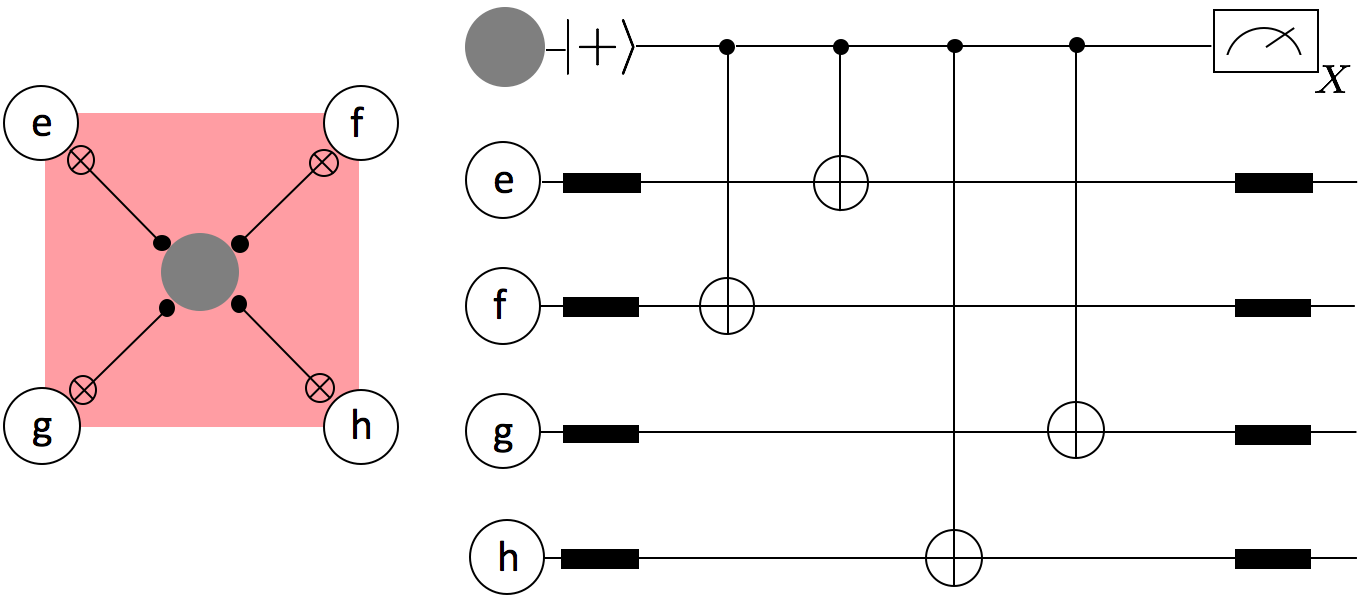}
\caption{}
\label{fig:XstabSurfaceCode}
\end{subfigure}
\begin{subfigure}{0.35\textwidth}
\includegraphics[width=\textwidth]{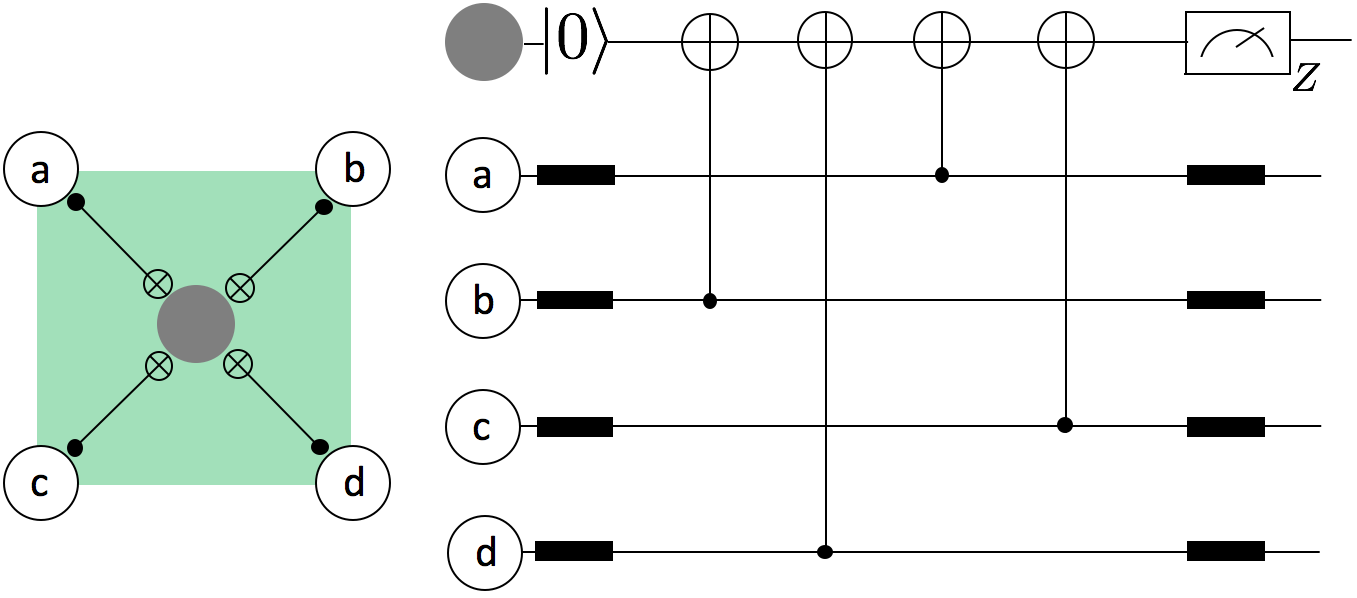}
\caption{}
\label{fig:ZstabSurfaceCode}
\end{subfigure}
\caption{\cref{fig:XstabSurfaceCode} illustrates the circuit used to measure the
stabilizer $X^{\otimes 4}$ and \cref{fig:ZstabSurfaceCode} illustrates the
circuit used to measure the stabilizer $Z^{\otimes 4}$. As can be seen, a full
surface code measurement cycle is implemented in six time steps.}
\label{fig:CircuitsForMeasuringAncillas}
\end{figure}

In this section we focus on the rotated surface code 
\cite{BK98, DKLP02, FMMC12, KITAEV97Surface, TS14, PhysRevLett.90.016803}. The
rotated surface code is a $\codepar{d^{2},1,d}$ stabilizer code with qubits
arranged on a 2-dimensional lattice as shown in
\cref{fig:Distance5SurfaceCodeLattice}. Any logical $X$ operator has $X$
operators acting on at least $d$ qubits with one $X$ operator in each row of the
lattice involving an even number of green faces. Similarly, any logical $Z$
operator has $Z$ operators acting on at least $d$ qubits with one $Z$ operator
in every column of the lattice involving an even number of red faces.

It is possible to measure all the stabilizer generators by providing only local
interactions between the data qubits and neighbouring ancilla qubits. The
circuits used to measure both $X$ and $Z$ stabilizers are shown in
\cref{fig:CircuitsForMeasuringAncillas}. Note that all stabilizer generators
have weight two or four regardless of the size of the lattice.

Several decoding protocols have been devised for topological codes. Ideally, we
would like decoders which have extremely fast decoding times to prevent errors
from accumulating in hardware during the classical processing time while also
having very high thresholds. The most common algorithm for decoding topological
codes is Edmond's perfect matching algorithm (PMA) \cite{Edmonds65}. Although
the best know thresholds for topological codes under circuit level noise have
been achieved using a slightly modified version of PMA \cite{FowlerAutotune},
the decoding algorithm has a worst case complexity of $\mathcal{O}(n^3)$. Recent
progress has shown that minimum weight perfect matching can be performed in
$\mathcal{O}(1)$ time on average given constant computing resources per unit
area on a 2D quantum computer \cite{FowlerO1}. With a single processing element
and given $n$ detection events, the runtime can be made $\mathcal{O}(n)$
\cite{FowlerSingleProcessor}. Renormalization group (RG)
decoders have been devised that can achieve $\mathcal{O}(\log{n})$ decoding
times under parallelization \cite{DuclosRG1,DuclosRG2,BravyiRG}. However such
decoders typically have lower thresholds than PMA. Wootton and Loss
\cite{WootonRG} use a Markov chain Monte Carlo method to obtain near optimal
code capacity noise thresholds of the surface code at the cost of slower
decoding times compared to other schemes. Recently, Delfosse and Nickerson
\cite{DF17} have devised a near linear time decoder for topological codes that
achieves thresholds slightly lower than PMA for the 2-dimensional toric code.

Here we construct a decoder for the surface code which has extremely fast
decoding times and achieves high pseudo-thresholds which will serve as a core
for our deep neural decoder construction of \cref{sec:DND}. Our decoder will be
based on a lookup table construction which 
could be used for distances $d \le 7$. Before describing the
construction of the lookup table, we point out that a single fault on the second
or third CNOT gates in \cref{fig:XstabSurfaceCode,fig:ZstabSurfaceCode} can
propagate to a data qubit error of weight-two. Thus for a surface code that can
correct $t = 2d+1$ errors, a correction $E'$ for an error $E$ resulting from $t$
faults, with $E' \sim E$, must be used when the syndrome $s(E)$ is measured. In other words, the minimum weight correction must not always be used for errors that result from faults occurring at the CNOT gates mentioned above. 

With the above in mind, the lookup table is constructed a follows. For every $1
\le m \le 2^{d^{2}-1}$,  use the lowest weight error $E' \sim E$ such that
converting the bit string $s(E)$ to decimal results in $m$. If $E$ is an error
that results from $v \le t=2d+1$ faults with $\text{wt}(E) > t$, then use $E'
\sim E$ instead of the lowest weight error corresponding to the syndrome $s(E)$.
Note that for this method to work, all errors $E$ with $\text{wt}(E) \le t$ must
have distinct syndromes from errors $E'$ that arise from $v \le t$ faults with
$\text{wt}(E') > t$. However this will always be the case for surface codes with
the CNOT ordering chosen in \cref{fig:CircuitsForMeasuringAncillas}.

Note that with the above construction, after measuring the syndrome $s$,
decoding simply consists of converting $s$ to decimal (say $m$) and correcting
by choosing the error on the $m$'th row of the lookup table. Note however that
this method is not scalable since the number of syndromes scales exponentially
with the code distance.

Lastly, the decoding scheme as currently stated is not fault-tolerant. The
reason is that if syndromes are measured only once, in some cases it would be
impossible to distinguish data qubit errors from measurement errors. For
instance, a measurement error occurring when measuring the green triangle of the
upper left corner of \cref{fig:Distance5SurfaceCodeLattice} would result in the
same syndrome as an $X$ error on the first data qubit. However, with a simple
modification, the surface code decoder can be made fault-tolerant. For 
distance 3 codes, the syndrome is measured three times and we decode using
the majority syndrome. If there are no majority syndromes, the syndrome from the
last round is used to decode. For instance, suppose that the syndromes
$s_1,s_2,s_2$ were obtained, then the syndrome $s_2$ would be used to decode
with the lookup table. However if all three syndromes $s_1,s_2,s_3$ were
different, then $s_3$ would be used to decode with the lookup table. This
decoder was shown to be fault-tolerant in \cite{CCBT17}.

For higher distance codes, we use the following scheme. First, we define the
counter $n_{\text{diff}}$ (used for keeping track of changes in consecutive
syndrome measurements) as

\vspace{10px}

\fbox{\begin{minipage}{23em}
\underline{\textbf{Decoding protocol -- update rules:}} 

Given a sequence of consecutive syndrome measurement outcomes $s_{k}$ and
$s_{k+1}$:
\begin{enumerate}
\item If $n_{\text{diff}}$ did not increase in the previous round, and 
$s_{k}\neq s_{k+1}$, increase $n_{\text{diff}}$ by one.
\end{enumerate}
\end{minipage}}

\vspace{10px}

We also define $E(s)$ to be the correction obtained from either the lookup table
decoder or naive decoder (described in section \cref{subsec:NaiveDecoder}) using
the syndrome $s$. With the above definition of $n_{\text{diff}}$, the decoding
protocol for a code that can correct any error $E$ with $\text{wt}(E) \le t =
\lfloor \frac{(d-1)}{2} \rfloor$ is implemented as

\vspace{10px}

\fbox{\begin{minipage}{23em}
\underline{\textbf{Decoding protocol  -- corrections:}} 

Set $n_{\text{diff}}=0$.

Repeat the syndrome measurement.

Update $n_{\text{diff}}$ according to the update rule above.
 
   \begin{enumerate}
   \item If at anytime $n_{\text{diff}}=t$, repeat the syndrome measurement
   yielding the syndrome $s$. Apply the correction $E(s)$.
   \item If the same syndrome $s$ is repeated $t-n_{\text{diff}}+1$ times in a
   row, apply the correction $E(s)$.
    \end{enumerate}  
\end{minipage}}

\vspace{10px}

Note that in the above protocol, the number of times the syndrome is repeated is
non-deterministic. The minimum number of syndrome measurement repetitions is
$t+1$ while in \cite{CB17} it was shown that the maximum number of syndrome
measurement repetitions is $\frac{1}{2}(t^{2}+3t+2)$. Further, a proof that the
above protocol satisfies both fault-tolerance criteria in
\cref{Def:FaultTolerantDef} is given in Appendix A of \cite{CB17}.

\subsection{Steane error correction}
\label{subsec:SteaneEC}

Calderbank-Shor-Steane (CSS) codes \cite{CS96,SteaneCSS} are quantum error
correcting codes which are constructed from two classical error correcting codes
$C_1$ and $C_2$ where $C_1^{\perp} \subseteq C_2$. The last condition guarantees
that by choosing the $X$ and $Z$ stabilizers to correspond to the parity check
matrices $H_X$ and $H_Z$ of $C_1$ and $C_2$, all operators in $H_X$ will commute
with those of $H_Z$.  Additionally, CSS codes are the only codes such that a
transversal CNOT gate performs a logical CNOT.

\begin{figure}
\centering
\begin{subfigure}{0.22\textwidth}
\includegraphics[width=\textwidth]{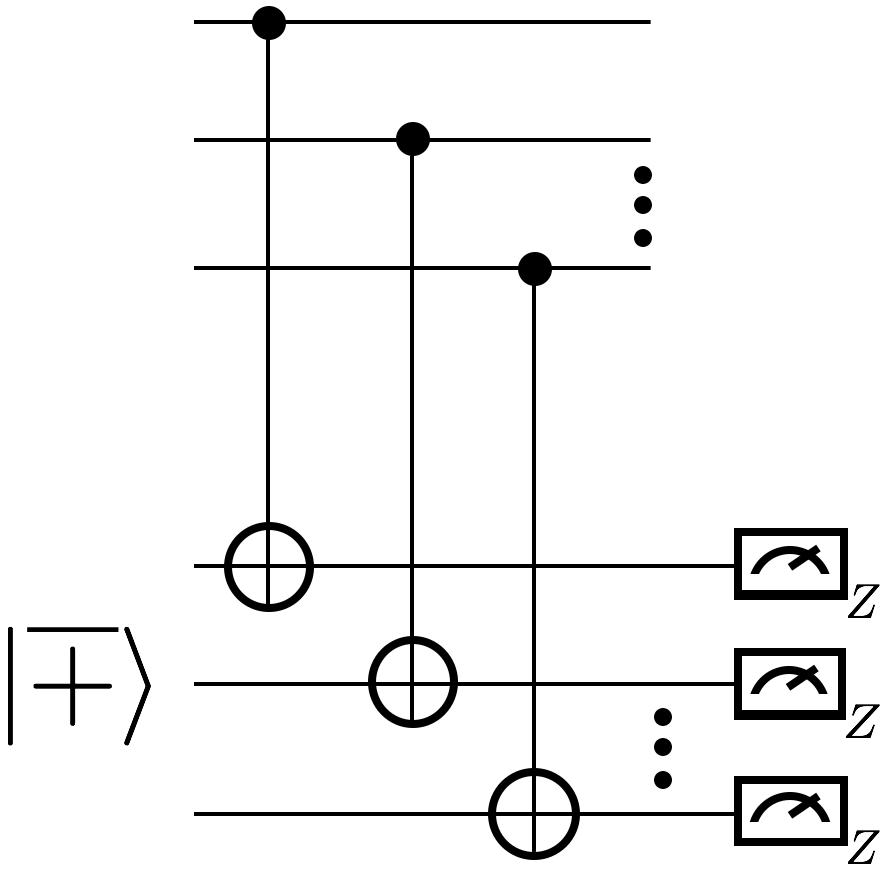}
\caption{}
\label{fig:XMeasCircuitSteane}
\end{subfigure}
\begin{subfigure}{0.22\textwidth}
\includegraphics[width=\textwidth]{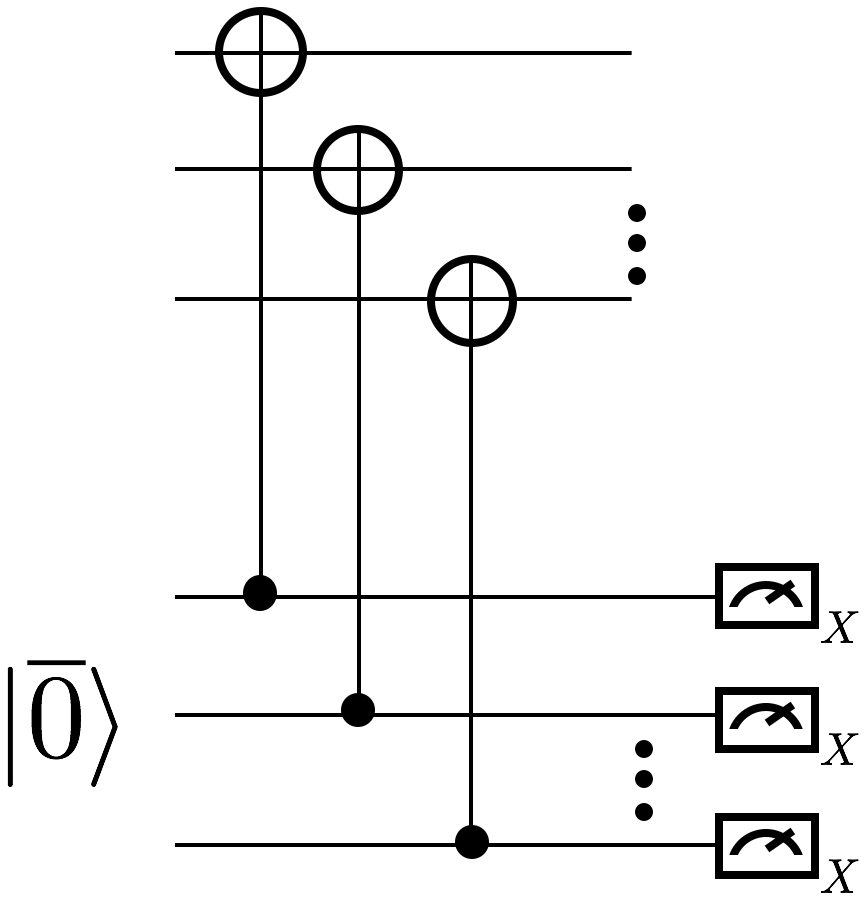}
\caption{}
\label{fig:ZMeasCircuitSteane}
\end{subfigure}
\caption{Circuits for measuring $X$ and $Z$ stabilizers in Steane-EC. The
circuit in \cref{fig:XMeasCircuitSteane} measures bit-flip errors whereas the
circuit in \cref{fig:ZMeasCircuitSteane} measures phase-flip errors. Note that
the first block consists of the data qubits encoded in a CSS code. The states
$\ket{\overline{0}}$ and $\ket{\overline{+}}$ represent logical $\ket{0}$ and
$\ket{+}$ states encoded in the same CSS code used to protect the data.}
\label{fig:CircuitSteaneXZMeas}
\end{figure}

Steane error correction \cite{Steane97} takes advantage of properties of CSS
codes to measure the $X$ and $Z$ stabilizers using transversal CNOT gates. To
see this, consider the circuit in \cref{fig:XMeasCircuitSteane}. The transversal
CNOT gate between the encoded data block $\ket{\overline{\psi}}$ and ancilla
$\ket{\overline{+}}$ acts trivially (i.e.
$\overline{\text{CNOT}}\ket{\overline{\psi}}\ket{\overline{+}} =
\ket{\overline{\psi}}\ket{\overline{+}}$). However, any $X$ errors afflicting
the data block would then be copied to the ancilla state. Furthermore, CSS codes
have the property that transversally measuring the codeword $\ket{\overline{+}}$
in the absence of errors would result in a codeword of $C_1$ chosen
uniformly at random. If $X$ errors are present on the codeword
$\ket{\overline{+}}$, then the transversal measurement would yield the classical
codeword $e+f+g$. Here, $(e|0)$ (written in binary symplectic form) are the $X$
errors on the data qubits, $(f|0)$ are the $X$ errors that arise during the
preparation of the $\ket{\overline{+}}$ state and $(g|0)$ are bit-flip errors
that arise during the transversal measurement. Applying the correction
$X_{e}X_{f}X_{g}$ on the data would result in an $X$ error of weight $f+g$. An
analogous argument can be made for $Z$ errors using the circuit of
\cref{fig:ZMeasCircuitSteane} (note that in this case we measure in the
$X$-basis which maps $C_1 \rightarrow C_2$ and $Z \rightarrow X$).

\begin{figure}
\center
\includegraphics[scale=.33]{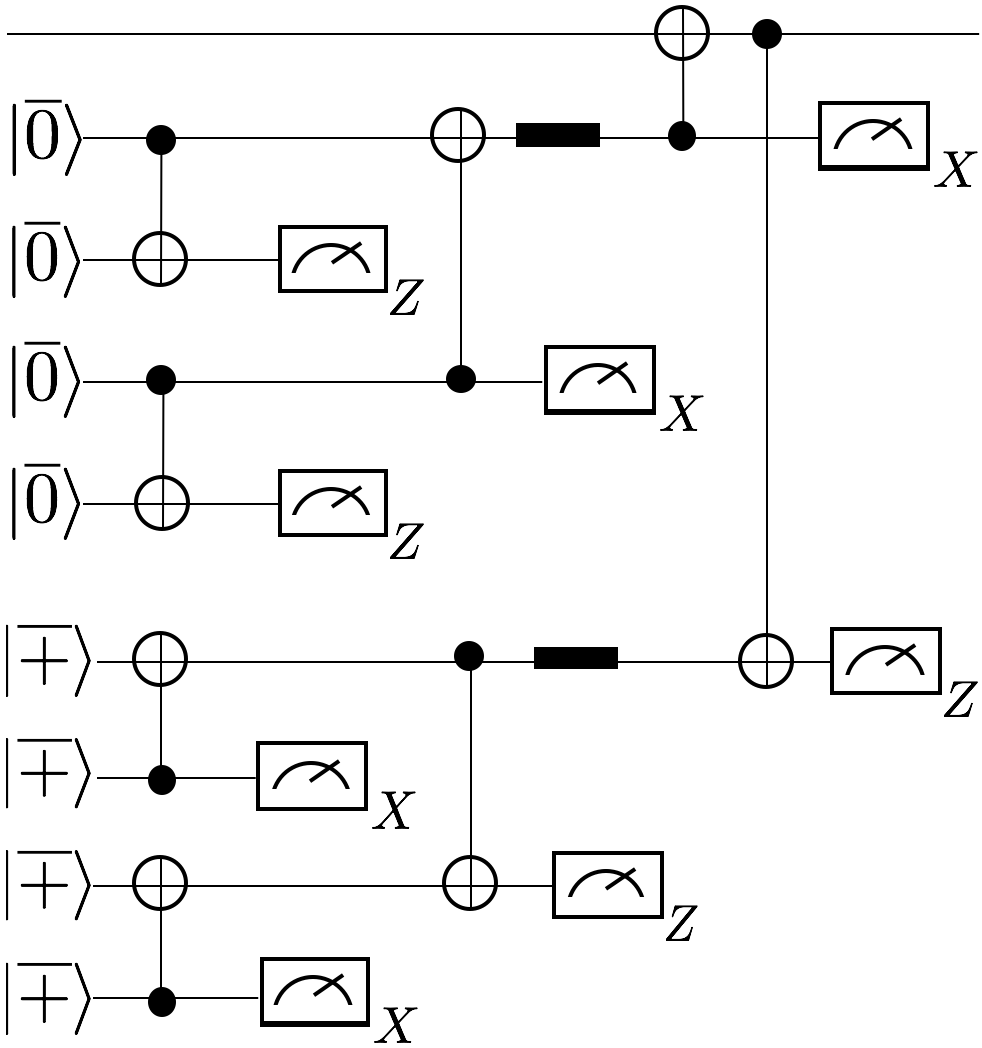}
\caption{Full Steane error correction circuit. Each line represents encoded data
qubits and all CNOT gates  and measurements are performed transversally. The
circuits used to prepare the encoded $\ket{\overline{+}}$ and
$\ket{\overline{0}}$ are in general not fault-tolerant. Consequently, extra
"verifier" ancilla states are used to detect errors arising during the
preparation of  $\ket{\overline{+}}$ and $\ket{\overline{0}}$. If the verifier
states measure a non-trivial syndrome or the $-1$ eigenvalue of a logical Pauli
is measured, the ancilla states are rejected and new ancilla states are brought
in until they pass the verification step.}
\label{fig:FullSteaneECCircuit}
\end{figure}

Note that the circuits used to prepared the encoded $\ket{\overline{+}}$ and
$\ket{\overline{0}}$ states are in general not fault-tolerant. In the case of
$\ket{\overline{+}}$, low weight errors can spread to high-weight $X$ errors
which can change the outcome of the measurement and $Z$ errors which can
propagate to the data block due to the transversal CNOT gates. However, by
preparing extra ``verifier" states encoded in $\ket{\overline{+}}$ and coupling
these states to the original $\ket{\overline{+}}$ ancilla as shown in
\cref{fig:FullSteaneECCircuit}, high weight $X$ and $Z$ errors arising from the
ancilla can be detected.  Furthermore, after a classical error correction step,
the eigenvalue of $\overline{X}$ and $\overline{Z}$ can be measured. Therefore
if a non-trivial syndrome is measured in the verifier states or the $-1$
eigenvalue of a logical operator is measured, the ancilla qubits are rejected
and new ancilla qubits are brought in to start the process anew.

We would like to point out that instead of verifying the ancilla qubits for
errors and rejecting them when a non-trivial syndrome is measured, it is also
possible to replace the verification circuit with a decoding circuit. By
performing appropriate measurements on the ancilla qubits and making use of
Pauli frames \cite{Knill05,Barbara15,CIP17}, any errors arising from $t$-faults
in the ancilla circuits can be identified and corrected \cite{DA07} (note that
DiVincenzo and Aliferis provided circuits for Steane's \codepar{7,1,3} code so
that $t=1$). However in this paper we will focus on ancilla verification
methods.

It can be shown that the Steane-EC circuit of \cref{fig:FullSteaneECCircuit}
satisfies both fault-tolerant conditions of \cref{Def:FaultTolerantDef} for
distance-three codes \cite{Gottesman2010}. It is possible to use the same
ancilla verification circuits in some circumstances for higher distance codes by
carefully choosing different circuits for preparing the logical
$\ket{\overline{0}}$ and  $\ket{\overline{+}}$ states (see \cite{PR12} for some
examples). In this paper, we will choose appropriate $\ket{\overline{0}}$ and
$\ket{\overline{+}}$ states such that the the decoding schemes will be 
fault-tolerant using the ancilla verification circuits in
\cref{fig:FullSteaneECCircuit}. We would like to add that although the order in
which transversal measurements to correct bit-flip and phase-flip errors does
not affect the fault-tolerant properties of Steane-EC, it does create an
asymmetry in the $X$ and $Z$ logical failure rates \cite{PR12,CJL16,CJL16b}. For
instance, an $X$ error arising on the target qubit of the logical CNOT used to
detect phase errors would be copied to the $\ket{\overline{+}}$ ancilla. However
a $Z$ error arising on the target of this CNOT or control of the CNOT used to
correct bit-flip errors would not be copied to any of the ancilla qubits.

\begin{figure}
\centering
\begin{subfigure}{0.18\textwidth}
\includegraphics[width=\textwidth]{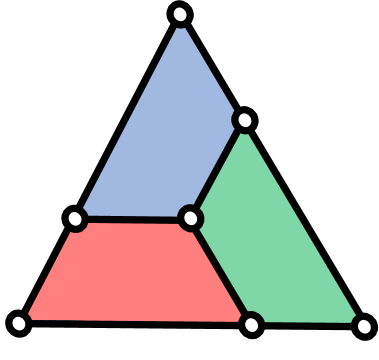}
\caption{}
\label{fig:SteaneCodeColor}
\end{subfigure}
\begin{subfigure}{0.2\textwidth}
\includegraphics[width=\textwidth]{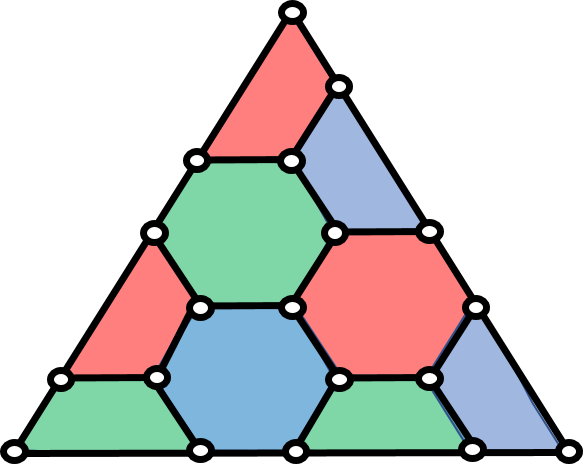}
\caption{}
\label{fig:19qubitColorLattice}
\end{subfigure}
\caption{\cref{fig:SteaneCodeColor} is a representation of the \codepar{7,1,3}
Steane code. The qubits are located at the white circles of the lattice. Each
face corresponds to both a $X^{\otimes 4}$ and $Z^{\otimes 4}$ stabilizer.
\cref{fig:19qubitColorLattice} is a representation of the \codepar{19,1,5} color
code. Like the Steane code, each face corresponds to an $X$ and $Z$ type
stabilizer. Notice that there are three weight-six stabilizers of each type.}
\label{fig:ColorCodeFigures}
\end{figure}

\begin{figure}
\center
\includegraphics[scale=.40]{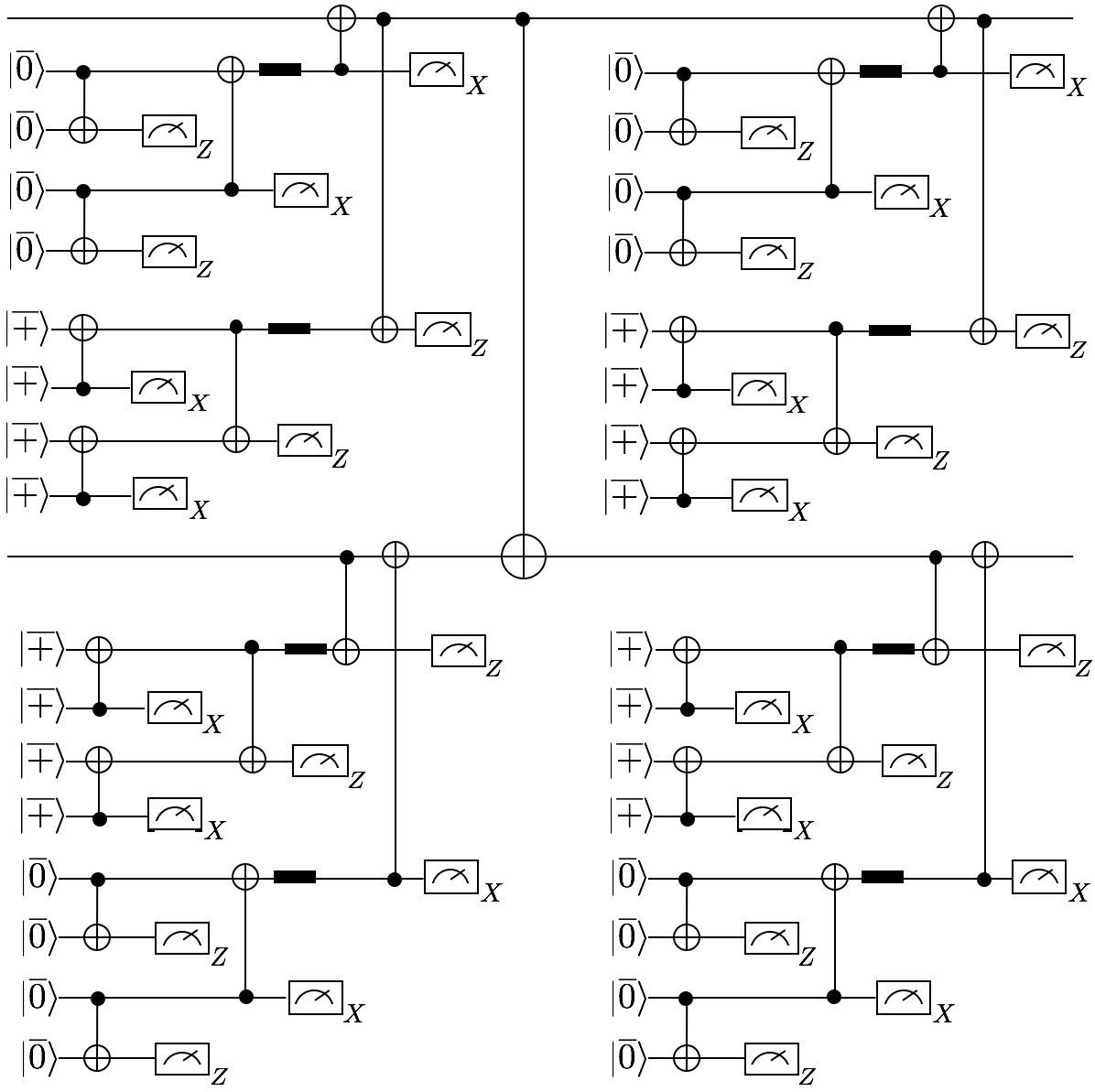}
\caption{CNOT-exRec for Steane-EC which contains four EC blocks. The CNOT-exRec
limits the pseudo-threshold of the \codepar{7,1,3} and \codepar{19,1,5} color
code due to the large number of locations and thus makes an ideal circuit to
optimize our decoding algorithm using machine learning.}
\label{fig:SteaneExRecCircuit}
\end{figure}

We conclude this section by describing the \codepar{7,1,3} and \codepar{19,1,5}
CSS color codes \cite{Bombin06TopQuantDist} which will be the codes used for
optimizing our decoding algorithms with machine learning applied to Steane and
Knill error correction (see \cref{subsec:KnillEC} for a description of Knill-
EC). A pictorial representation for both of these codes is shown in
\cref{fig:ColorCodeFigures}. Both the Steane code and the 19-qubit color code
are self-dual CSS codes (meaning that the $X$ and $Z$ stabilizers are
represented by the same parity check matrix). The Steane code has three $X$ and
$Z$ stabilizer generators while the 19-qubit color code has nine $X$ and $Z$
stabilizer generators. Since these codes are small, it is possible to use a
lookup table decoder similar to the one presented in
\cref{subsec:RotatedSurfaceCode} to correct errors. The only difference is that
we do not have to consider weight-two errors arising from a single fault (since
all gates in Steane and Knill-EC are transversal). We will also analyze the
performance of both codes using the naive decoder described in
\cref{subsec:NaiveDecoder}.

To obtain a pseudo-threshold for both of these codes, we will consider the
CNOT-exRec since it is the logical gate with the largest number of locations and
thus will limit the performance of both codes \cite{AGP06} (here we are
considering the universal gate set generated by 
$\langle \text{CNOT}, T, H \rangle$ where
$T= \text{diag}(1,e^{i\pi /4})$ and $H$ is the Hadamard gate \cite{BMPRV99}).
The full CNOT-exRec circuit for Steane-EC is shown in
\cref{fig:SteaneExRecCircuit}. Note that the large number of CNOT gates will
result in a lot of correlated errors which adds a further motivation to consider
several neural networks techniques to optimize the decoding performance.

\subsection{Knill error correction}
\label{subsec:KnillEC}

\begin{figure}
\center
\includegraphics[scale=.30]{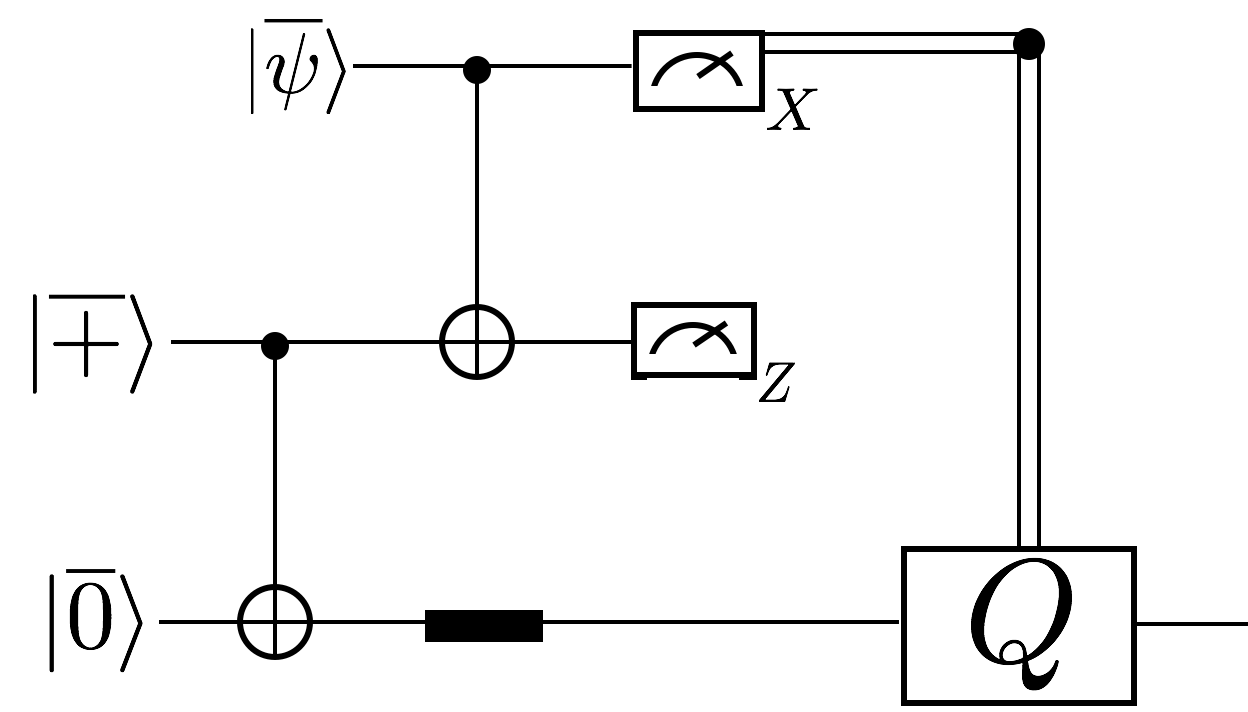}
\caption{Knill error correction circuit. As with Steane-EC, all CNOT gates and
measurements are performed transversally. The logical $\ket{\overline{0}}$ and
$\ket{\overline{+}}$ states are also encoded using the same code that protects
the data. A transversal CNOT gate is applied between them to form a logical Bell
state. The operator $Q$ is used to complete the teleportation protocol of the
logical state as well as to correct errors which were on the original data
block.}
\label{fig:KnillSnapShot}
\end{figure}

Steane error correction described in \cref{subsec:SteaneEC} only applies to
stabilizer codes which are CSS codes. Further, the protocol requires two transversal CNOT gates
between the data and ancilla qubits. In this section we will give an overview of
Knill error correction \cite{Knill04,Knill05} which is applicable to any
stabilizer code. As will be shown Knill-EC only requires a single transversal CNOT gate between the
data qubits and ancilla qubits.

Consider a Pauli operator $P$ acting on the data block of the circuit in
\cref{fig:KnillSnapShot}. Consider the same Pauli $P$ (but with a possibly
different sign) acting on the first ancilla block of the logical Bell pair. $P$
can be any Pauli but in the argument that follows we will be interested in cases
where $P \in N(\mathcal{S})$. Taking into account the sign of $P$ and writing it
as a product of $X$ and $Z$, we have that

\begin{align}
(-1)^{b_i}P = i^{c(P_{X},P_{Z})}(-1)^{b_i}P_{X}P_{Z}.
\label{eq:Pauli1}
\end{align}
The function $c(P_{X},P_{Z}) = 0$ if $P_{X}$ and $P_{Z}$ commute and one
otherwise. The phase $ i^{c(P_{X},P_{Z})}$ comes from the $Y$ operators in $P$
and $(-1)^{b_i}$ indicates the sign of the Pauli where $i=0$ for the data block
and $i=1$ for the ancilla block.

Applying the transversal CNOT's between the ancilla and data block performs the
following transformations
\begin{align}
(-1)^{b_0}P \otimes I 
\rightarrow i^{c(P_{X},P_{Z})}(-1)^{b_0}P_{X}P_{Z} \otimes P_{X}, \\
(-1)^{b_1}I \otimes P 
\rightarrow i^{c(P_{X},P_{Z})}(-1)^{b_1}P_{Z} \otimes P_{X}P_{Z},
\label{eq:TransformPauli}
\end{align}
and therefore
\begin{align}
(-1)^{b_{0}+b_{1}}P \otimes P 
\rightarrow (-1)^{b_{0}+b_{1}+c(P_{X},P_{Z})}P_{X} \otimes P_{Z}. 
\label{eq:CombinePauli}
\end{align}
From \cref{eq:CombinePauli}, we can deduce that a subsequent measurement of $X$
on each physical data qubit and measurement of $Z$ on each physical qubit in the first ancilla
block lets us deduce the eigenvalue of $P$ (since $c(P_{X},P_{Z})$ is known, we
learn $b_{0}+b_{1}$).

Since the above arguments apply to any Pauli, if $P$ is a stabilizer we learn
$s_{0}+s_{1}$ where $s_{0}$ is the syndrome of the data block and $s_{1}$ is the
error syndrome of the first ancilla block. Furthermore, the measurements also
allow us to deduce the eigenvalues of the logical Pauli's $\overline{X}_{i}
\otimes \overline{X}_{i}$ and  $\overline{Z}_{i} \otimes \overline{Z}_{i}$ for
every logical qubit $i$. This means that in addition to error correction we can
also perform the logical Bell measurement required to teleport the encoded data
to the second ancilla block.

Note that pre-existing errors on the data or ancilla block can change the
eigenvalue of the logical operator $\overline{P} \otimes \overline{P}$ without
changing the codeword that would be deduced using an ideal decoder. For
instance, if $E_d$ is the error on the data block and $E_a$ the error on the
ancilla block with $\text{wt}(E_d) + \text{wt}(E_a) \le t$, then if $(-1)^b$ is
the eigenvalue of $\overline{P} \otimes \overline{P}$, we would instead measure
$(-1)^{b'}$ where $b' = b + c(E_d,\overline{P}) + c(E_a,\overline{P})$. The same
set of measurements also let's us deduce the syndrome $s(E_d) + s(E_a) =
s(E_{d}E_{a})$. But since $\text{wt}(E_{d}E_{a} \le t)$, from $ s(E_{d}E_{a})$
we deduce the error $E' = E_{a}E_{d}M$ where $M \in \mathcal{S}$. Hence once
$E'$ is deduced, we also get the correct eigenvalue of  $\overline{P} \otimes
\overline{P}$ thus obtaining the correct  outcome for the logical Bell
measurement.

There could also be faults in the CNOT's and measurements when performing
Knill-EC. We can combine the errors from the CNOT's and measurements into
the Pauli $G$ on the data block and $F$ on the ancilla block where the weight of
$GF$ is less than or equal to the number faults at the CNOT and measurement
locations. Given the basis in which the measurements are performed, we can
assume that $G$ consists only of $Z$ errors and $F$ of $X$ errors. Consequently,
for a full circuit level noise model, the final measured syndrome is
$s(E_{d}E_{a}GF)$.

\begin{figure}
\center
\includegraphics[scale=.37]{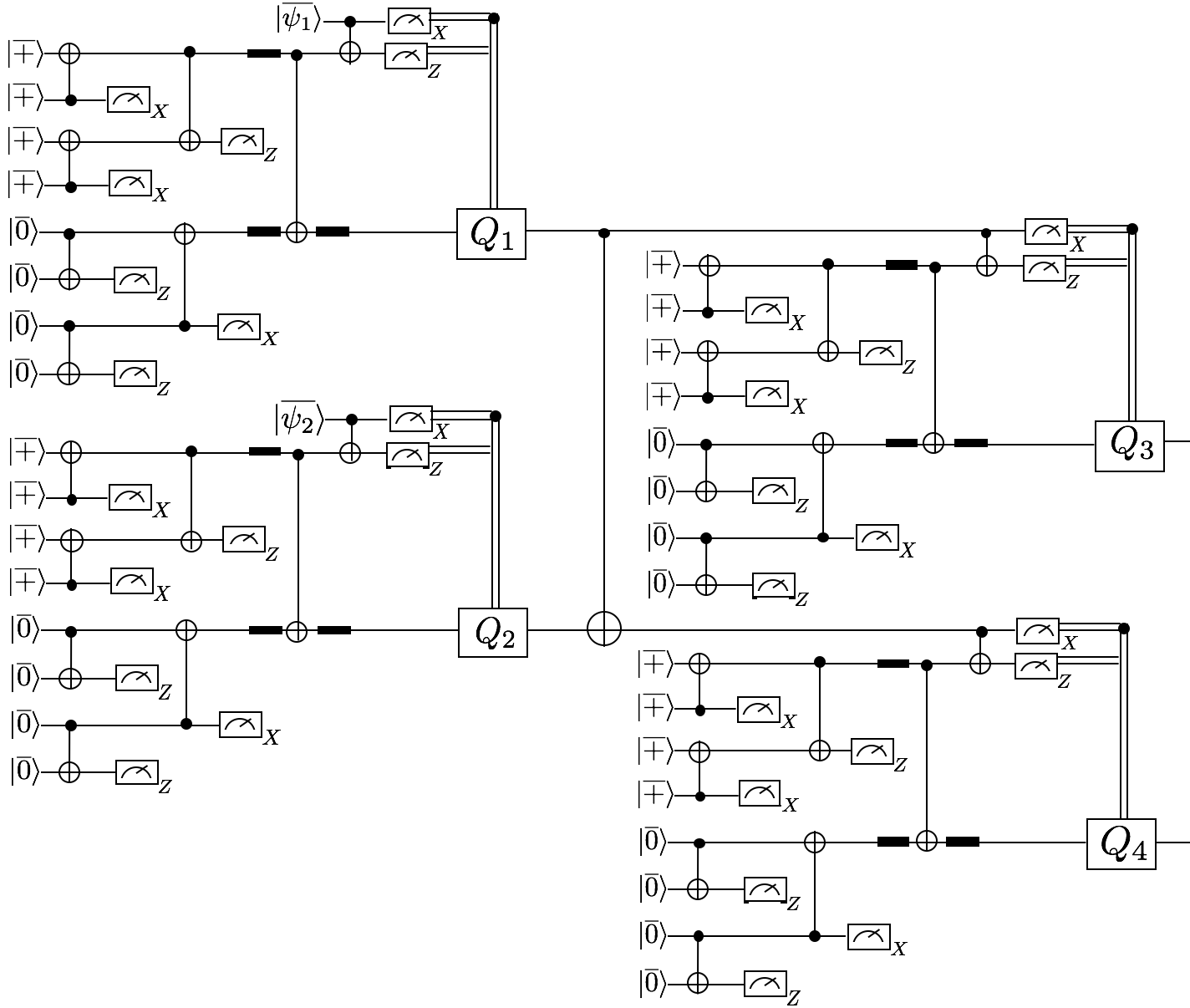}
\caption{Full CNOT-exRec circuit using Knill error correction. Each Pauli
operator $Q_{1},Q_{2},Q_{3}$ and $Q_{4}$ is used to correct errors in the
initial data blocks as well as the complete teleportation protocol of
the logical Bell measurement.}
\label{fig:KnillExRecCircuit}
\end{figure}

As in Steane-EC, the circuits for preparing the logical $\ket{\overline{0}}$ and
$\ket{\overline{+}}$ states are not fault-tolerant and can result in high weight
errors on the data. However, if the error correcting code is a CSS code, then we
can use the same ancilla verification method presented in \cref{subsec:SteaneEC}
to make the full Knill-EC protocol fault-tolerant. In
\cref{fig:KnillExRecCircuit} we show the full CNOT-exRec circuit using Knill-EC.
Note that for each EC unit, there is an extra idle qubit location compared to
Steane-EC.

Lastly, we point out that another motivation for using Knill-EC is it's ability
to handle leakage errors. A leakage fault occurs when the state of a two-level
system, which is part of a higher dimensional subspace, transitions outside of
the subspace. In \cite{ATLeakage06}, it was shown how leakage faults can be
reduced to a regular fault (which acts only on the qubit subspace) with the use
of Leakage-Reduction Units (LRU's). One of the most natural ways to implement
LRU's is through quantum teleportation \cite{MochonTeleportation04}. Since
Knill-EC teleports the data block to the ancilla block, unlike in Steane-EC,
LRU's don't need to be inserted on the input data block. However, LRU's still
need to be inserted after the preparation of every $\ket{\overline{0}}$ and
$\ket{\overline{+}}$ states.

\subsection{Naive decoder}
\label{subsec:NaiveDecoder}

Since the lookup table decoder scheme presented in previous sections is not
scalable, it would be desirable to have a scalable and fast decoding scheme that
can achieve competitive thresholds when paired with a deep neural decoder. In
this section we provide a detailed description of a naive decoder which can
replace the lookup table scheme in all of the above protocols.

We first note that the recovery operator $R_s$ for a measured syndrome $s$ can
be written as \cite{Poulin06,DuclosRG1}
\begin{align}
R_s = \mathcal{L}(s)\mathcal{T}(s)\mathcal{G}(s)
\label{eq:LSTDecomp}
\end{align}
which we will refer to as the LST decomposition of $E$. In \cref{eq:LSTDecomp},
$\mathcal{L}(s)$ is a product of logical operators (operators in $N(\mathcal{S})
\setminus \mathcal{S}$), $\mathcal{G}(s)$ is a product of stabilizers (operators
in $\mathcal{S}$) and $\mathcal{T}(s)$ is a product of pure errors. Pure errors
form an abelian group with the property that $T_{i}$ appears in $\mathcal{T}(s)$
if and only if the $i$'th syndrome bit is 1 (i.e. $[T_{i},T_{j}] = 0$ and
$[T_{j},g_{k}] = \delta_{j,k}$ where $g_{k}$ is the $k$'th stabilizer
generator). Thus pure errors can be obtained from Gaussian elimination. Note
that the choice of operators in $\mathcal{G}(s)$ will not effect the outcome of
the recovered state. Consequently, given a measured syndrome $s$, decoding can
be viewed as finding the most likely logical operator in  $ \mathcal{L}(s)$.

For a measured syndrome $s$, a naive decoding scheme is to always choose the
recovery operator $R_l = \mathcal{T}(s)$ which is clearly suboptimal. However,
for such a decoder, the decoding complexity results simply from performing the
matrix multiplication $\mathbf{s}\ T$ where $\mathbf{s} = (s_{1},s_{2},\cdots ,
s_{n-k})$ is the syndrome written as a $1 \times (n-k)$ vector and $T$ is a
$(n-k) \times n$ matrix where the $j$'th row corresponds to $T_{j}$. The goal of
all neural networks considered in \cref{sec:DND} will then be to find the most
likely operator $ \mathcal{L}(s)$ from the input syndrome $l$.

The set of stabilizer generators, logical operators and pure errors for all the
codes considered in this paper are provided in \cref{tab:CodeParameters}.
Lastly, we point out that a version of the above decoding scheme was implemented
in \cite{CrigerNN17} for the distance-three surface code.

\subsection{Lookup table and naive decoder complexity}
\label{sec:LookpNaiveComplexity}

From a complexity theoretic point of view, read-out of an entry of an array or a
hash table requires constant time. In hash tables, a hash function is calculated
to find the address of the entry inquired. The hash function calculation takes
the same processing steps for any entry, making this calculation $O(1)$.  In the
case of an array, the key point is that the array is a sequential block of the
memory with a known initial pointer. Accessing any entry requires calculating
its address in the memory by adding its index to the address of the beginning of
the array. Therefore, calculating the address of an entry in an array also takes
$O(1)$.

It remains to understand that accessing any location in the memory given its
address is also $O(1)$ as far as the working of the memory hardware is
concerned. This is the assumption behind \emph{random access memory} (RAM) where
accessing the memory comprises of a constant time operation performed by the
multiplexing and demultiplexing circuitry of the RAM. This is in contrast  with
direct-access memories (e.g. hard disks, magnetic tapes, etc) in which the time
required to read and write data depends on their physical locations on the
device and the lag resulting from disk rotation and arm movement.

Given the explanation above, a decoder that relies solely on accessing recovery
operators from an array operates in $O(1)$ time. This includes the lookup table
and the inference mapping method of \cref{sec:inference} below.

For the naive decoder of \cref{subsec:NaiveDecoder}, we may also assume that the
table of all pure errors (denoted as $T$ in \cref{subsec:NaiveDecoder}) is stored
in a random access memory. However, the algorithm for generating a recovery from
the naive decoder is more complicated than only accessing an element of $T$.
With $n$ qubits and $n - k$ syndromes, for every occurrence of $1$ in the
syndrome string, we access an element of $T$. The elements accessed in this
procedure have to be added together. With parallelization, we may assume that a
tree adder is used which, at every stage, adds two of the selected pure error
strings to each other. Addition of every two pure error strings is performed
modulo 2 which is simply the XOR of the two strings, which takes $O(1)$ time
assuming parallel resources. The entire procedure therefore has a time
complexity of $O( (n-k) \log(n-k) )$, again assuming parallel digital resources.

\section{Deep neural decoders}
\label{sec:DND}

In most quantum devices, fully characterizing the noise model afflicting the
system can be a significant challenge. Furthermore, for \textit{circuit level}
noise models which cannot be described by Pauli channels, efficient simulations
of a codes performance in a fault-tolerant implementation cannot be performed
without making certain approximations (a few exceptions for repetition codes can
be found in \cite{YasunariEfficientRep17}). However, large codes are often
required to achieve low failure rates such that long quantum computations can be
performed reliably. These considerations motivate fast decoding schemes which
can adapt to unknown noise models encountered in experimental settings.

Recall from \cref{subsec:NaiveDecoder} that decoding can be viewed as
finding the most likely operator $L \in \mathcal{L}(\mathbf s)$ given a measured
syndrome $\mathbf s$. Since all codes considered in this paper encode a single
logical qubit, the recovery operator for a measured syndrome $\mathbf s$ can be
written as
\begin{align}
R_{\mathbf s} = 
X_{L}^{b_{1}(\mathbf s)}Z_{L}^{b_{2}(\mathbf s)}
\mathcal{T}(\mathbf s)\mathcal{G}(\mathbf s)
\label{eq:RecoveryRsingleQubit}
\end{align}
where $X_{L}$ and $Z_{L}$ are the codes logical $X$ and $Z$ operators and
$b_{1}(\mathbf s),b_{2}(\mathbf s) \in \mathbb{Z}_{2}$. In \cite{CWBL16}, a
decoding algorithm applicable to general Markovian channels was presented for
finding the coefficients $b_{1}(\mathbf s)$ and $b_{2}(\mathbf s)$ which
optimized the performance of error correcting codes. However, the algorithm
required knowledge of the noise channel and could not be directly applied to
circuit level noise thus adding further motivation for a neural network decoding
implementation.

In practice, the deep learning schemes described in this section can be trained
as follows. First, to obtain the training set, the data qubits are 
fault-tolerantly prepared in a known logical $\ket{\overline{0}}$ or
$\ket{\overline{+}}$ state followed by a round of fault-tolerant error
correction (using either the lookup table or naive decoders). The encoded
\textit{data} is then measured in the logical $Z$ or $X$ basis yielding a -1
eigenvalue if a logical $X$ or $Z$ error occurred. The training set is
constructed by repeating this sequence several times both for states prepared in
$\ket{\overline{0}}$ or $\ket{\overline{+}}$. For each experiment, all syndromes
are recorded as well as the outcome of the logical measurement. Given the most
likely error $E$ with syndrome $s(E) = \mathbf s$ (in general $E$ will not be
known), the neural network must then find the vector $\bold{b} = (b_{1}(\mathbf
s),b_{2}(\mathbf s))$ such that $X_{L}^{b_{1}(\mathbf s)}Z_{L}^{b_{2}(\mathbf
s)}R_{\mathbf s}E = I$ where $R_{\mathbf s}$ was the original recovery operator
obtained from either the lookup table or naive decoders described in
\cref{sec:FaultTolerantProtocols}.

Once the neural network is trained, to use it in the inference mode (as
explained in Section \ref{sec:inference}), a query to the network simply
consists of taking as input all the measured syndromes and returning as output
the vector $\mathbf b$. For Steane and Knill EC, the syndromes are simply the
outcomes of the transversal $X$ and $Z$ measurements in the leading and trailing
EC blocks. For the surface code, the syndromes are the outcomes of the ancilla
measurements obtained from each EC round until the protocols presented in
\cref{subsec:RotatedSurfaceCode} terminate.

Lastly, we note that a similar protocol was used in 
\cite{Baireuther2018machinelearning} which also
used the outcome of the final measurement on the data qubits to decode. However
by using our method, once the neural network is trained, it only takes as input
the measured syndromes in an EC round to compute the most likely $\bold{b}$.

\subsection{Deep learning}
\label{subsec:DLOverview}

Here we explain the generic framework of our deep learning experiments. We
refer the reader to \cite{Goodfellow-et-al-2016} for an introduction to deep
learning and to \cite{bishop2013pattern} for machine learning methods in
classification tasks.

Let $D \subseteq \mathcal D$ be a data set. In our case, $\mathcal D = S \times
B$ is the set of all pairs of syndromes and \emph{error labels}. Every element
in $D$ and $\mathcal D$ is therefore a pair $(\mathbf s, \mathbf e)$ of measured
syndromes $\mathbf s \in S$ and error labels $\mathbf e \in B$. The error labels
can be different depending on how we model the learning problem. For instance,
every $\mathbf e \in B$ can be a bit string carrying a prescription of recovery
operators: 
$$B = \{I, X, Y, Z\}^{\#\text{physical qubits}}.$$

There is however a major drawback in modelling the errors in the above fashion.
For deep learning purposes the elements $\mathbf e \in B$ are represented in
their \emph{1-hot encoding}, i.e. a bit string consisting of only a single 1,
and zeros everywhere else. The 1-hot encoding therefore needs $|E|$ bits of
memory allocated to itself which by the definitions above, grows exponentially
in either the number of physical qubits.

Our solution for overcoming this exponentially growing model is to take
advantage of the decomposition (\cref{eq:RecoveryRsingleQubit}) of
the recovery operator and only predict vectors $\mathbf b = (b_1 (\ell) , b_2
(\ell))$ as explained earlier. In other words, the elements of $B$ contain
information about the logical errors remaining from the application of another
auxiliary encoding scheme:
$$B = \{I, X, Y, Z\}^{\# \text{logical qubits}}.$$

\subsubsection{The objective function}
As customary in machine learning, the occurrences $x = (\mathbf s, \mathbf b)
\in D$ are viewed as statistics gathered from a conditional probability
distribution function $p(x) = \mathbb P(\mathbf b\,|\,\mathbf s)$ defined over
$S \times E$. The goal is then to approximate $p$ by another distribution $p_w$
which is easy to compute from a set of real-valued parameters $w$. The
\emph{training} phase in machine learning consists of optimizing the parameter
vector $w$ such that $p_w$ is a good approximation of $p$. The optimization
problem to solve is therefore
\begin{align}\label{eq:ml-optimization}
\min_{w} \quad &  \Delta(p, p_w).
\end{align}
Here $\Delta$ is some notion of distance in the space of probability
distribution functions which, when applied to machine learning, is also called
\emph{the loss function}. In our case, the distance is the \emph{softmax} cross
entropy as explained here. The softmax function with respect to $p$ is given via
\begin{equation}
\label{eq:softmax}
\rho (x) = \frac{e^{p(x)}}{\sum_{x \in \mathcal D} e^{p(x)}}.
\end{equation}
From this definition it is obvious that no normalization of the dataset $D$ is
needed since softmax already results in a probability distribution function. The
cross entropy function
\begin{align}\label{eq:crossentropy}
H(\pi_1, \pi_2) 
= H(\pi_1) + D_{KL} (\pi_1 |\!| \pi_2) = - \sum_x \pi_1(x) \log \pi_2 (x)
\end{align}
is then applied after softmax. This turns \eqref{eq:ml-optimization} into
\begin{align}\label{eq:opt-softmaxcrossentropy}
\min_w \quad &  h(w) = H(\rho(p), \rho(p_w)).
\end{align}
Optimization of the softmax cross-entropy is a common practice in classification
problems.

\subsubsection{The neural network} A neural network is a directed graph equipped
with a random variable assigned to each of its nodes. The elements of the
parameter vector $w$ are assigned either to an edge of the graph or a node of
the graph (in the former case they are called \emph{weights} and in the latter
case they are called \emph{biases}). The roll of the neural network in solving
\eqref{eq:opt-softmaxcrossentropy} is to facilitate a gradient descent direction
for the vector $w$ in \eqref{eq:opt-softmaxcrossentropy}. This is achieved by
imposing the random variables of each node to be a function of the random
variables with incoming edges to the former one. The common choice for such a
functional relationship is an affine transformation composed with a nonlinear
function (called the \emph{activation} function) with an easy to compute
derivative. Given every node $v$ of the neural network, we define:
\begin{equation}
\label{eq:layers}
X_v= a_v \left(\sum_{u \rightarrow v} w_{uv} X_u + w_v\right).
\end{equation}
The simplest activation function is of course the identity. Historically, the
sigmoid function $\sigma(x) = \frac{1}{1 + e^{-x}}$ was the most commonly used
activation function and is motivated by its appearance in training restricted
Boltzmann machines. By performing a change of variables, one obtains the trigonometric
activation function $\tanh(x)$. These activation functions can cause the learning rate to slow down due to vanishing gradients in the early layers of
deep neural networks, and this is the motivation for other proposed activation
functions such as the rectified linear unit $\relu(x)$. Design and analysis of
activation functions is an important step in machine learning
\cite{LeCun1998,Nair:2010:RLU:3104322.3104425, pmlr-v9-glorot10a}.

The first and last layers of the network are known as the \emph{visible} layers
and respectively correspond to the input and output data (in our case the tuples
$(\mathbf s, \mathbf b) \in S \times B$ as explained above). Successive
applications of \cref{eq:layers} restricts the conditional distribution
$p_w(\mathbf b\,|\,\mathbf s)$ into a highly nonlinear function $f (w, \mathbf
s, \mathbf b)$, for which the derivatives with respect to the parameters $w$ are
easy to compute via the chain rule. We may therefore devise a gradient descent
method for solving \cref{eq:opt-softmaxcrossentropy} by successive choices of
descent directions starting from the deep layers and iterating towards the input
nodes. In machine learning, this process is known as \emph{back-propagation}.

\begin{rmk} The softmax function (\cref{eq:softmax}) is in other words the
activation function between the last two layers of the neural network.
\end{rmk}

\subsubsection{Layouts}
Although deep learning restricts the approximation of $p_w (\mathbf b | \mathbf
s )$ to functions of the form $f(w, \mathbf s, \mathbf b)$ as explained above,
the latter has tremendous representation power, specially given the freedom in
choice of the layout of the neural network. Designing efficient layouts for
various applications is an artful and challenging area of research in machine
learning. In this paper, we discuss three such layouts and justify their usage
for the purposes of our deep neural decoding.

\vskip1.5mm
{\noindent\it Feedforward neural network.} By this we mean a multi-layer neural
network consisting of consecutive layers, each layer fully connected to the next
one. Therefore, the underlying undirected subgraph of the neural network
consisting of the neurons of two consecutive layers is a complete bipartite
graph. In the case that the neural network only consists of the input and output
layers (with no hidden layers), the network is a generalization of logistic
regression (known as the softmax regression method).

\vskip1.5mm
{\noindent\it Recurrent neural network (RNN).} RNNs have performed incredibly
well in speech recognition and natural language processing tasks 
\cite{Fernandez:2007:ARN:1778066.1778092,
Sutskever:2014:SSL:2969033.2969173,45446,2015arXiv151200103G}. 
The network is designed to resemble a temporal sequence of input data,
with each input layer connecting to the rest of the network at a corresponding
temporal epoch. The \emph{hidden cell} of the network could be as simple as a
single feedforward layer or more complicated. Much of the success of RNNs is
based on peculiar designs of the hidden cell such as the Long-Short Term Memory
(LSTM) unit as proposed in \cite{Hochreiter:1997:LSM:1246443.1246450}.

\vskip1.5mm
{\noindent\it Convolutional neural network (CNN).} CNNs have been successfully
used in image processing tasks
\cite{Schmidhuber:2012:MDN:2354409.2354694,ILSVRC15}. The network is designed to
take advantage of local properties of an image by probing a kernel across the
input image and calculating the cross-correlation of the kernel vector with the
image. By applying multiple kernels, a layer of \emph{features} is constructed.
The features can then be post-processed via downsizing (called 
\emph{max-pooling}) or by yet other feedforward neural networks.

In sections \ref{subsec:DNDSteaneKnill} and \ref{subsec:DNDSurfaceCode}, we
present further details about applications of these neural networks to the
error-decoding task.

\subsubsection{Stochastic gradient descent}

Since the cross-entropy in \cref{eq:crossentropy} is calculated by a weighted
sum over all events $x \in \mathcal D$, it is impractical to exactly calculate
it or its derivatives as needed for backpropagation. Instead, one may choose
only a single sample $x = (\mathbf s, \mathbf b)$ as a representative of the
entire $\mathcal D$ in every iteration. Of course, this is a poor approximation
of the true gradient but one hopes that the occurrences of the samples according
to the true distribution would allow for the descent method to `average out'
over many iterations. This method is known as stochastic gradient descent (SGD)
or \emph{online learning}. We refer the reader to
\cite{Nemirovski:2009:RSA:1654243.1654247} and \cite{NIPS2003_2365} and the
references therein for proofs of convergences and convergence rates of online
learning. In practice, a middle ground between passing through the entire
dataset and sampling a single example is observed to perform better for machine
learning tasks \cite{LeCun1998}: we fix a batch size and in every iteration
average over a batch of the samples of this size. We call this approach
\emph{batch gradient descent} (also called mini-batch gradient descent for
better contrast). The result is an update rule for the parameter vector of the
form $w_{t+1} \leftarrow w_t + \Delta_t$ where $\Delta_t$ is calculated as
$$\Delta_t =  - \eta_t \nabla_{t-1},$$ 
for some step size $\eta_t$, where $\nabla_{t-1}= \nabla_{w_{t-1}} \tilde
h(w_{t-1})$ to simplify the notation. Here $\tilde h$ is an approximation
of $h$ in \eqref{eq:opt-softmaxcrossentropy} 
by the partial sum over the training batch. Finding a good
schedule for $\eta_t$ can be a challenging engineering task that will be
addressed in \cref{subsec:Hypertuning}. Depending on the optimization landscape,
SGD might require extremely large numbers of iterations for convergence. 
One way to improve the convergence rate of SGD is to add a \emph{momentum} term
\cite{Rumelhart:1986:LIR:104279.104293}:
$$\Delta_t = p \Delta_{t-1} - \eta_t \nabla_{t-1}.$$

On the other hand, it is convenient to 
have the schedule of $\eta_t$ be determined through the
training by a heuristic algorithm that adapts to the frequency of every event.
The method AdaGrad was developed to allow much larger updates for infrequent
samples \cite{Pennington14glove:global}: 
$$\Delta_t =  -
\diag\left({\frac{\eta}{\sqrt{\Sigma_{ti} + \epsilon}}}\right) \nabla_{t-1}.$$
Here $\Sigma_{ti}$ is the sum of the squares of all previous values of the
$i$-th entry of the gradient. The quantity $\epsilon$ is a small (e.g. $10^{-8}$) smoothening
factor in order to avoid dividing by zero. The denominator in this formula is
called the \emph{root mean squared} (RMS). An important advantage of AdaGrad is
the fact that the freedom in the choice of the step-size schedule is restricted
to choosing one parameter $\eta$, which is called \emph{the learning rate}.

Finally RMSProp is an improvement on AdaGrad in order to slow down the
aggressive vanishing rate of the gradients in AdaGrad \cite{tielemanH12}. This
is achieved by adding a momentum term to the root mean squared:
$$\diag(\Sigma_{t}) 
= p \diag(\Sigma_{t-1}) + (1-p) \nabla_{t-1} \nabla_{t-1}^T.$$

\subsubsection{Hyperparameter tuning}
\label{subsec:Hypertuning}

From the above exposition, it is apparent that a machine learning framework
involves many algorithms and design choices. The performance of the framework
depends on optimal and consistent choices of the free parameters of each piece,
the \emph{hyperparameters}. For example, while a learning rate of $10^{-3}$
might be customary for a small dataset such as that of MNIST digit recognition,
it might be a good choice for a small feedforward network and a bad choice
for the RNN used in our problem scenario. In our case, the hyperparameters
include the decay rate, the learning rate, the momentum in RMSProp, the
number of hidden nodes in each layer of the network, the number of hidden layers
and filters, and some categorical variables such as the activation function of
each layer, the choice of having peepholes or not in the RNN.

It would be desirable if a metaheuristic can find appropriate choices of
hyperparameters. The challenges are
\begin{enumerate}
\item Costly function evaluation: the only way to know if a set of
hyperparameters is appropriate for the deep learning framework, is to run the
deep learning algorithm with these parameters;
\item Lack of a gradient-based solution: the solution of the deep learning
framework does not have a known functional dependence on the hyperparameters.
Therefore, the metaheuristic has no knowledge of a steepest descent direction.
\end{enumerate}
It is therefore required for the metaheuristic to be (1) sample efficient and
(2) gradient-free. Having a good metaheuristic as such is extremely desirable,
since:
\begin{enumerate}
\item The performance of the ML framework might be more sensitive to some
parameters than to others. It is desirable for the metaheuristic to identify
this.
\item Compatibility of the parameters: leaving the hypertuning job to a
researcher can lead to search in very specific regimes of hyperparameters that
are expected to be good choices individually but not in combination.
\item Objectivity of the result: a researcher might spend more time tuning the
parameters of their proposal than on a competing algorithm. If the same
metaheuristic is used to tune various networks, such as feedforward networks,
RNNs and CNNs, the result would be a reliable comparison between all
suggestions.
\end{enumerate}

{\noindent\it Bayesian optimization.} Bayesian optimization
\cite{mockus1989bayesian,mockus1989bayesian} is a nonlinear optimization
algorithm that associates a surrogate model to its objective function and
modifies it at every function evaluation. It then uses this surrogate model to
decide which point to explore next for a better objective value
\cite{JMLR:v15:martinezcantin14a}. Bayesian optimization is a good candidate for
hypertuning as it is sample efficient and can perform well for multi-modal
functions without a closed formula. A disadvantage of Bayesian optimization to
keep in mind is that it relies on design choices and parameters of its own that
can affect its performance in a hyperparameter search.

\subsection{Steane and Knill EC deep neural decoder for the CNOT-exRec}
\label{subsec:DNDSteaneKnill}

The simplest deep neural decoder for any dataset is a feedforward network with
none or many hidden layers, each layer fully connected to the next one. The
input layer receives the bit strings of $X$ and $Z$ syndromes. And the output
layer corresponds to the $X$ and $Z$ recovery operators on the physical qubits
of the code. Since multiple physical qubits might be used to encode a single
logical operator, a better choice is for the output layer to encode whether an
auxiliary (but efficient) decoding scheme is causing logical faults or not. The
goal would be to predict such logical faults by the deep neural decoder and when
the deep neural decoder predicts such a fault, we will impose a logical Pauli
operator after the recovery suggested by the auxiliary decoder. The 1-hot
encoding in two bits, $10$ and $01$, respectively stand for $I$ and $X$ for the
$X$-errors, and it stands for $I$ and $Z$ for the $Z$ errors.

\begin{figure}
\center{}
\includegraphics[scale=.5]{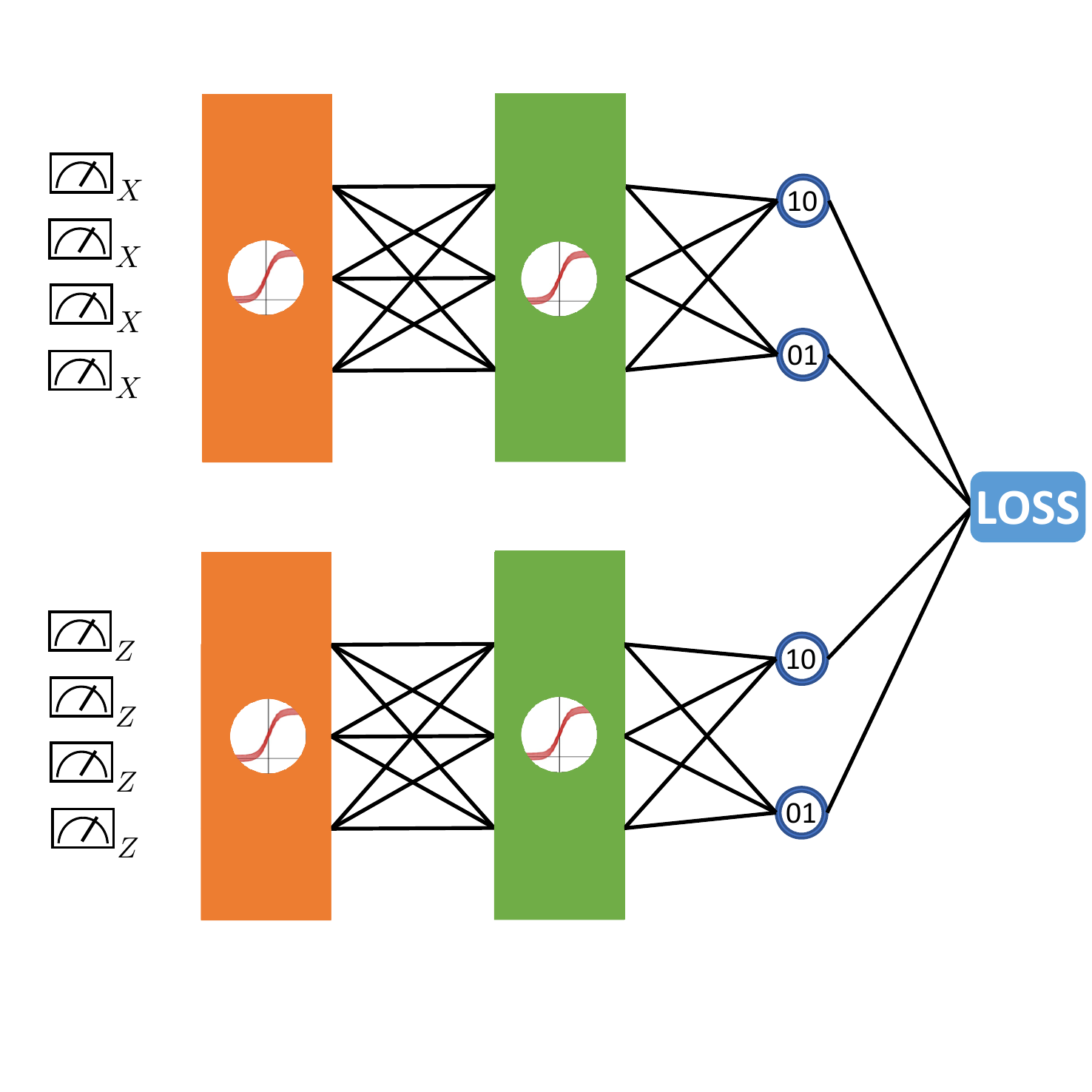}
\caption{Schematics of a feedforward network consisting of disjoint $X$ and $Z$
networks. There may be none, one or multiple hidden layers with different
activation functions. The output layers correspond to logical $I$- and
$X$-errors for the $X$ network and to logical $I$- and $Z$-errors for the $Z$
network. The activation function of the last layer before the error layer is the
identity since in the softmax cross entropy loss function, the activation (by
softmax) is already included.}
\label{fig:FF}
\end{figure}

From our early experiments it became apparent that it is beneficial to half
separate $X$ and $Z$ neural networks that share a loss function, that is the sum
of the soft-max cross entropies of the two networks. \cref{fig:FF} shows
the schematics of such a feedforward network.

\begin{figure}
\center
\includegraphics[scale=.5]{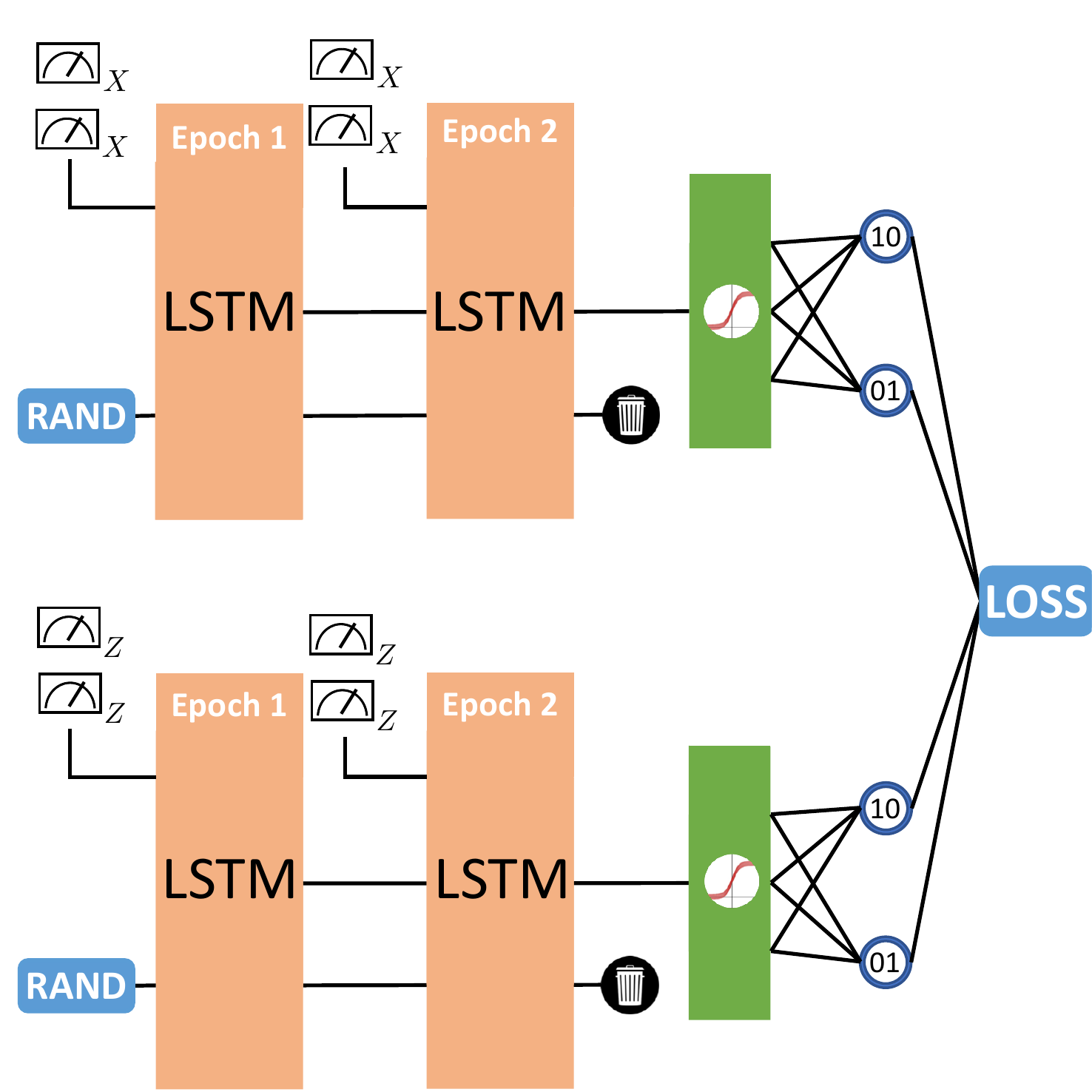}
\caption{Schematics of a network consisting of two disjoint $X$ and $Z$ RNNs.
Each RNN receives the syndromes of leading and trailing EC rounds as inputs for
two epochs of its LSTM unit. The internal state of the first copy is initialized
randomly and the internal state of the last copy is garbage-collected. The
hidden state of the last copy of the LSTM unit is then fully connected to a
hidden layer with user-defined activation function. This hidden unit is then
fully connected to output nodes denoted by $01$ and $10$ which are respectively
the one-hot encoding of the prediction as to whether an $X$-recovery or a 
$Z$-recovery operation is needed on the output qubits from exRec-CNOT. The loss
function is the sum of the loss functions of the two networks.}
\label{fig:RNN}
\end{figure}

\subsubsection{The CNOT-exRec RNN} In the case of the CNOT-exRec, the leading EC
rounds have temporal precedence to the trailing EC rounds. Therefore a plausible
design choice for the deep neural decoder would be to employ an RNN with two
iterations on the hidden cell. In the first iteration, the syndrome data from
the leading EC rounds are provided and in the second iteration the syndrome data
from the trailing EC rounds are provided. A demonstration of this network is
given in \cref{fig:RNN}.

\begin{figure}
\includegraphics[scale=.4]{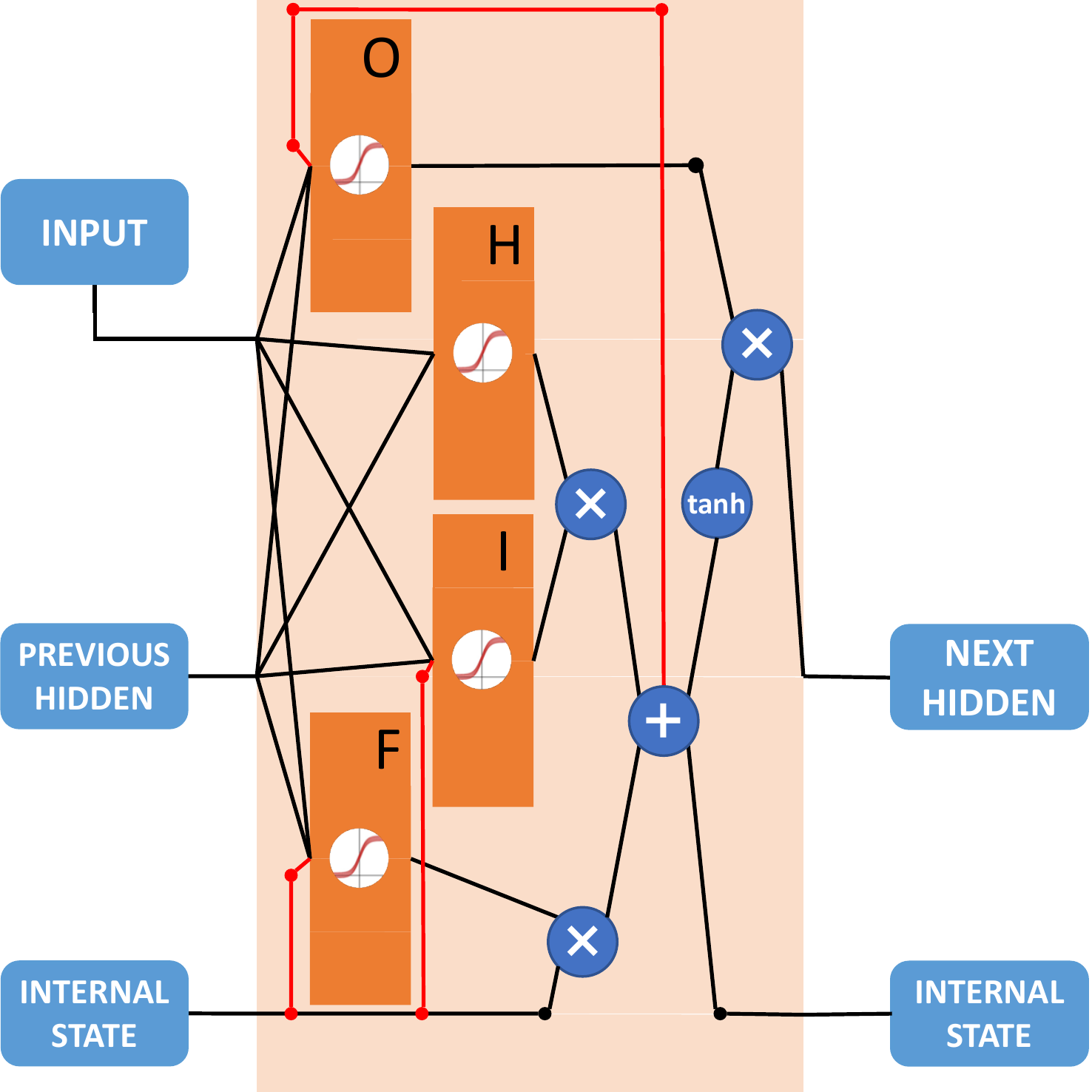}
\caption{Schematics of a long-short term memory (LSTM) cell. Without the red
circuits, this neural network is called a simple LSTM unit. The red circuit is
called peepholes. An LSTM cell with peepholes can outperform a simple LSTM cell
in some tasks. There are four hidden layers with user-defined activation 
functions in an LSTM unit known as the forget layer (F), input layer (I), 
hidden layer (H) and the output layer (O). There are four 2 to 1 logical gates 
in the unit that depending on the sign written on them applies an element-wise
operation between the vectors fed into the logical gates. There is also a 1 to 1
logical gate that applies an element-wise $tanh$ function on its input vector.
The \emph{internal state} of an LSTM
unit serves as the backbone of a sequence of replications of the LSTM unit.
The roll of the internal state is to capture temporal features of the sequence
of input data. }
\label{fig:LSTMPeep}
\end{figure}

The hidden cell of the RNN may be an LSTM, or an LSTM with peepholes as shown in
\cref{fig:LSTMPeep}. An LSTM cell consists of an
\emph{internal state} which is a vector in charge of carrying temporal
information through the unrolling of the LSTM cell in time epochs. There are 4
hidden layers. The layer $H$ is the `actual' hidden layer including the input
data of the current epoch with the previous hidden layer from the previous
epoch. The activation of $H$ is usually $\tanh$. The `input' layer $I$ is
responsible for learning to be a bottleneck on how important the new input is,
and the `forget' layer $F$ is responsible for creating a bottleneck on how much
to forget about the previous epochs. Finally the `output' layer $O$ is
responsible for creating a bottleneck on how much data is passed through from
the new internal state to the new hidden layer. The peepholes in
\cref{fig:LSTMPeep} allow the internal state to also contribute in the hidden
layers $F$, $I$ and $O$.

\subsection{Surface code deep neural decoder}
\label{subsec:DNDSurfaceCode}

Other than the multi-layer feedforward network of \cref{fig:FF}, there are two
other reasonable designs for a deep neural network when applied to the surface
code.

\subsubsection{The surface code RNN} In the fault-tolerant scheme of the rotated
surface code, multiple rounds of error correction are done in a sequence as
explained in Sec. \ref{subsec:RotatedSurfaceCode}. It is therefore encouraging
to consider an RNN with inputs as syndromes of the consecutive EC rounds. The
network looks similar to that of \cref{fig:RNN} except that the number of
epochs is equal to the maximum number of EC rounds. In particular, the fault
tolerant scheme for the distance-three rotated surface code consists of three EC
rounds. In the case of the distance-five surface code, the maximum number of EC
rounds through the
algorithm of Sec. \ref{subsec:RotatedSurfaceCode} is six. If the rounds of EC
stop earlier, then the temporal input sequence of syndrome strings is padded by
repeating the last syndrome string. As an example, if after three rounds the fault
tolerant scheme terminates, then the input syndromes of epochs three to six of the
RNN are all identical and equal to the third syndrome string.

\subsubsection{The surface code CNN} The errors, syndromes and recovery
operators of the surface code are locally affected by each other. It is
therefore suggestive to treat the syndromes of the surface code as a 2-dimensional
array, the same way pixels of an image are treated in image processing tasks.
The multiple rounds of EC would account for a sequence of such images, an
\emph{animation}. Therefore a 3-dimensional CNN appears to be appropriate. This
means that the kernels of the convolutions are also 3-dimensional, probing the
animation along the two axes of each image and also along the third axis
representative of time.

\begin{figure*}[t]
\center
\includegraphics[scale=.35]{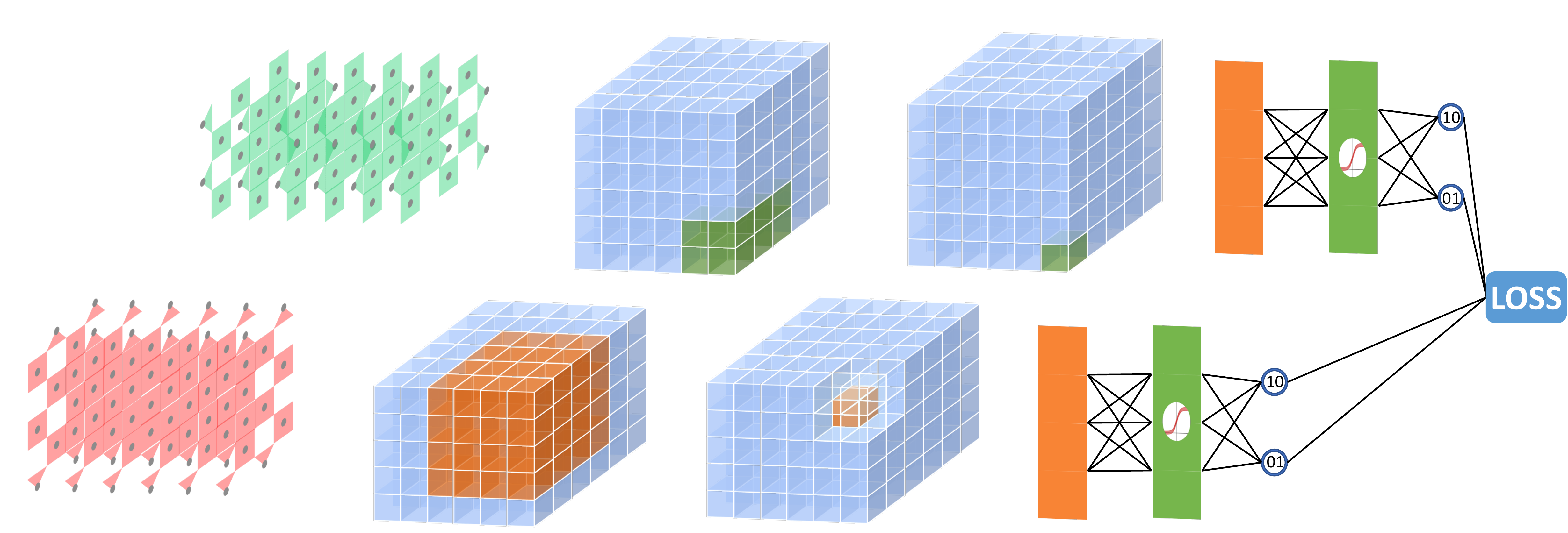}
\caption{
Schematics of a deep neural decoder for the distance-five rotated surface code.
The network consists of two disjoint neural networks contributing to the same
loss function via softmax cross entropy. Each neural network consists of two
layers of 3D CNNs. The first layer consists of a number of filters, each filter
performing a convolution of a $3 \times 3 \times 3$ kernel by the input
syndromes. The second 3D CNN layer uses $4 \times 4 \times 4$ kernels. The
colored boxes demonstrate how each layer is padded in order for the size of the
3D layers to be preserved. When the kernel dimension is even for instance, the
padding from the top and left are of size 1, and the padding from the bottom and
right are of size 2.}
\label{fig:CNN}
\end{figure*}

Through our test-driven design, it became obvious that treating the $X$ and $Z$
syndromes as channels of the same 3-dimensional input animation is not a good
choice. Instead, the $X$ and $Z$ syndromes should be treated as disjoint inputs
of disjoint networks which in the end contribute to the same loss function.
Notice that in the case of the distance-five rotated surface code, the $X$
network receives a 3D input of dimensions $3 \times 4 \times 6$ and the $Z$
network
receives a 3D input of dimensions $4 \times 3 \times 6$. To create edge features,
the inputs were padded outwards \emph{symmetrically}, i.e. with the same binary
values as their adjacent bits. This changes the input dimensions to $4 \times 5
\times 6$ and $5 \times 4 \times 6$ respectively for the $X$ and $Z$ animations.
Via similar experiments, we realized that two convolutional layers do a better
job in capturing patterns in the syndromes data. The first convolutional layer
is probed by a $3 \times 3 \times 3$ kernel, and the second layer is probed by
a $4 \times 4 \times 4$ kernel. After convolutional layers, a fully connected
feedforward layer with dropouts and relu activations is applied to the extracted
features and then the softmax cross-entropy is measured. The schematic of such a
neural network is depicted in \cref{fig:CNN}.

\section{Numerical experiments}
\label{sec:NumericalExperiments}

In the experimental results reported in this section, multiple data sets were
generated by various choices of physical error rates ranging between $p = 1.0
\times 10^{-4}$ to $p = 2.0 \times 10^{-3}$. Every data set consisted of
simulating the circuit-level depolarizing channel (see
\cref{sec:FaultTolerantProtocols} for a detailed description of the noise model)
for the corresponding circuit, and included the syndrome and resulting error bit
strings in the data set. Note that the error strings are only used as part of
the simulation to compute the vector $\bold{b}$ of logical faults. In an actual
experiment, $\bold{b}$ would be given directly (see the discussion above
\cref{subsec:DLOverview}). We excluded the cases were both the syndrome and
error strings were all zeros. The simulation was continued until a target number
of \textit{non-zero} training samples were gathered. The target size of the
training data set was chosen as $2 \times 10^6$ for distance-three codes, and as
$2 \times 10^7$ for distance-five codes.

Hypertuning was performed with the help of BayesOpt
\cite{JMLR:v15:martinezcantin14a}. In every hypertuning experiment, each query
consisted of a full round of training the deep learning network on 90\% of the
entire dataset and cross-validating on the remaining 10\%. It is important to
add randomness to the selection of the training and cross-validating data sets
so that the hyperparameters do not get tuned for a fixed choice of data entries.
To this aim, we uniformly randomly choose an initial element in the entire data
set, take the 90\% of the dataset starting from that initial element (in a
cyclic fashion) as the training set, and the following 10\% as the test dataset.

The cross-entropy of the test set is returned as the final outcome of one query
made by the hypertuning engine. For all hypertuning experiments, 10 initial
queries were performed via Latin hypercube sampling. After the initial queries,
50 iterations of hypertuning were performed.

For each fault-tolerant error correction scheme, hypertuning was performed on
only a single data set (i.e. only for one of the physical error rates). A more
meticulous investigation may consist of hypertuning for each individual physical
error rate separately but we avoided that, since we empirically observed that
the results are independent of the choice of hypertuning data set. At any rate,
the data chosen for distance-three codes was the one corresponding to $p = 4
\times 10^{-4}$. For the distance-five rotated surface code, $p= 6.0 \times
10^{-4}$ and for the 19-qubit color code using Steane and Knill-EC, $p= 1.0
\times 10^{-3}$ were chosen for hypertuning.

Hyperparameters chosen from this step were used identically for training all
other data sets. For every data set (i.e. every choice of physical fault rate
$p$) the deep learning experiment was run 10 times and in the diagrams reported
below the average and standard deviations are reported as points and error bars.
In every one of the 10 runs, the training was done on 90\% of a data set, and
cross validation was done on the remaining 10\%. All the machine learning
experiments were implemented in Python 2.7 using TensorFlow
1.4\cite{tensorflow2015-whitepaper} on top of CUDA 9.0 running installed on
TitanXp and TitanV GPUs produced by
NVIDIA\cite{Nickolls:2008:SPP:1365490.1365500}.

All experiments are reported in \cref{fig:SteaneD3LU}--\cref{fig:SurfaceD5PE}.
Before continuing with detailed information on each experiment, we refer the
reader to \cref{table:Summary} where we provide the largest ratios of the
pseudo-thresholds obtained using a neural network decoder to pseudo-thresholds
obtained from bare lookup table decoders of each fault-tolerant protocol
considered in this paper.

\begin{table}[h!]
  \begin{tabular}[t]{c|c|c|c}
    {\bf FTEC} & {\bf Lookup} & {\bf DND} & {\bf Ratio}\\
    \hline
    $d=3$ Steane 
    & $p_\text{th} = 2.10 \times 10^{-4}$ 
    & $p_\text{th} = 3.98 \times 10^{-4}$ 
    & 1.90 \\
    $d=5$ Steane 
    & $p_\text{th} = 1.43 \times 10^{-3}$ 
    & $p_\text{th} = 2.17 \times 10^{-3}$ 
    & 1.52 \\
    $d=3$ Knill 
    & $p_\text{th} = 1.76 \times 10^{-4}$ 
    & $p_\text{th} = 2.22 \times 10^{-4}$ 
    & 1.26 \\
    $d=5$ Knill 
    & $p_\text{th} = 1.34 \times 10^{-3}$ 
    & $p_\text{th} = 1.54 \times 10^{-3}$ 
    & 1.15 \\
    $d=3$ Surface code 
    & $p_\text{th} = 2.57 \times 10^{-4}$ 
    & $p_\text{th} = 3.18 \times 10^{-4}$ 
    & 1.24 \\
    $d=5$ Surface code 
    & $p_\text{th} = 5.82 \times 10^{-4}$ 
    & $p_\text{th} = 7.11 \times 10^{-4}$ 
    & 1.22 
  \end{tabular}
\captionof{table}{
Pseudo-thresholds for the 6 fault-tolerant error correction protocols considered
in the experiments. The second column corresponds to the highest 
pseudo-thresholds obtained from a bare lookup table decoder whereas the third
column gives the highest pseudo-thresholds using neural network decoders. The
last column corresponds to the ratio between the pseudo-thresholds obtained from
the best neural network decoders and the lookup table decoders.}
\label{table:Summary}
\end{table}

\subsubsection{Steane-EC CNOT-exRec for the \codepar{7,1,3} code} 

\begin{table}[h!]
  \begin{tabular}[t]{ l|l|l}
    {\bf parameter} & {\bf lower bound} & {\bf upper bound} \\ \hline
    decay rate & $0.0$ & $1.0 - 10^{-6.0}$\\ 
    momentum & $0.0$ & $1.0 - 10^{-6.0}$\\ 
    learning rate & $10^{-5.0}$ & $10^{-1.0}$\\
    initial std & $10^{-3.0}$ & $10^{-1.0}$ \\ 
    num hiddens & $100$& $1000$
  \end{tabular} 
\captionof{table}
{Bayesian optimization parameters for the CNOT-exRec of the \codepar{7,1,3} code
using Steane and Knill-EC and the distance-three rotated surface code. Here the
decay rate, momentum and learning rate pertain to
the parameters of RMSProp. The row `initial std' refers to the standard
deviation of the initial weights in the neural networks, the mean of the weights
was set to zero. The initial biases of the neural networks were set to zero. The
row `num hiddens' refers to the number of hidden nodes in the layers of neural
network. This parameter is optimized for each layer of the neural network
independently (e.g. for a feedforward network consisting of 3 hidden layers,
there are 3 numbers of hidden nodes to be tuned). For an RNN this number
indicates the number of hidden nodes in every one of the 4 hidden layers of the
LSTM unit (all of the same size). }
\label{table:D3Hyperparameters}
 \label{Tab:ContinuousHyper}
\end{table}

The considered continuous and integer hyperparameters are given in
\cref{Tab:ContinuousHyper}.

We also tuned over the categorical parameters of
\cref{table:SteaneD3Categorical}.
\begin{table}[h!]
  \begin{tabular}[t]{ l|l}
    {\bf parameter} & {\bf values} \\ \hline
    activation functions & relu, tanh, sigmoid, identity\\ 
    numbers of hidden layers & 0, 1, 2, \ldots 
  \end{tabular}
\captionof{table}{Categorical hyperparameters. Optimizations over activation
functions was only performed for the distance-three Steane code. Since rectified
linear units showed better results, we committed to this choice for all other
error correction schemes. However, for the second categorical hyperparameter
(the numbers of hidden layers), the search was performed for all error
correction schemes separately and was stopped at the numbers of hidden layers
where the improvements in the results discontinued.}
\label{table:SteaneD3Categorical}
\end{table}
The categorical parameters are tuned via grid-search. We observed that for all
choices of neural networks (feedforward networks with various numbers of hidden
layers and recurrent neural networks with or without peepholes), the rectified
linear unit in the hidden layers and identity for the last layer resulted in the
best performance. We accepted this choice of activation functions in all other
experiments without repeating a grid-search.

\cref{fig:SteaneD3LU,fig:SteaneD3PE} compare the performance of the feedforward
and RNN decoders that respectively use the lookup table and naive-decoder as
their underlying decoders, respectively referred to as LU-based deep neural
decoders (LU-DND) and PE-based deep neural decoders (PE-DND). We use PE since
naive-decoders correct by applying pure errors. We
observe that softmax regression (i.e. zero hidden layers) is enough to get
results on par with the lookup table method in the LU-based training method,
this was not the case in the PE-based method. The RNNs perform well but they are
outperformed by two-hidden-layer feedforward networks. Additional hidden layers
improve the results in deep learning. However, since this is in expense for a
cross-entropy optimization in higher dimensions, the training of deeper networks
is significantly more challenging. This trade-off is the reason the feedforward
networks improve up to two hidden layers, but passing to three and higher
numbers of hidden layers gave worse results (not reported in these diagrams).

We finally observe that PE-DND with even a single hidden layer feedforward
network is almost on par with the LU-DND with two hidden layers. This is an
impressive result given the fact that a table of pure errors grows linearly in
the number of syndromes, but a lookup table grows exponentially. We believe
this is a result of the fact that logical faults are much more likely to occur
when using recovery operators which only consist of products of pure-errors, the
training sets are less sparse and therefore deep learning is able to
capture more patterns for the classification task at hand.

\subsubsection{Knill-EC CNOT-exRec for the \codepar{7,1,3} code}

The hypertuning of continuous variables was done using the same bounds as in
\cref{table:D3Hyperparameters}. \cref{fig:KnillD3LU,fig:KnillD3PE}
respectively show the results of LU-DND and PE-DND methods. The best results
were obtained by feedforward networks with respectively 3 and 2 hidden layers,
in both cases slightly outperforming RNNs.

\subsubsection{Distance-three rotated surface code}

Similar to the previous distance-three codes, we compared using RNNs with
feedforward networks with multiple hidden layers. We observed that the
feedfoward network with a single hidden layer achieves the best performance and
RNNs do not improve the results. Also consistent with the distance-three CNOT-
exRec results, the PE-based DND can perform as good as the LU-based one (and
slightly improves upon it). Results of these experiments are reported in
\cref{fig:SurfaceD3LU,fig:SurfaceD3PE}.

\subsubsection{Steane-EC CNOT-exRec for the \codepar{19,1,5} code}

As the size of the input and output layers of DNNs grow, the ranges of the
optimal hyperparameters change. For the distance-five Steane exRec circuit
applied to the \codepar{19,1,5} color code, the considered hyperparameter ranges
(allowing smaller orders of magnitudes for the initial weight standard
deviations and much smaller learning rates) are given in \cref{Tab:D5Hypers}.
 
\begin{table}[h!]
  \begin{tabular}[t]{ l|l|l}
    {\bf parameter} & {\bf lower bound} & {\bf upper bound} \\ \hline
    decay rate & $0.0$ & $1.0 - 10^{-6.0}$\\ 
    momentum & $0.0$ & $1.0 - 10^{-6.0}$\\ 
    learning rate & $10^{-7.0}$ & $10^{-3.0}$\\
    initial std & $10^{-5.0}$ & $10^{-3.0}$ \\ 
    num hiddens & $100$& $1000$
  \end{tabular}
\captionof{table}{Bayesian optimization parameters for $d=5$ Steane and Knill
CNOT-exRecs. Given the larger size of the training sets and longer input
strings, for these datasets, smaller orders of magnitudes for the initial weight
standard deviations and much smaller learning rates were
explored.}
\label{Tab:D5Hypers}
\end{table}

\cref{fig:SteaneD5PE,fig:SteaneD5LU} show that the PE-DNDs has a slightly harder
time with pattern recognition compared to the LU-DNDs. Nevertheless, both
methods significantly improve the pseudo-thresholds of the distance-five
Steane-EC scheme, with no advantage obtained from using an RNN over a 2-hidden
layer feedforward network. In both experiments, the 3-hidden layer feedforward
networks also did not result any improvements.

\subsubsection{Knill-EC CNOT-exRec for the \codepar{19,1,5} code}

The hyperparameter ranges used for hypertuning were similar to those obtained
for the Steane-EC CNOT-exRec applied to the \codepar{19,1,5} code. Given the
effectiveness of the 2-hidden layer feedforward network, this feedforward neural
network was chosen for the Knill exRec $d=5$ experiment. We see a similar
improvement on the pseudo-threshold of the error correction scheme using either
of LU-DND and PE-DND.

\subsubsection{Distance-five rotated surface code}

For rotated surface codes, we only considered numerical simulations using one EC
rather than the full exRec. This choice was made to be consistent with previous
analyses of the surface codes performance.

The hyperparameter ranges used for hypertuning the feedforward neural networks
were chosen according to \cref{Tab:D5SurfaceHypers}.
\begin{table}[h!]
  \begin{tabular}[t]{ l|l|l}
    {\bf parameter} & {\bf lower bound} & {\bf upper bound} \\ \hline
    decay rate & $0.0$ & $1.0 - 10^{-6.0}$\\ 
    momentum & $0.0$ & $1.0 - 10^{-6.0}$\\ 
    learning rate & $10^{-6.0}$ & $10^{-2.0}$\\
    initial std & $10^{-6.0}$ & $10^{-2.0}$ \\ 
    num hiddens & $100$& $1000$
  \end{tabular}
\captionof{table}{
Bayesian optimization parameters for the distance-five rotated surface code. The
parameter search is in a slightly tighter domain than in the case of the
distance-five Knill and Steane CNOT-exRecs in view of the empirical initial
tests performed.}
\label{Tab:D5SurfaceHypers}
\end{table}

As explained in the previous section, a CNN engineered appropriately could be a
viable layout design for large surface codes. Beside previous hyperparameters,
we now also need to tune the number of filters, and drop-out rate. A summary of
the settings for Bayesian optimization are given in \cref{Tab:CNNsurfaceD5}.

\begin{table}[h!]
  \begin{tabular}[t]{ l|l|l}
    {\bf parameter} & {\bf lower bound} & {\bf upper bound} \\ \hline
    decay rate & $0.0$ & $1.0 - 10^{-6.0}$\\ 
    momentum & $0.0$ & $1.0 - 10^{-6.0}$\\ 
    learning rate & $10^{-6.0}$ & $10^{-2.0}$\\
    initial std & $10^{-6.0}$ & $10^{-2.0}$\\ 
    num hiddens & $100$& $1000$\\
    keep rate & $0.0$& $1.0$\\
    num filters & $5$& $10$
  \end{tabular}
\captionof{table}{
Bayesian optimization parameters for a 3-dimensional CNN. The filters were fixed
to be $3 \times 3 \times 3$ and $4 \times 4 \times 4$ but their quantities 
were tuned. Since CNNs are larger and deeper than other networks considered in
this paper, they are more prone to vanishing gradients. Therefore it is
beneficial to consider drop-outs in the hidden layer after feature extraction.
The hyperparameter corresponding to drop-outs is `keep rate' allowing more
drop-outs when it is smaller.}
\label{table:CNNHyperparameters}
\label{Tab:CNNsurfaceD5}
\end{table}

We compare the PE-based and LU-based feedforward networks with the CNN proposed
in \cref{subsec:DNDSurfaceCode}. \cref{fig:SurfaceD5LU,fig:SurfaceD5PE} show
that feedforward networks with 2 hidden layers result in significant
improvements both using the PE-based and LU-based DNDs. The 3D-CNN is slightly
improving the results of the feedforward network in PE-DND but is only slightly
better than the lookup table based method in the LU-DND case. The best overall
performance is obtained by using a feedfoward network with 2 hidden layers for the
LU-DND. A slightly less performant result can also be obtained if the PE-DND 
method is used in conjunction with either of the 2-hidden layer feedforward
network or the 3D convolutional neural network.

\begingroup
\leftskip-5cm
\begin{figure*}
\centering
\begin{minipage}[t]{.49\textwidth}
\includegraphics[trim=2cm 0 0 0, scale=.45]{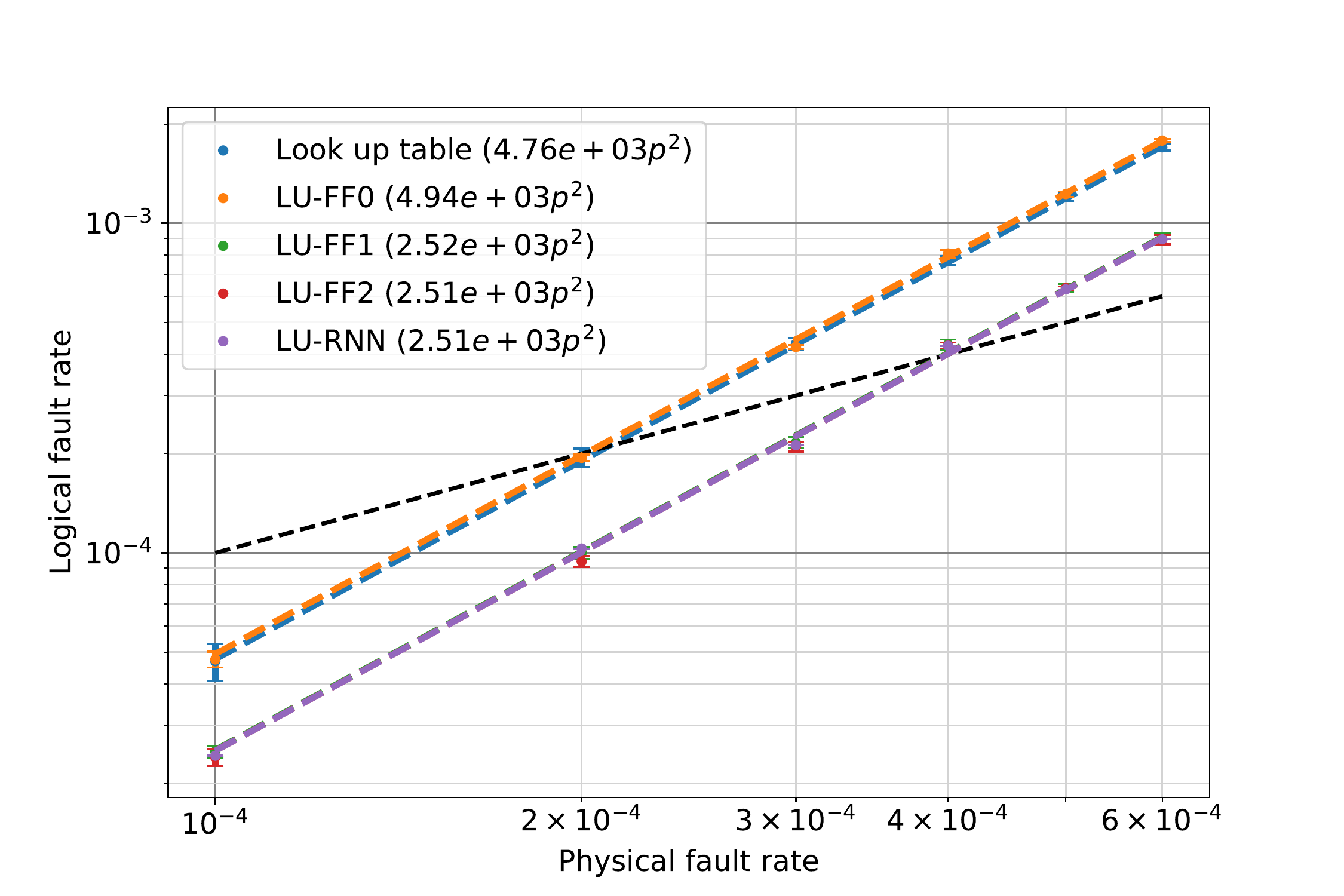}
\caption{LU-DND for the distance-three Steane CNOT-exRec.}
\label{fig:SteaneD3LU}
\end{minipage}
\begin{minipage}[t]{.49\textwidth}
\includegraphics[trim=1cm 0 0 0, scale=.45]{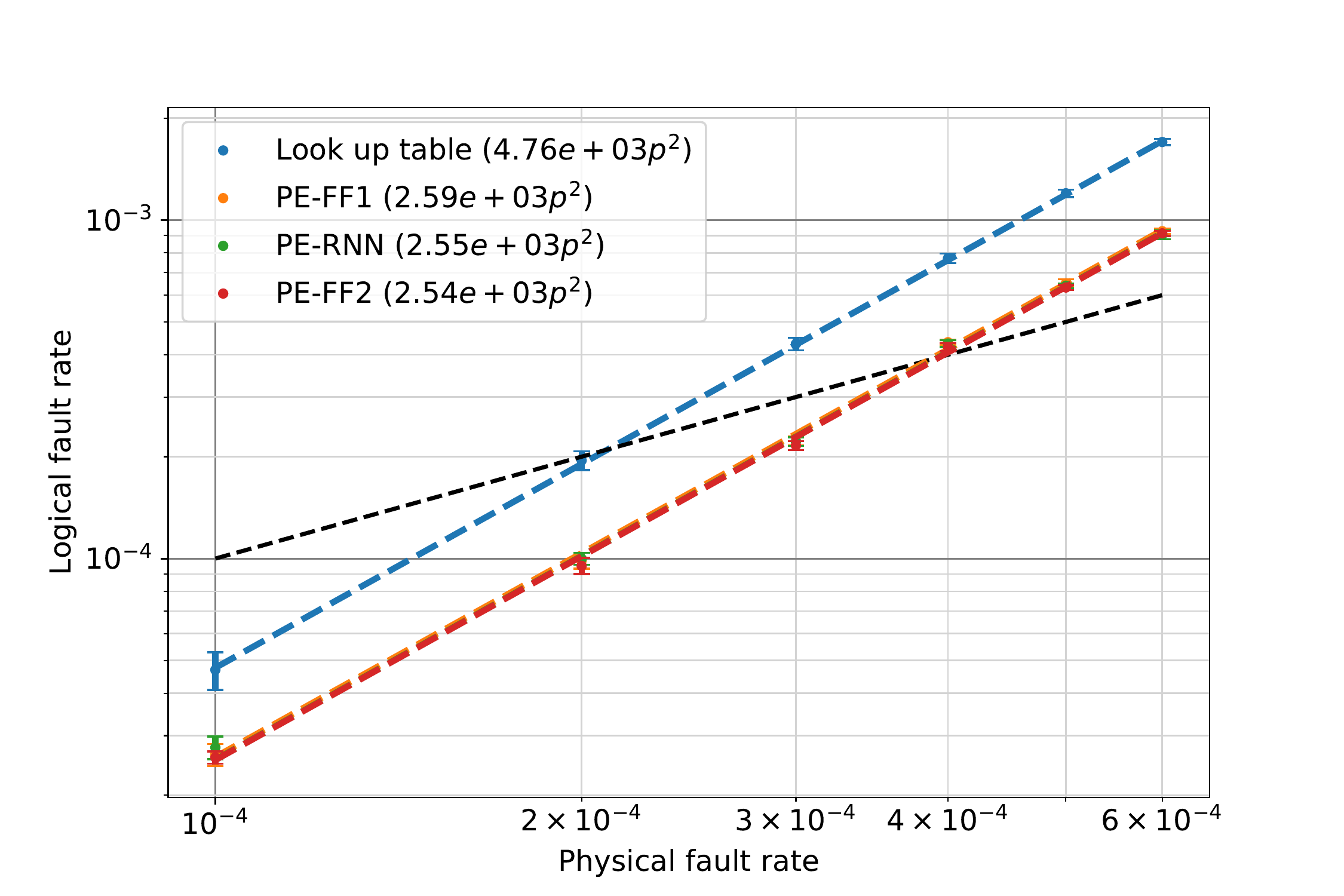}
\caption{PE-DND for the distance-three Steane CNOT-exRec.}
\label{fig:SteaneD3PE}
\end{minipage}
\centering
\begin{minipage}[t]{.49\textwidth}
\includegraphics[trim=2cm 0 0 0, scale=.45]{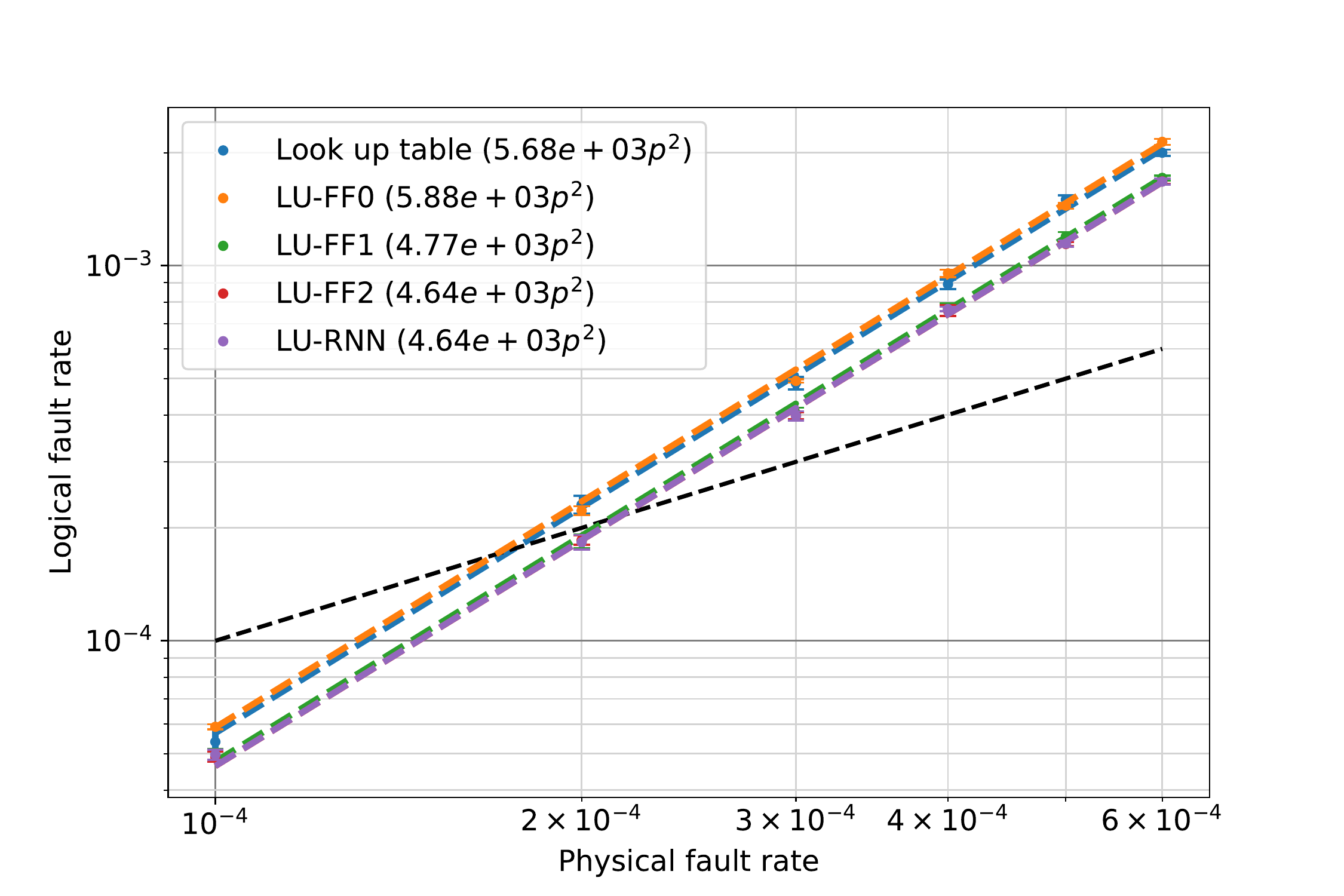}
\caption{LU-DND for the distance-three Knill CNOT-exRec.}
\label{fig:KnillD3LU}
\end{minipage}
\begin{minipage}[t]{.49\textwidth}
\includegraphics[trim=1cm 0 0 0, scale=.45]{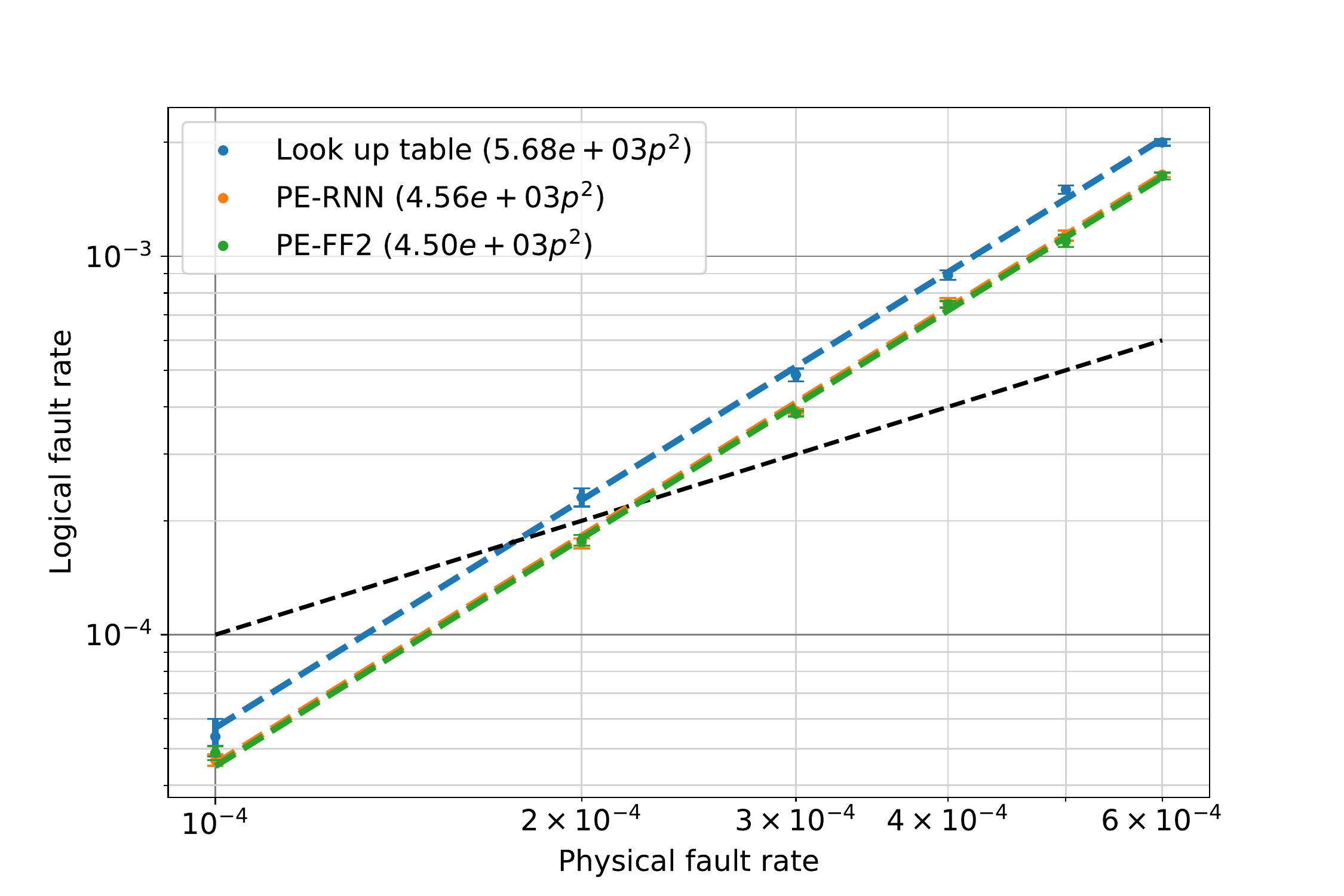}
\caption{PE-DND for the distance-three Knill CNOT-exRec.}
\label{fig:KnillD3PE}
\end{minipage}
\centering
\begin{minipage}[t]{.49\textwidth}
\includegraphics[trim=2cm 0 0 0, scale=.45]{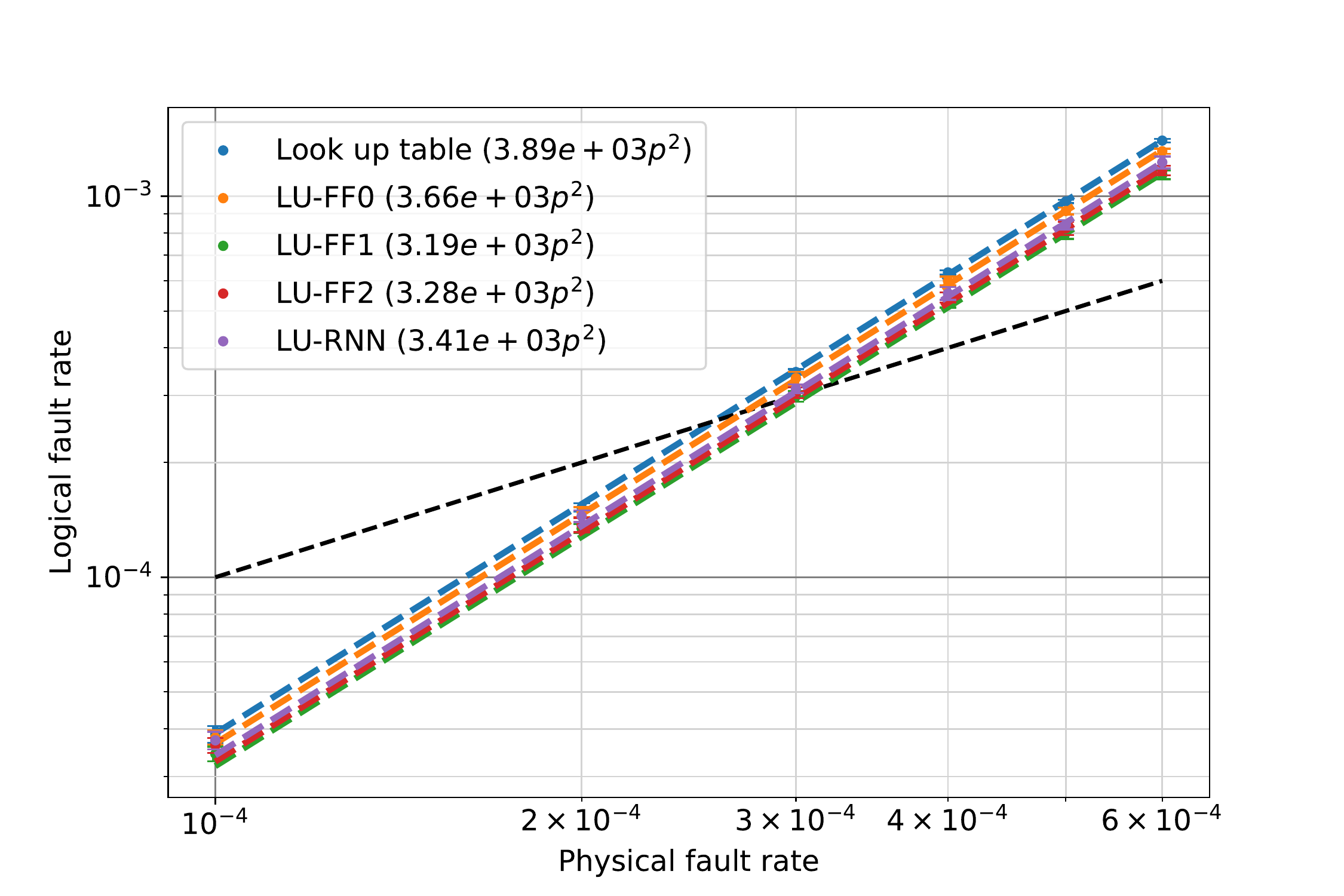}
\caption{LU-DND for the distance-three surface code.}
\label{fig:SurfaceD3LU}
\end{minipage}
\begin{minipage}[t]{.49\textwidth}
\includegraphics[trim=1cm 0 0 0, scale=.45]{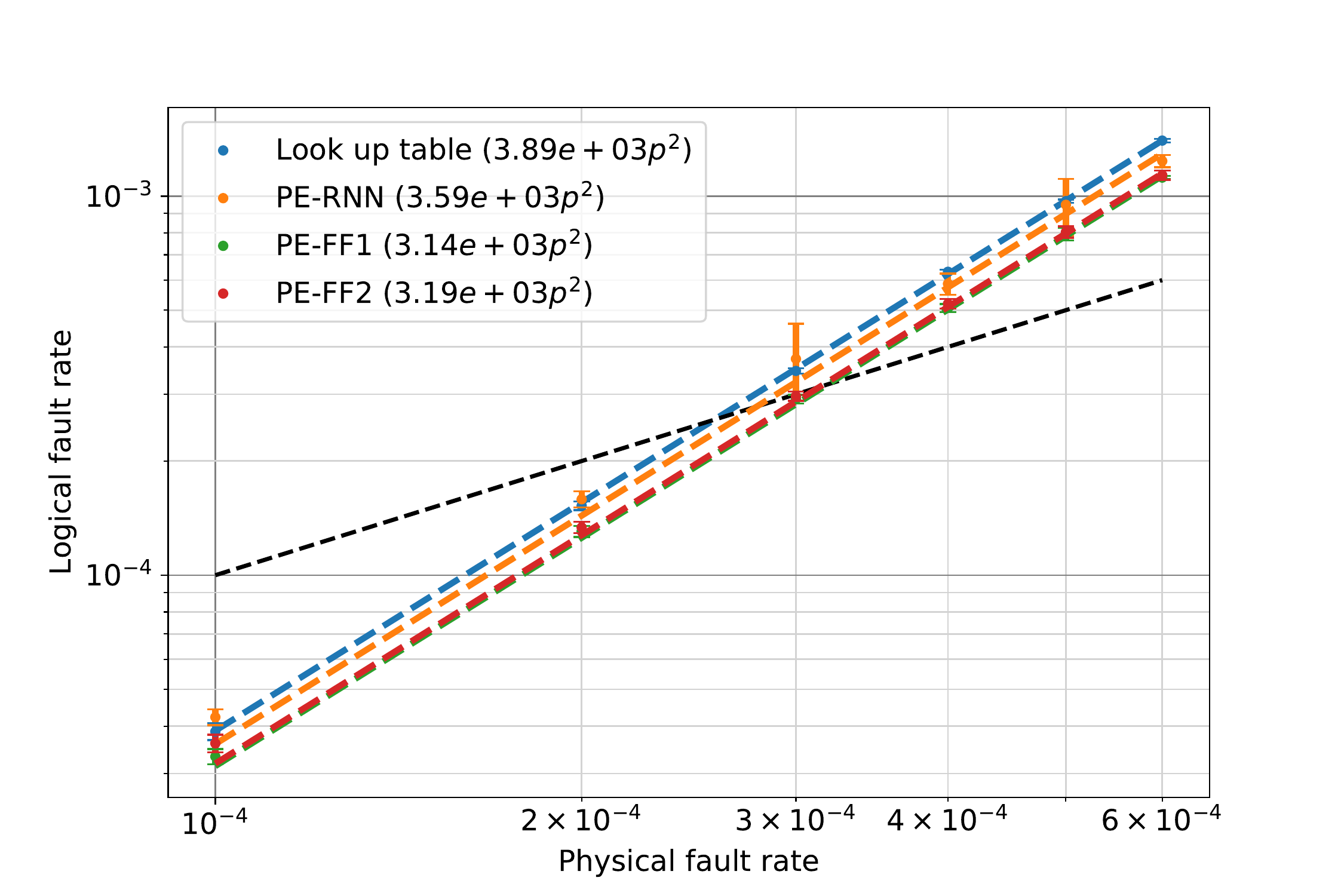}
\caption{PE-DND for the distance-five surface code.}
\label{fig:SurfaceD3PE}
\end{minipage}
\caption*{In \cref{fig:SteaneD3LU}--\cref{fig:SurfaceD3PE} each data point has
the height on the vertical axis being the average of 10 logical fault rates 
collected for each physical fault rate $p$ specified on the horizontal axis.
Error bars represent the standard deviation from these average values.
For each DND-based decoder, the cuve-fitting method used is a non-linear least 
square fitting between the average logical fault rates as a function of the
physical fault rates, and a quadratic monomial.}
\end{figure*}
\endgroup
\begingroup
\leftskip-5cm
\begin{figure*}
\begin{minipage}[b]{.49\textwidth}
\includegraphics[trim=2cm 0 0 0, scale=.45]{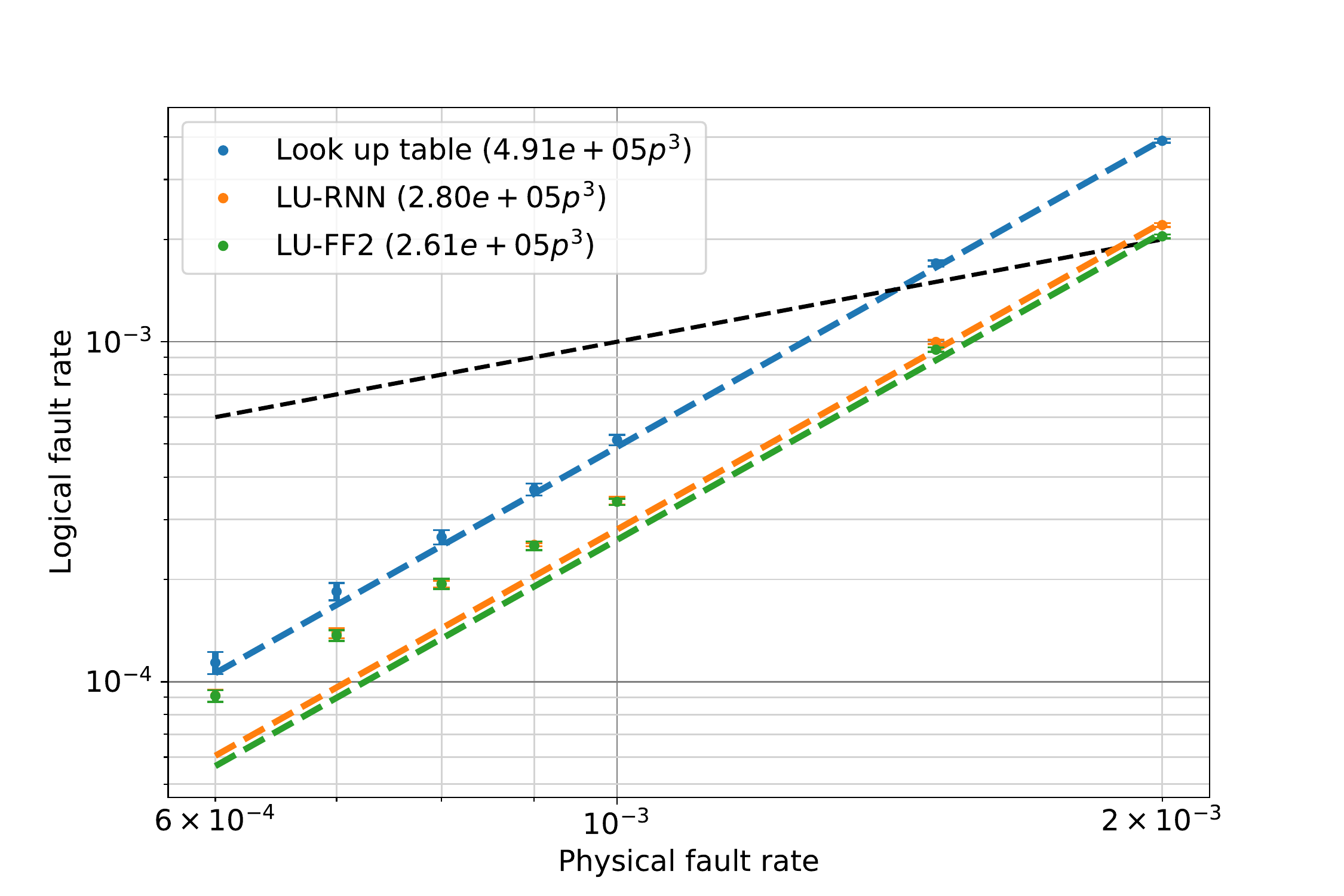}
\caption{LU-DND for the distance-five Steane CNOT-exRec.}
\label{fig:SteaneD5LU}
\end{minipage}
\begin{minipage}[b]{.49\textwidth}
\includegraphics[trim=1cm 0 0 0, scale=.45]{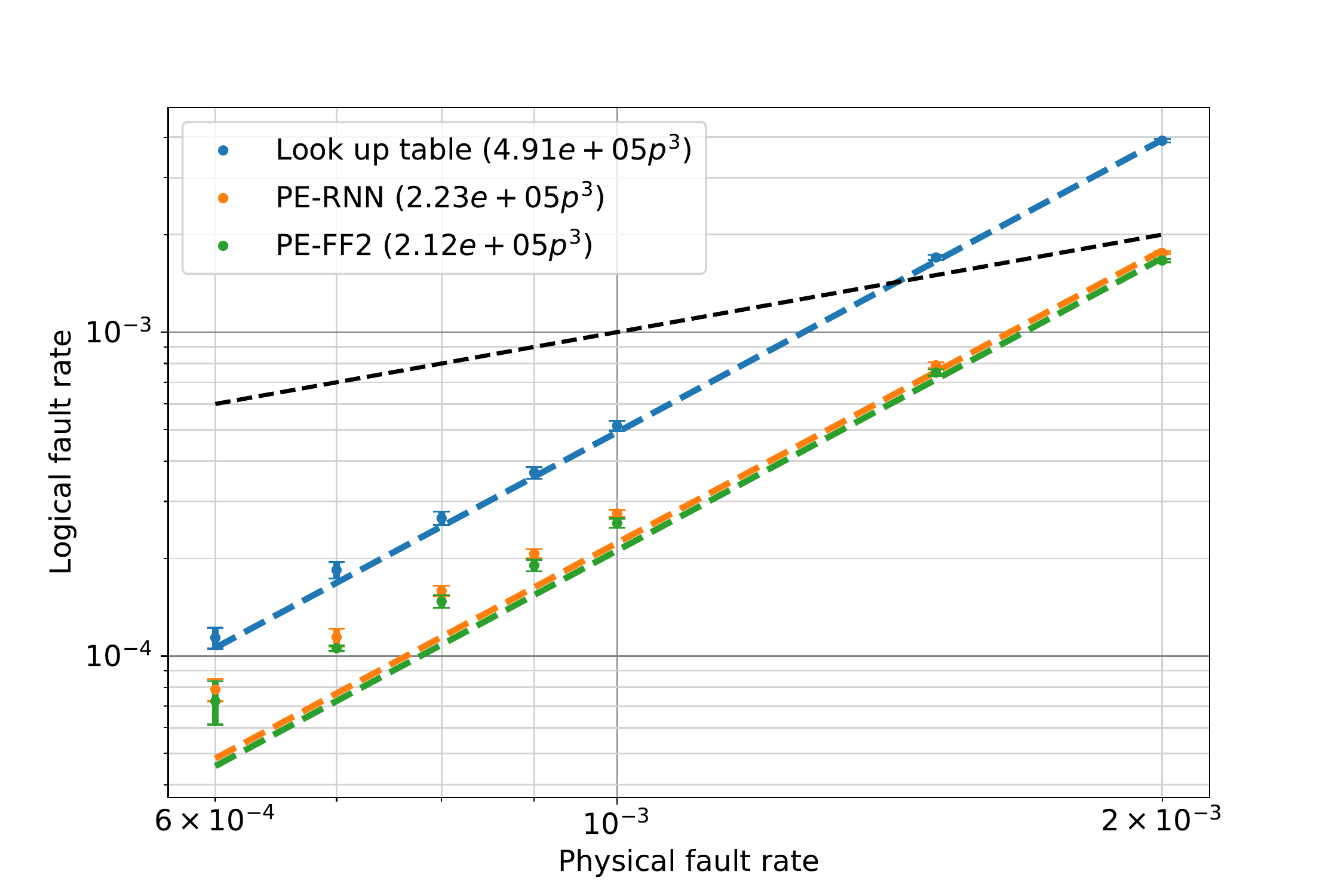}
\caption{PE-DND for the distance-five Steane CNOT-exRec.}
\label{fig:SteaneD5PE}
\end{minipage}
\centering
\begin{minipage}[b]{.49\textwidth}
\includegraphics[trim=2cm 0 0 0, scale=.45]{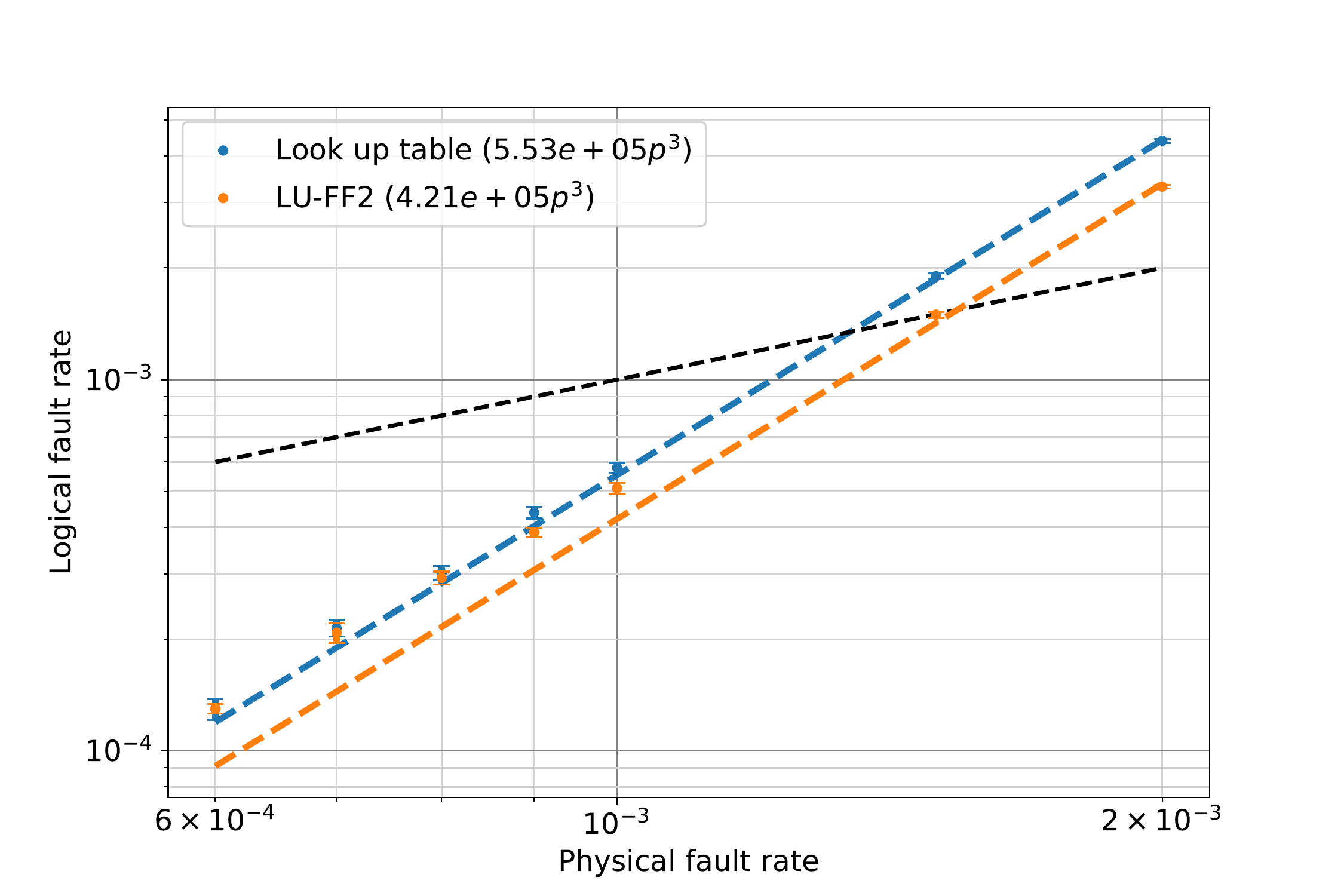}
\caption{LU-DND for the distance-five Knill CNOT-exRec.}
\label{fig:KnillD5LU}
\end{minipage}
\begin{minipage}[b]{.49\textwidth}
\includegraphics[trim=1cm 0 0 0, scale=.45]{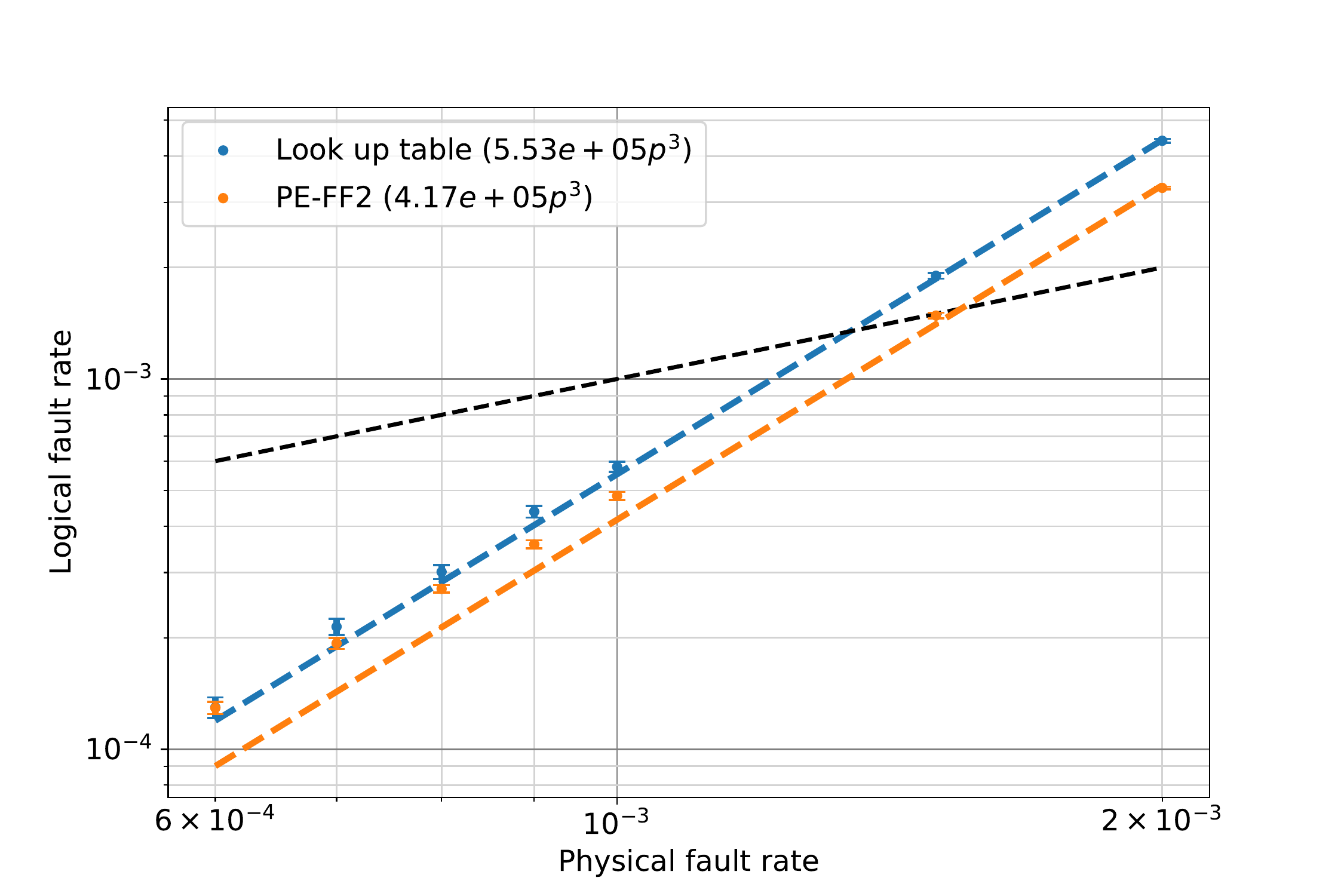}
\caption{PE-DND for the distance-five Knill CNOT-exRec.}
\label{fig:KnillD5PE}
\end{minipage}
\centering
\begin{minipage}[b]{.49\textwidth}
\includegraphics[trim=2cm 0 0 0, scale=.45]{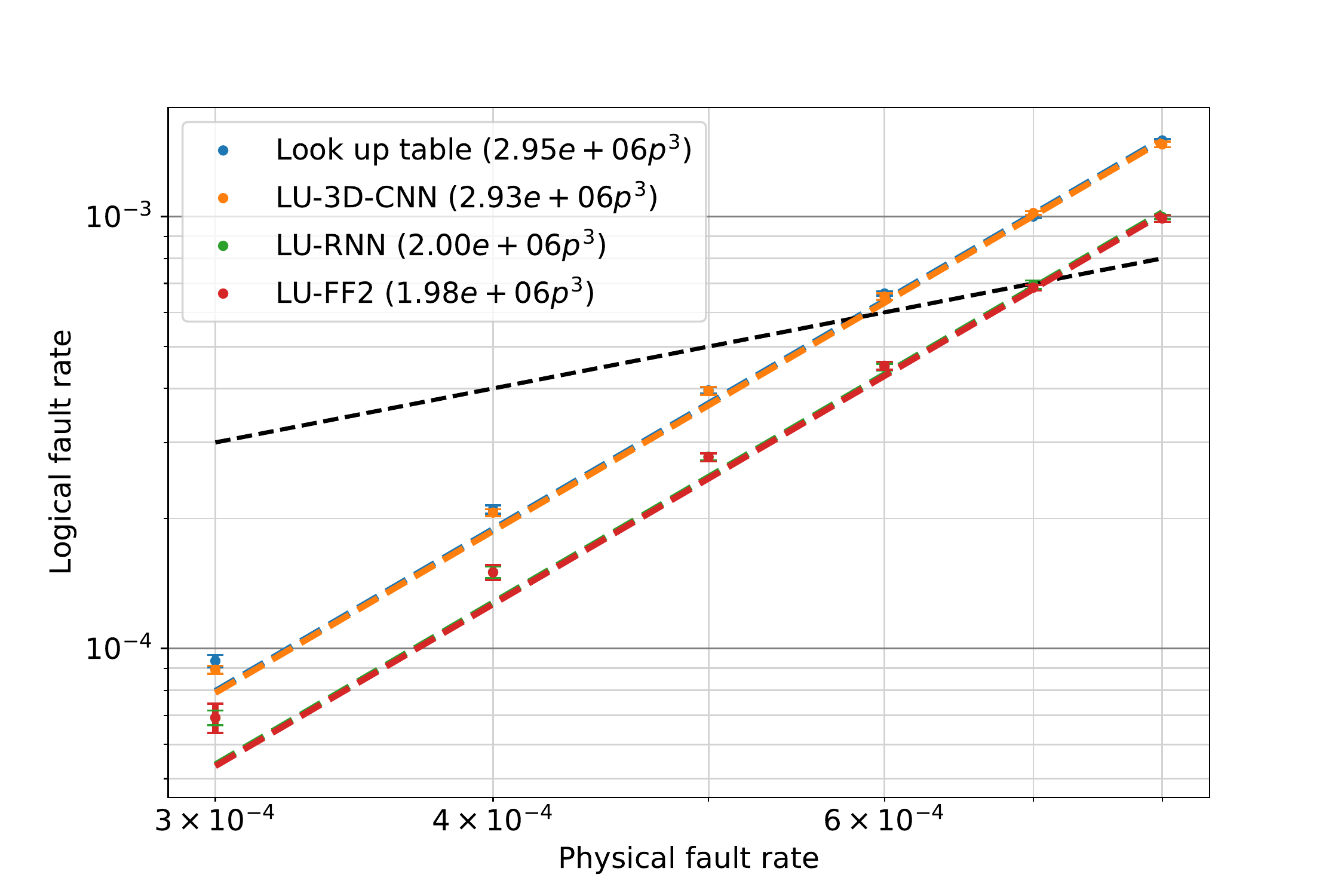}
\caption{LU-DND for the distance-five surface code.}
\label{fig:SurfaceD5LU}
\end{minipage}
\begin{minipage}[b]{.49\textwidth}
\includegraphics[trim=1cm 0 0 0, scale=.45]{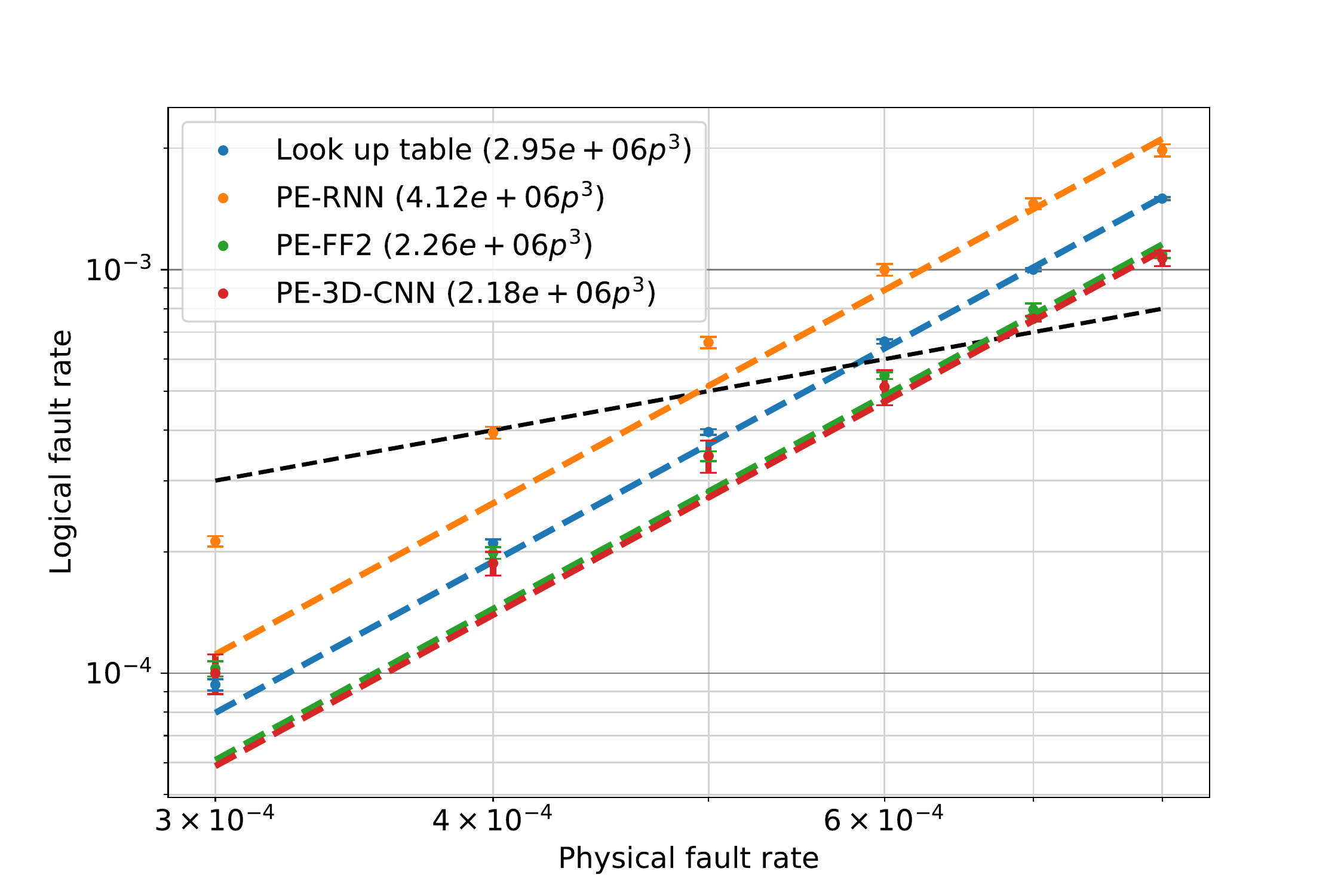}
\caption{PE-DND for the distance-five surface code.}
\label{fig:SurfaceD5PE}
\end{minipage}
\caption*{In \cref{fig:SteaneD5LU}--\cref{fig:SurfaceD5PE} data points, averages 
and error bars are obtaines in a similar fashion to 
\cref{fig:SteaneD3LU}--\cref{fig:SurfaceD3PE}. The cuve-fitting method is also
a non-linear least square method, this time fitting a cubic monomial through the
data points.}
\end{figure*}
\endgroup

\section{Performance analysis}
\label{sec:performance}

In this section we consider the efficiency of the deep neural decoders in
comparison to the lookup table decoders described in
\cref{subsec:RotatedSurfaceCode,subsec:SteaneEC}. The size of a lookup table
grows exponentially in the number of syndromes therefore making lookup table
based decoding intractable as the codes grow. However, it is important to note
that as long as the size of the lookup table allows for storage of the entire
table in memory, as described in \cref{sec:LookpNaiveComplexity}, the lookup
from an array or a hash table happens effectively in
$O(1)$ time. Therefore a lookup table based decoding scheme would be the most
efficient decoder by far. A similar approach to a lookup table decoder is
possible by making an inference mapping from all the possible input strings of
a trained neural decoder. This method is discussed in \cref{sec:neurallookup}.
For larger codes, neither a lookup table decoder, nor an inference mapping
decoder is an option due to exponentially growing memory usage. 

More complicated decoders such as minimum weight perfect matching can be
extremely inefficient solutions for decoding despite polynomial asymptotic
complexity. With gates operating at 100Mhz (that is $10$ns gate times) 
\cite{0034-4885-80-10-106001}, 
which is much faster than the state of the art\footnote{In fact, existing
prototypes of quantum computers have much longer gate delays. Typical gate times
in a superconducting system are $130$ns for single-qubit and $250-450$ns for
2-qubit gates. For a trapped-ion system, gate times are even longer, reaching
$20\mu $s for single-qubit gates and $250\mu$s for 2-qubit gates
\cite{QISKit,Linke3305}.}, the
simplest quantum algorithms foreseen to run on near term devices would require
days of runtime on the system \cite{2017PNAS..114.7555R}. With the above
gate times, the CNOT-exRec using Steane and Knill EC units as well as the
multiple rounds of EC for surface codes would take as small as a hundred
nanoseconds. In order to perform active error correction, we require classical
decoding times to be implemented on (at worst) a comparable time scale as the EC
units, and therefore merely a complexity theoretic analysis of a decoding
algorithm is not enough for making it a viable solution. Alternatively, given
a trained DND, inference of new recovery operations from it is a simple
algorithm requiring a sequence of highly parallelizable matrix multiplications.
We will discuss this approach in \cref{sec:inference} and \cref{sec:digital}.

\subsection{Inference mapping from a neural decoder}
\label{sec:neurallookup}

For codes of arbitrary size, the most time-performant way to use a
deep neural decoder is to create an array of all inputs and outputs of the DNN
in the test mode (i.e. an inference map which stores all possible syndromes
obtained from an EC unit and assigns each combination to a recovery
operator\footnote{For the CNOT-exRec, the inference map would map syndromes from
all four EC units to a recovery operator. For the surface code, the inference
map would map syndromes measured in each round to a recovery operator.}). This
is possible for distance-three fault-tolerant EC schemes such as Steane, Knill
and surface codes
(as well as other topological schemes such as those used for color codes). For
all these codes, the memory required to store the inference map is 2.10
megabytes. This method is not feasible for larger distance codes. For the Knill
and Steane-EC schemes applied to the \codepar{19,1,5} color code, the memory
required is 590 exabytes and for the distance-five rotated surface code it is
$2.79 \times 10^{24}$ exabytes.

\subsection{Fast inference from a trained neural network}
\label{sec:inference}

An advantage of a deep neural decoder is that the complications of decoding are
to be dealt with in the training mode of the neural network. The trained network
is then used to suggest recovery operations. The usage of the neural network in
this passive step, i.e. without further training, is called the \emph{inference
mode}. Once the neural network is trained, usage of it in the inference mode
requires only a sequence of few simple arithmetic operations between the
assigned valued of its input nodes and the trained weights. This makes
inference an extremely simple algorithm and therefore a great candidate for
usage as a decoder while the quantum algorithm is proceeding.

However, even for an algorithm as simple as inference, further hardware and
software optimization is required. For example, \cite{CrigerNN17} predicts that
on an FPGA (field-programmable gate array) every inference from a single layer
feedforward network would take as long as 800ns. This is with the optimistic
assumption that float-point arithmetic (in 32 and 64-bit precision) takes 2.5 to
5 nanoseconds and only considering a single layer feedforward network.

In this section, we consider alternative optimization techniques for fast
inference. We will consider a feedforward network with two hidden layers given
their promising performance in our experiments.

\subsubsection{Network quantization} Fortunately, quantum error correction is not
the only place where fast inference is critical. Search engines, voice and
speech recognition, image recognition, image tagging, and many more applications
of machine learning are nowadays critical functions of smart phones and many
other digital devices. As the usage grows, the need for efficient inference from
the trained models of these applications grow. It is also convenient to move
such inference procedures to the usage platforms (e.g. the users smart phones
and other digital devices) than merely a cloud based inference via a data
centre. Recent efforts in high performance computing has focused on fabricating
ASICs (Application Specific Integrated Circuits) specifically for inference from
neural networks. Google's TPU (Tensor Processing Unit)
\cite{2017arXiv170404760J} is being used for inference in Google Search, Google
Photos and in DeepMind's AlphaGo against one of the the world's top Go player,
Lee Sedol.

It is claimed that the reduction in precision of a trained neural network from
32-bit float point precision in weights, biases, and arithmetic operations, to only 8-bit fixed point preserves the quality of inference from trained
models \cite{37631}. This procedure is called \emph{network quantization}. There
is no mathematical reason to believe that the inference quality should hold up
under network quantization. However, the intuitive explanation has been that
although the training mode is very sensitive to small variations of parameters
and hyperparameters, and fluctuations of the high precision weights of the
network in individual iterations of training is very small, the resulting
trained network is in principle robust to noise in data and weights.

The challenge in our case is that in quantum error correction, the input data is
already at the lowest possible precision (each neuron attains 0 or 1, therefore
only using a single bit). Furthermore, an error in the input neurons results in
moving from one input syndrome to a completely different one (for instance, as
opposed to moving from a high resolution picture to a low resolution, or poorly
communicated one in an image processing task). We therefore see the need to
experimentally verify whether network quantization is a viable approach to 
high-performance inference from a DND.

\begin{figure}[t]
\center{}
\includegraphics[scale=.4]{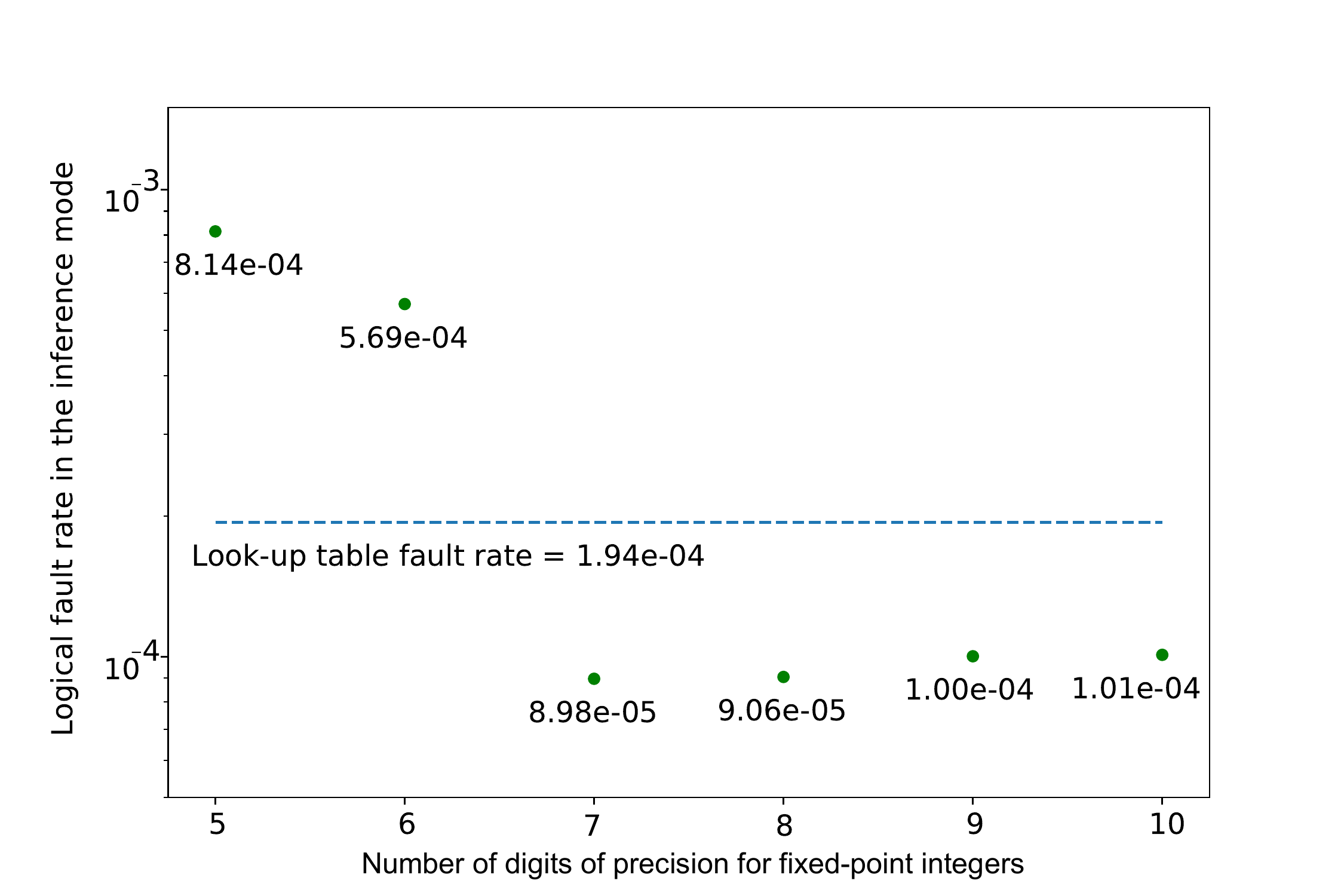}
\caption{
Quantization of the feedforward neural network with 2 hidden layers, trained on
the Steane EC dataset at a physical error rate of $p= 2 \times 10^{-4}$. Each
point is calculated as the average logical error rate obtained from 10 rounds of
training and cross-validating similar to the experiments in
\cref{sec:NumericalExperiments}.}
\label{fig:Quantization}
\end{figure}

\cref{fig:Quantization} demonstrates an experiment to validate network
quantization on a trained DND. Using 32-bit float-point precision, the results
of \cref{fig:SteaneD3LU} show that the trained DND improves the logical fault
rate from $1.95 \times 10^{-4}$ obtained by lookup table methods to $9.45
\times 10^{-5}$ obtained by the LU-DND with 2 hidden layers. We observe that
this improvement is preserved by the quantized networks with 8 bits and even 7
bits of precision using fix-point arithmetic.

We now explain how the quantized network for this experiment was constructed.
Let us assume the available precision is up to $k$ bits. First, the weights and
biases of the network are rescaled and rounded to nearest integers such that the
resulting parameters are all integers between $-2^{k-1}+1$ and $2^{k-1}$ stored
as signed $k$-bit integers. Each individual input neuron only requires a single
bit since they store zeros and ones. But we also require that the result of
feedforward obtained by multiplications and additions and stored in the hidden
layers is also a $k$-bit signed integer. Unlike float-point arithmetic, fixed
point arithmetic operations can and often overflow. The result of multiplication
of two $k$-bit fixed-point integers can span $2k$ bits in the worst case.
Therefore the results of each hidden layer has to be shifted to a number of
significant digits and the rightmost insignificant digits have to be forgotten.
For instance, in the case of the CNOT-exRec with Steane EC units, each input
layer has 12 bits, which get multiplied by 12 signed integers each with $k$-bit
fixed point precision. A bias with $k$-bit fixed point precision is then added
to the result.
We therefore need at most $k + \lceil \log_2 (13) \rceil$-bits to store the
result. Therefore the rightmost $\lceil \log_2 (13) \rceil$ bits have to be
forgotten. If the weights of the trained neural network are symmetric around
zero, it is likely that only a shift to the right by $2$ bits is needed in this
case. Similarly, if each hidden layer has $L$ nodes, the largest shift needed
would be $\lceil \log_2 (L + 1) \rceil$ but most likely $\lceil \log_2 (L +
1)\rceil - 1$ shifts suffices. In the experiment of \cref{fig:Quantization},
each hidden layer had 1000 nodes and the feedforward results were truncated in
their rightmost 9 digits.

\subsection{Classical arithmetic performance}
\label{sec:digital}

In the previous section we showed that 8-bit fixed point arithmetic is all that
is needed for high quality inference from the trained deep neural decoder. We
now consider a customized digital circuit for the inference task and estimate
how fast the arithmetic processing units of this circuit have to be in order for
the inference to be of practical use for active quantum error correction.

The runtime of a digital circuit is estimated by considering the time 
that is required
for the electric signal to travel through the \emph{critical path} of the
logical circuit \cite{Rabaey:2008:DIC:1817056}, the path with the longest
sequence of serial digital operations.

\begin{figure}[t]
\center
\includegraphics[scale=.16]{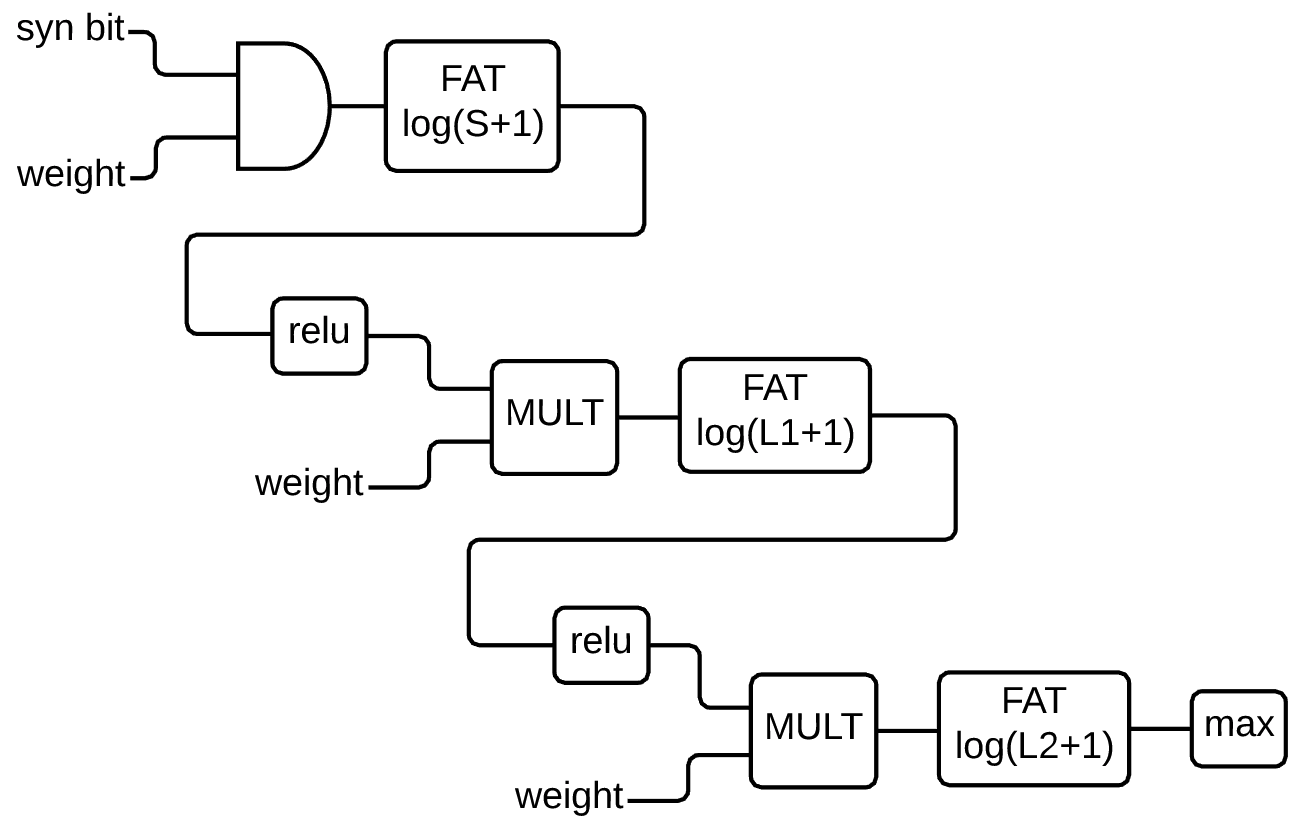}
\caption{The critical path of a custom inference circuit. Every syndrome bit
represents an input node of the neural network and is multiplied by 8-bit
integer weights. A set of such products are added together and together with an
8-bit bias integer to find the activation on a node of the first hidden layer.
Given $S$ input syndromes, this amounts to the addition of $S + 1$ integers
which can be done with a tree of 8-bit integer full-adders (Full-Adder Tree or
FAT for short) of depth $\log(S+1)$. After the quantized rectified linear unit,
a similar procedure is iterated for the first hidden layer with the full-adder
tree of depth $\log(L_1 + 1)$ where $L_1$ is the number of neurons in the first
hidden layer. This pattern continues for other hidden layers. The MAX unit
compares two 8-bit integers and outputs 0 if the first one is bigger and 1 if
the second one is bigger.}
\label{fig:inference}
\end{figure}

\cref{fig:inference} shows the critical path of a circuit customized to
carry inference in a feedforward network with 2 hidden layers. Since the input
neurons represent syndrome bits, multiplying them with the first set of
weights can be done with parallel AND between the syndrome bit and the weight
bits. The rectified linear unit is efficient since it only requires a NAND
between the sign of the 8-bit signed integer with the other 7 bits of
it. The most expensive units in this circuit are the $8\times 8$ multipliers and
adders. Every $8 \times 8$ multiplier gives a 16-bit fixed point integer which
is then shifted 8-bits to the right by ignoring the first 8-bits. The total time
delay $t_{\mathrm{TOT}}$ of this path in the circuit is
\begin{multline}
t_{\mathrm{TOT}} 
= t_{\mathrm{AND}} 
+ \lceil \log (S+1) \rceil \ t_{\mathrm{ADD}} + t_{\mathrm{MAX}}\\
+ \sum_{i= 1}^H (t_{\mathrm{NOT}} + t_{\mathrm{AND}} + t_{\mathrm{MULT}} 
+ \lceil \log(L_i + 1) \rceil \ t_{\mathrm{ADD}})
\end{multline}
where $H$ is the number of hidden layers and $L_i$ is the number of neurons in
the $i$-th hidden layer. From a complexity theoretic point of view this is
promising since it shows that the cost of inference is logarithmic in the number
of syndromes and the size of hidden layers, and linear in the number of hidden
layers. For a feedforward network with two hidden layers and at most $1000$
neurons in each hidden layer,
\begin{multline}
t_{\mathrm{TOT}} = 3\ t_{\mathrm{AND}} + 2\ t_{\mathrm{NOT}} 
+ 2\ t_{\mathrm{MULT}} + t_{\mathrm{MAX}}\\
+ \left( \lceil \log (S+1)\rceil + 20 \right) \ t_{\mathrm{ADD}}.
\end{multline}
Since the adders contribute the most in the above time delay, let us give an
upper bound on how fast the adder units need to be in order for the total time
delay to be comparable to the runtime of the fault-tolerant quantum error
correction protocols considered in this paper.
 
In \cref{table:TimeDelay} we compute upper bounds on the adder units for the
fault-tolerant error correction protocols considered in this paper. We emphasize
that this estimation is by the
optimistic assumption that all independent arithmetic operations are done in
parallel. In reality, this is not possible due to limitations in area and power
consumption of the ASIC. Also considering that multiple rounds of inference have
to happen, a pipeline architecture should be considered for independent batches
of inference on the ASIC.
Lastly, the time for multiplier unit and the comparator
are ignored since (if all independent jobs are done in parallel) there are only
two serial multipliers in the critical path. With all of these considerations,
the last row of this table should be interpreted as an optimistic allowed time
for the adder units and that the actual adder delays should be \emph{well below}
these numbers.

\begin{table}[t]
  \begin{tabular}[t]{c|c|c|c|c}
    {\bf FTEC} & {\bf FTEC} & {\bf Syn} & {\bf Num}
    & {\bf Adder}\\
    {\bf Circuit} & {\bf Depth} & {\bf Size} & {\bf Adders} & {\bf Leniency}\\
    \hline
    $d=3$ Steane & 6 & 12 & 24 & 2.5ns \\
    $d=5$ Steane & 6 & 36 & 26 & 2.3ns \\
    $d=3$ Knill & 8 & 12 & 24 & 3.3ns \\
    $d=5$ Knill & 8 & 36 & 26 & 3.1ns \\
    $d=3$ Surface code & 18 & 12 & 24 & 7.5ns \\
    $d=5$ Surface code & 36 & 72 & 27 & 13.3ns
  \end{tabular}
\captionof{table}{FTEC depth is the depth of the FTEC circuit. For Steane and
Knill EC, this is the depth of the CNOT-exRec circuit (excluding the ancilla
verification steps) and in the surface code, it is the depth of the circuit for
multiple rounds of syndrome measurement (note that for the distance 5 surface
code we considered the worst case of 6 syndrome measurement rounds). The
syndrome size is only that of one of $X$ and $Z$ since the inference for $X$ and
$Z$ logical errors can happen in parallel and independently. The adder time
leniency is calculated based on 10ns quantum gate delays. Therefore, it 
is the depth of the FTEC multiplied by 10ns and divided by the number of
adders.}\label{table:TimeDelay}
\end{table}

In particular we conclude that in order to perform active error correction
with the methods summarized in \cref{table:TimeDelay} on a quantum computer with
$10$ns gate delays, the classical control unit of the quantum computer has to
comprise of arithmetic units that are fast enough to perform arithmetic 
operations well below the time limits reported in the last column of this table.
In hardware engineering, there are many approaches to implementation of
arithmetic and logical units \cite{Parhami:1999:CAA:318930}. Without going into
the details of the circuit designs we mention that the adder leniencies
in \cref{table:TimeDelay} are in reach of high performance VLSI
\cite{25695,488718}, but could be challenging to achieve using FPGAs
\cite{Wolf:2004:FSD:983326,Xing,831434}.

\subsection{Limitations of deep neural decoders}
\label{subsec:Limits}

We interpret the results of this section to suggest that, once implemented on a
high performance computing platform, inference can be computed efficiently from
a trained deep neural decoder. Further, the results of
\cref{sec:NumericalExperiments} show that with a large enough training set,
neural network decoders achieve lower logical failure rates compared to the
lookup table schemes presented in this paper. However, this does not imply that
deep neural decoders are scalable. As the size of the codes grow, training the
neural decoders becomes much more daunting. This is due to the fact that deep
learning classifiers are not viable solutions for \textit{sparse} classification
problems. As the codes become better and/or physical error rates become smaller,
the training samples become more and more sparse, providing less and less
effective training samples for the neural network. Without nontrivial training
samples, the neural networks learn ``zeros" rather than capturing significant
patterns in the data set.

\begin{figure}[t]
\center
\includegraphics[scale=.4]{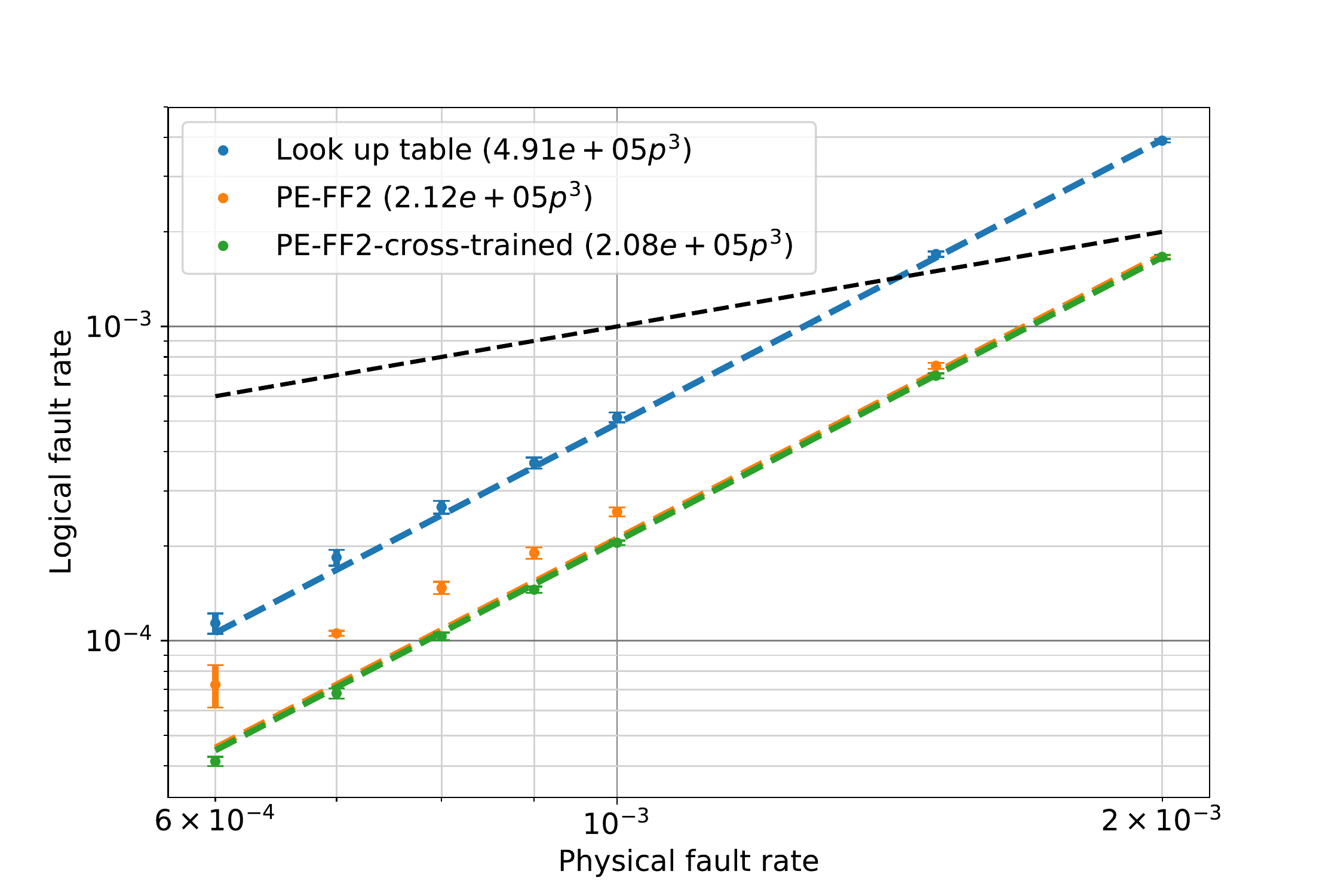}
\caption{
A comparison between two training procedures for the CNOT-exRec of the 
\codepar{19,1,5} color code using Steane-EC units.
The orange dots are the results of training a feedforward network
with 2 hidden layers as reported also in \cref{fig:SteaneD5LU}. In this case,
the DND is trained on a given physical error rate $p$ and tested on a test
dataset for the same physical error rate. We observe that the logical error rate
does not exactly follow a cubic growth since the training is less successful
when the physical error rate is small. The green line, demonstrates the
performance of the same DND if trained only for the largest physical error rate
$p = 2 \times 10^{-3}$, and later on tested on test datasets from every other
physical error rate. The neural network captured syndrome and recovery patterns
occurring in the CNOT-exRec that are valid for all values of physical error rate.
As previously explained, such a training scenario is not possible for real-world
experiments, or on physical realizations of quantum computers.}
\label{fig:crosstraining}
\end{figure}

As evidence for the effect of sparsity of the dataset on successful training
of the deep neural decoding we refer the reader to an experiment reported in
\cref{fig:crosstraining}. In this experiment, the DND is trained on the dataset
corresponding to the highest physical fault rate $p= 2 \times 10^{-3}$. The same
trained DND is used to cross-validate on test datasets for all other physical
fault rates. We observe that this DND is more successful in recovery inference
for smaller physical error rates, even though it is trained on a ``wrong''
dataset. It is important to note that this experiment does not provide an
improved method for training a neural network for error correction on a physical
realization of a quantum processor. Firstly, in any manufactured quantum device
the error model will not be entirely known (and is not necessarily close to a
theoretic noise model such as the depolarizing channel). And secondly, the error
of the device cannot be intensified intentionally for the purpose of training a
deep neural decoder, to  be later used on a less noisy device.

\section{Conclusion}
\label{sec:Conclusion} 

To conclude, the main contributions of this paper were considering multiple
fault-tolerant schemes and using several neural network architectures to train
decoders in a full circuit-level noise framework. Although our analysis was
done for Pauli channels, we expect that for non-Pauli noise models, the
improvements could be even more significant than what was observed in our work.
Evidence of this can be found in \cite{CWBL16} where decoders were adapted to
non-Pauli noise channels.

From a machine learning point of view, we applied state-of-the-art techniques
used in training neural networks. While considering many network designs, we
used the same hyperparameter tuning methodology to achieve unbiased and reliable
results. Consequently, we successfully observed a clear advantage in using deep
networks in comparison with single hidden layer networks and regression methods.
On the other hand, we provided clear evidence of the realistic limitations of
deep learning in low noise rate regimes. In particular, scaling the neural
network to large distance codes appears to be a significant challenge. For large
scale quantum computations, decoders that work less well than neural decoders
trained on small distance codes but which are scalable would clearly be the
better option. Lastly, we gave a rigorous account of the digital hardware
resources needed for inference and runtime analysis of the critical path of the
customized digital circuitry for high performance inference.

There remains many interesting future directions for designing improved and
efficient decoders which work well in fault-tolerant regimes.  One such avenue
would be to tailor machine learning algorithms specifically designed for
decoding tasks. In particular, finding machine learning algorithms which work
well with sparse data would be of critical importance. It would also be
interesting to apply the methods introduced in this work to actual quantum
devices that are currently being developed. It most certainly will be the case
that fault-tolerant designs will be tailored to a particular quantum
architecture. This would lead to further areas in which machine learning could
be extremely useful for finding improved decoders.

\section{Acknowledgements} 
Both authors contributed equally to this work. We acknowledge Steve~G.~Weiss for
providing the necessary computing resources. The authors would also like to
thank Ben~Criger, Raymond~Laflamme, Thomas~O'Brien, Xiaotong~Ni, Barbara~Terhal,
Giacomo~Torlai, Tomas~Jochym-O'Connor, Aleksander~Kubica and Ehsan~Zahedinejad
for useful
discussions. C.C. acknowledges the support of NSERC through the PGS D
scholarship. P.R. acknowledges the support of the government of Ontario and
Innovation, Science and Economic Development Canada.
 
\newpage 

\bibliographystyle{ieeetr}
\bibliography{bibliography.bib}

\begin{table*}[t]
	\begin{tabular}{c|c|c|c}
		\codepar{7,1,3} Steane code &\codepar{9,1,3} (Surface-17) code & \codepar{19,1,5} color code &\codepar{25,1,5} (Surface-49) code   \\ \hline
		$g^{(x)}_{1} = X_{4}X_{5}X_{6}X_{7}$&$g^{(x)}_{1} = X_{1}X_{2}X_{4}X_{5}$&$g^{(x)}_{1} = X_{1}X_{2}X_{3}X_{4}$ & $g^{(x)}_{1} = X_{1}X_{2}X_{6}X_{7}$ \\
		$g^{(x)}_{2} = X_{2}X_{3}X_{6}X_{7}$&$g^{(x)}_{2} = X_{7}X_{8}$&$g^{(x)}_{2} = X_{1}X_{3}X_{5}X_{7}$ & $g^{(x)}_{2} = X_{11}X_{12}X_{16}X_{17}$ \\
		$g^{(x)}_{3} = X_{1}X_{3}X_{5}X_{7}$&$g^{(x)}_{3} = X_{2}X_{3}$&$g^{(x)}_{3} = X_{5}X_{7}X_{8}X_{11}X_{12}X_{13}$ & $g^{(x)}_{3} = X_{21}X_{22}$ \\
		$g^{(z)}_{1} = Z_{4}Z_{5}Z_{6}Z_{7}$&$g^{(x)}_{4} = X_{5}X_{6}X_{8}X_{9}$&$g^{(x)}_{4} = X_{1}X_{2}X_{5}X_{6}X_{8}X_{9}$ & $g^{(x)}_{4} = X_{2}X_{3}$ \\
		$g^{(z)}_{2} = Z_{2}Z_{3}Z_{6}Z_{7}$&$g^{(z)}_{1} = Z_{1}Z_{4}$&$g^{(x)}_{5} = X_{6}X_{9}X_{16}X_{19}$ & $g^{(x)}_{5} = X_{7}X_{8}X_{12}X_{13}$\\
		$g^{(z)}_{3} = Z_{1}Z_{3}Z_{5}Z_{7}$&$g^{(z)}_{2} = Z_{2}Z_{3}Z_{5}Z_{6}$&$g^{(x)}_{6} = X_{16}X_{17}X_{18}X_{19}$ & $g^{(x)}_{6} = X_{17}X_{18}X_{22}X_{23}$ \\
		&$g^{(z)}_{3} = Z_{4}Z_{5}Z_{7}Z_{8}$&$g^{(x)}_{7} = X_{8}X_{9}X_{10}X_{11}X_{16}X_{17}$ & $g^{(x)}_{7} = X_{3}X_{4}X_{8}X_{9}$ \\
		&$g^{(z)}_{4} = Z_{6}Z_{9}$&$g^{(x)}_{8} = X_{10}X_{11}X_{12}X_{15}$ & $g^{(x)}_{8} = X_{13}X_{14}X_{18}X_{19}$  \\ 
		&& $g^{(x)}_{9} = X_{12}X_{13}X_{14}X_{15}$ & $g^{(x)}_{9} = X_{23}X_{24}$ \\ 
		&& $g^{(z)}_{1} = Z_{1}Z_{2}Z_{3}Z_{4}$ & $g^{(x)}_{10} = X_{4}X_{5}$\\ 
		&& $g^{(z)}_{2} = Z_{1}Z_{3}Z_{5}Z_{7}$ & $g^{(x)}_{11} = X_{9}X_{10}X_{14}X_{15}$\\ 
		&& $g^{(z)}_{3} = Z_{5}Z_{7}Z_{8}Z_{11}Z_{12}Z_{13}$ & $g^{(x)}_{12} = X_{19}X_{20}X_{24}X_{25}$\\ 
		&& $g^{(z)}_{4} = Z_{1}Z_{2}Z_{5}Z_{6}Z_{8}Z_{9}$& $g^{(z)}_{1} = Z_{1}Z_{6}$\\ 
		&& $g^{(z)}_{5} = Z_{6}Z_{9}Z_{16}Z_{19}$& $g^{(z)}_{2} = Z_{2}Z_{3}Z_{7}Z_{8}$\\ 
		&& $g^{(z)}_{6} = Z_{16}Z_{17}Z_{18}Z_{19}$& $g^{(z)}_{3} = Z_{4}Z_{5}Z_{9}Z_{10}$ \\ 
		&& $g^{(z)}_{7} = Z_{8}Z_{9}Z_{10}Z_{11}Z_{16}Z_{17}$& $g^{(z)}_{4} = Z_{6}Z_{7}Z_{11}Z_{12}$\\ 
		&& $g^{(z)}_{8} = Z_{10}Z_{11}Z_{12}Z_{15}$& $g^{(z)}_{5} = Z_{8}Z_{9}Z_{13}Z_{14}$\\ 
		&& $g^{(z)}_{9} = Z_{12}Z_{13}Z_{14}Z_{15}$& $g^{(z)}_{6} = Z_{10}Z_{15}$\\ 
		&&&$g^{(z)}_{7} = Z_{11}Z_{16}$ \\ 
		&&&$g^{(z)}_{8} = Z_{12}Z_{13}Z_{17}Z_{18}$ \\ 
		&&&$g^{(z)}_{9} = Z_{14}Z_{15}Z_{19}Z_{20}$ \\
		&&&$g^{(z)}_{10} = Z_{16}Z_{17}Z_{21}Z_{22}$ \\  
		&&&$g^{(z)}_{11} = Z_{18}Z_{19}Z_{23}Z_{24}$ \\
		&&&$g^{(z)}_{12} = Z_{20}Z_{25}$ \\ \hline
		    
                $T^{(x)}_{1} = X_{4}$&$T^{(x)}_{1} = X_{1}$&$T^{(x)}_{1} = X_{4}$& $T^{(x)}_{1} = X_{1}$ \\
	        $T^{(x)}_{2} = X_{2}$&$T^{(x)}_{2} = X_{3}$&$T^{(x)}_{2} = X_{3}X_{4}$& $T^{(x)}_{2} = X_{3}$ \\
		$T^{(x)}_{3} = X_{1}$&$T^{(x)}_{3} = X_{8}$&$T^{(x)}_{3} = X_{13}X_{14}$& $T^{(x)}_{3} = X_{5}$ \\
		$T^{(z)}_{1} = Z_{3}Z_{7}$&$T^{(x)}_{4} = X_{9}$&$T^{(x)}_{4} = X_{5}X_{7}$& $T^{(x)}_{4} = X_{3}X_{7}$ \\
		$T^{(z)}_{2} = Z_{5}Z_{7}$&$T^{(z)}_{1} = Z_{4}$&$T^{(x)}_{5} = X_{18}X_{19}$& $T^{(x)}_{5} = X_{5}X_{9}$ \\
		$T^{(z)}_{3} = Z_{6}Z_{7}$&$T^{(z)}_{2} = Z_{7}$&$T^{(x)}_{6} = X_{6}X_{9}X_{17}$ & $T^{(x)}_{6} = X_{5}X_{10}$ \\
		&$T^{(z)}_{3} = Z_{2}Z_{4}$&$T^{(x)}_{7} = X_{6}X_{9}$ & $T^{(x)}_{7} = X_{3}X_{7}X_{11}$ \\
		&$T^{(z)}_{4} = Z_{6}$&$T^{(x)}_{8} = X_{6}X_{9}X_{10}$ & $T^{(x)}_{8} = X_{18}X_{24}$  \\ 
		&& $T^{(x)}_{9} = X_{6}X_{9}X_{10}X_{15}$ & $T^{(x)}_{9} = X_{5}X_{10}X_{15}$ \\
		&& $T^{(z)}_{1} = Z_{2}Z_{5}Z_{7}$ & $T^{(x)}_{10} = X_{17}X_{18}X_{24}$ \\ 
		&& $T^{(z)}_{2} = Z_{1}Z_{2}$& $T^{(x)}_{11} = X_{24}$ \\ 
		&& $T^{(z)}_{3} = Z_{13}Z_{14}$& $T^{(x)}_{12} = X_{5}X_{10}X_{15}X_{20}$ \\ 
		&& $T^{(z)}_{4} = Z_{5}Z_{7}$ & $T^{(z)}_{1} = Z_{6}$ \\ 
		&& $T^{(z)}_{5} = Z_{18}Z_{19}$ & $T^{(z)}_{2} = Z_{16}$\\ 
		&& $T^{(z)}_{6} = Z_{6}Z_{9}Z_{16}Z_{18}Z_{19}$& $T^{(z)}_{3} = Z_{21}$\\ 
		&& $T^{(z)}_{7} = Z_{6}Z_{9}$& $T^{(z)}_{4} = Z_{2}Z_{6}$ \\ 
		&& $T^{(z)}_{8} = Z_{5}Z_{7}Z_{8}Z_{11}$& $T^{(z)}_{5} = Z_{12}Z_{16}$ \\ 
		&& $T^{(x)}_{9} = Z_{6}Z_{9}Z_{11}Z_{12}$& $T^{(z)}_{6} = Z_{21}Z_{22}$ \\ 
		&&& $T^{(z)}_{7} = Z_{8}Z_{12}Z_{16}$ \\    
		&&& $T^{(z)}_{8} = Z_{19}Z_{25}$ \\ \  
		&&& $T^{(z)}_{9} = Z_{21}Z_{22}Z_{23}$ \\    
		&&& $T^{(z)}_{10} = Z_{4}Z_{8}Z_{12}Z_{16}$ \\    
		&&& $T^{(z)}_{11} = Z_{14}Z_{19}Z_{25}$ \\   
		&&& $T^{(z)}_{12} = Z_{25}$ \\ \hline  
		
		 $X_{L} = X^{\otimes 7}$,$Z_{L}= Z^{\otimes 7}$&$X_{L} =X_{3}X_{5}X_{7}$,$Z_{L} =Z_{1}Z_{5}Z_{9}$& $X_{L} = X^{\otimes 19}$,$Z_{L}= Z^{\otimes 19}$&$X_{L} =X_{5}X_{9}X_{13}X_{17}X_{21}$,$Z_{L} =Z_{1}Z_{7}Z_{13}Z_{19}Z_{25}$  \\ \hline 
		                 
	\end{tabular}
	\caption{Table containing a list of the stabilizer generators (second row), pure errors (third row) and logical operators (fourth row) for all the codes considered in this article.}
	\label{tab:CodeParameters}
\end{table*}

\end{document}